\documentclass[12pt]{article}

\usepackage[total={6.5in,9in},top=1in,headsep=0.1in,headheight=1in]{geometry}
\usepackage[utf8]{inputenc}
\usepackage{enumitem,amssymb}
\usepackage{ragged2e}
\usepackage{longtable}
\usepackage{graphicx}
\usepackage{capt-of}
\usepackage[normalem]{ulem}
\graphicspath{{Figures/}}
\usepackage{charter}
\usepackage{fullpage}
\usepackage[svgnames]{xcolor}

\usepackage[format=plain,labelfont=bf,font=scriptsize]{caption}
\usepackage{lipsum}
\usepackage[sort&compress,numbers]{natbib}
\usepackage{hyperref}
\hypersetup{
        colorlinks=true,
        citecolor=SteelBlue,
        filecolor=LimeGreen,
        linkcolor=SlateBlue,
        urlcolor=DarkOrchid
}

\usepackage{aas_macros}

\newcommand\snowmass{
\begin{center}
\end{center}
}
\newcommand\conveners{
\begin{center}
  \vskip 2.4in
  {\scshape Conveners \\}
\end{center}
}
\newcommand\reps{
\begin{center}
  \vskip 1.8in
  {\scshape Early Career Representatives \\ 
  }
  \vskip 0.3cm
  {\small Tiffany R. Lewis$^{10}$ and Kristi Engel$^{11,12}$}
  \vskip 0.25cm
  {\scriptsize $^{10}$ NASA Postdoctoral Program Fellow, Astroparticle Physics Lab, Goddard Space Flight Center, Greenbelt, MD 20771, USA}
  \vskip -0.15cm
  {\scriptsize $^{11}$ University of Maryland, College Park, College Park, MD, 20742, USA}
  \vskip -0.15cm
  {\scriptsize $^{12}$ Physics Division, Los Alamos National Laboratory, Los Alamos, NM, 87545, USA}\\
  \vskip 0.18in
  {\scshape  Contributors \\}
  \vskip 0.8cm
   \end{center}
 }

\usepackage[firstpage=true]{background}
\backgroundsetup{contents={\parbox{6.5in}{\snowmass\conveners\reps}}, scale=1,placement=top,opacity=1,color=black,position={3.25in,1.2in}}

\usepackage{fancyhdr}
\fancypagestyle{plain}{%
  \fancyhf{}%
  \fancyhead[C]{}
  \fancyfoot[C]{\thepage}
}

\title{Report of the Topical Group on Cosmic Probes of Fundamental Physics for for Snowmass 2021 
\vspace{1cm}}
\date{}

\usepackage{authblk}

\author[1]{Rana X. Adhikari}
\author[2,3,4]{Luis A. Anchordoqui}
\author[5]{Ke Fang}
\author[6,7,8]{B. S. Sathyaprakash}
\author[9]{Kirsten Tollefson}

\affil[1]{Division of Physics, Math, and Astronomy, California Institute of Technology, Pasadena, CA 91125, USA}
\affil[2]{Department of  Physics and Astronomy, Lehman College, City University of New York,  Bronx, NY 10468, USA}
\affil[3]{Department of Physics, Graduate Center, City University of New York, NY 10016, USA}
\affil[4]{Department of Astrophysics, American Museum of Natural History, NY 10024, USA}
\affil[5]{Department of Physics, Wisconsin IceCube Particle Astrophysics Center, University of Wisconsin, Madison, WI, 53706} 
\affil[6]{Institute for Gravitation and the Cosmos, Department of Physics, Pennsylvania State University, University Park, PA, 16802, USA}
\affil[7]{Department of Astronomy \& Astrophysics, Pennsylvania State University, University Park, PA, 16802, USA}
\affil[8]{School of Physics and Astronomy, Cardiff University, Cardiff, UK, CF24 3AA}
\affil[9]{Department of Physics \& Astronomy, Michigan State University, East Lansing, MI 48823, USA}

\begin{document}
\pagenumbering{roman}
\maketitle

\vspace{2.8cm}

\noindent {\small
Amin Aboubrahim (University of M\"{u}nster, Germany),
\"{O}zg\"{u}r Akarsu (Istanbul Technical University, Turkey),
Yashar Akrami (Case Western Reserve University and Imperial College London),
Roberto Aloisio (Gran Sasso Science Institute, Italy),
Rafael Alves Batista (Instituto de Física Teórica UAM-CSIC, Spain),
Mario Ballardini (University of Ferrara, Italy),
Stefan W. Ballmer (Syracuse University),
Ellen Bechtol (University of Wisconsin-Madison),
David Benisty (University of Cambridge, UK),
Emanuele Berti (Johns Hopkins University),
Simon Birrer (Stanford University),
Alexander Bonilla (Universidade Federal de Juiz de Fora, Brazil),
Richard Brito (Instituto Superior T\'ecnico, Portugal),
Mauricio Bustamante (Niels Bohr Institute, University of Copenhagen),
Robert Caldwell (Dartmouth College),
Vitor Cardoso (T\'ecnico Lisbon and Niels Bohr Institute, Denmark),
Sukanya Chakrabarti (University of Alabama, Huntsville),
Thomas Y. Chen (Columbia University),
Michele Cicoli (Universit\`a di Bologna, Italy),
Sebastien Clesse (Universit\'e Libre de Bruxelles, Belgium),
Alan Coleman (University of Delaware),
Yanou Cui (University of California, Riverside),
Giulia Cusin (Institut d'Astrophysique de Paris, France and University of Geneva, Switzerland),
Tansu Daylan (Princeton University),
Keith R.~Dienes (University of Arizona),
Eleonora Di Valentino (University of Sheffield, UK),
Cora Dvorkin (Harvard University), 
Celia Escamilla-Rivera (Universidad Nacional Aut\'{o}noma de M\'{e}xico, Mexico),
Glennys R. Farrar (New York University),
Jonathan L.~Feng (University of California, Irvine),
Noemi Frusciante (Universit\`a degli Studi  di Napoli  Italy),
Juan Garc\'ia-Bellido (Universidad Aut\'onoma de Madrid, Spain),
Carlos Garc{\'\i}a Canal (Universidad Nacional de La Plata, Argentina),
Maria~Vittoria~Garzelli (Universit\"at Hamburg, Germany),
Jonas Glombitza (Friedrich-Alexander-Universit\"{a}t Erlangen-N\"{u}rnberg, Germany),
Geraldina Golup (CONICET and Centro Atomico Bariloche, Argentina),
Maria Gritsevich (Finnish Geospatial Research Institute, Finland),
Zoltan Haiman (Columbia University),
Jaume Haro (Politechnic University of Catalonia, Spain),
Dhiraj Kumar Hazra (The Institute of Mathematical Sciences, India),
Alan Heavens (Imperial College London, UK),
Daniel Holz (University of Chicago),
J\"org R. H\"orandel (Radboud University, The Netherlands),
Mustapha Ishak (University of Texas at Dallas),
Mikhail M. Ivanov (Institute for Advanced Study, Princeton), 
Shahab Joudaki (University of Waterloo), 
Karl-Heinz Kampert (Wuppertal University),
Christopher M. Karwin (Clemson University),
Ryan Keeley (University of California Merced),
Michael Klasen (University of M\"{u}nster, Germany),
Rostislav Konoplich (Manhattan College and New York University),
John F. Krizmanic (Laboratory for Astroparticle Physics, NASA/GSFC),
Suresh Kumar (Indira Gandhi University Meerpur, India),
Noam Levi (Tel-Aviv University, Israel),
Benjamin L'Huillier (Sejong University, Korea),
Vuk Mandic (University of Minnesota Twin Cities),
Valerio Marra (Universidade Federal do Espírito Santo, Brazil),
C. J. A. P. Martins (CAUP, Porto, Portugal),
Sabino Matarrese (University of Padova, Italy),
Eric Mayotte (Colorado School of Mines),
Sonja Mayotte (Colorado School of Mines),
Laura Mersini-Houghton (UNC-Chapel Hill),
Joel Meyers (Southern Methodist University),
Andrew L. Miller (Université catholique de Louvain, Belgium),
Emil Mottola (University of New Mexico),
Suvodip Mukherjee (Tata Institute of Fundamental Research, India),
Kohta Murase (Pennsylvania State University),
Marco Stein Muzio (Pennsylvania State University),
Pran Nath (Northeastern University),
Ken~K.~Y.~Ng (Massachusetts Institute of Technology),
Jose Miguel No (Instituto de Física Teórica UAM-CSIC, Spain),
Rafael C. Nunes (Universidade Federal do Rio Grande do Sul, Brazil),
Angela V. Olinto (University of Chicago),
Francesco Pace (Universit\`a di Torino, Italy; 
Istituto Nazionale di Fisica Nucleare, Torino, Italy),
Supriya Pan (Presidency University, India),
Santiago E. Perez Bergliaffa (Universidade do Estado de Rio de Janeiro, Brazil), 
Levon Pogosian (Simon Fraser University, Canada),
Jocelyn Read (California State University Fullerton),
Maximilian Reininghaus (Karlsruher Institut für Technologie, Germany),
Mary Hall Reno (University of Iowa),
Adam G. Riess (Johns Hopkins University, USA),
Mairi Sakellariadou (King's College, London, UK),
Alexander S. Sakharov (Manhattan College and CERN),
Paolo Salucci (SISSA, INFN-TS),
Marcos Santander (University of Alabama),
Eva Santos (Institute of Physics of the Czech Academy of Sciences, Czech Republic),
Fred Sarazin (Colorado School of Mines),
Emmanuel N. Saridakis (National Observatory of Athens, Greece),
Sergio J. Sciutto (Universidad Nacional de La Plata, Argentina),
Arman Shafieloo (Korea Astronomy and Space Science Institute, Korea),
David H. Shoemaker (Kavli Institute, Massachusetts Institute of Technology),
Kuver Sinha (University of Oklahoma),
Dennis Soldin (University of Delaware),
Jorge F. Soriano (Lehman College, City University of New York),
Denitsa Staicova (Institute for Nuclear Research and Nuclear Energy, Bulgaria),
Ling Sun (The Australian National University, Australia),
D.~A.~Steer (APC, Universit\'e Paris Cit\'e, France),
Brooks Thomas (Lafayette College),
John A. Tomsick (UC Berkeley),
V{\'\i}ctor B. Valera (University of Copenhagen, Denmark),
J. Alberto Vazquez (Universidad Nacional Aut\'{o}noma de M\'{e}xico, Mexico),
Tonia~M.~Venters (Astrophysics Science Division, NASA Goddard Space Flight Center), 
Luca Visinelli (Tsung-Dao Lee Institute \& Shanghai Jiao Tong University, P.R.~China),
Scott Watson (Syracuse University),
John K. Webb (Clare Hall, Cambridge University, UK),
Amanda Weltman (University of Cape Town, South Africa),
Graham White (IPMU),
Stephanie Wissel (Pennsylvania State University),
Anil Kumar Yadav (United College of Engineering and Research, India),
Fengwei Yang (University of Utah),
Weiqiang Yang (Liaoning Normal University, Dalian, P.R.~China),
Nicol\'as Yunes (University of Illinois Urbana-Champaign),
Alexey Yushkov (Institute of Physics of the Czech Academy of Sciences, Czech Republic),
Haocheng Zhang (NASA Goddard Space Flight Center)}

\clearpage

\section*{Executive Summary}

Cosmic Probes of Fundamental Physics take two primary forms:  Very high energy particles and gravitational waves (GWs).  Already today, these probes give access to fundamental physics not available by any other means, helping elucidate the underlying theory that completes the Standard Model~\cite{ParticleDataGroup:2020ssz}. The last decade has witnessed a revolution of exciting discoveries such as the detection of high-energy neutrinos~\cite{IceCube:2013low, IceCube:2014stg}  and  gravitational waves~\cite{LIGOScientific:2016aoc}. The scope for major developments in the next decades is dramatic, as detailed in this report. For example, precise measurements of the cosmic microwave background (CMB)~\cite{Planck:2018vyg} and large scale structure hint at complications in the concordance model of cosmology, which will be subjected to independent clarification within a few years thanks to the new cosmic probe of gravitational waves~\cite{Abdalla:2022yfr}.  Another cosmic probe still in the incubation stage is the discovery and exploration of the cosmic neutrino background. 

The very high energy particles we exploit include cosmic rays, gamma rays, and neutrinos.  Their energies enable study of particle collisions at energies far above those accessible with laboratory measurements, and their enormous propagation distances enable extremely sensitive constraints to be placed on fundamental physics, including Grand Unified Theories (GUT)  and Planck-scale phenomena such as Lorentz Invariance Violation and GUT-scale dark matter~\cite{Coleman:2022abf,Ackermann:2022rqc, Engel:2022bgx, Arguelles:2022xxa, Abraham:2022jse}.  Ultra-high-energy cosmic rays (UHECRs) are observed with energies above $10^{11}~{\rm GeV}$~\cite{ParticleDataGroup:2020ssz}.  When a $10^{11}~{\rm GeV}$ nucleus collides with an air nucleus in the upper atmosphere, the total center-of-mass (CM) energy is $1,700~{\rm TeV}$ -- nearly twice that in Pb-Pb collisions at the Large Hadron Collider (LHC).  Of order 10\% of these UHECRs may be protons and, as techniques to constrain the nature of individual UHECRs improve and proton-induced collisions can be separately identified, $p$N collisions at $1,700~{\rm TeV}$ can be isolated.  This is 120~TeV in the $p$-nucleon CM even without collective effects, and a host of new phenomena can be studied in the ultra-high-energy air showers, including black hole production, quark gluon plasma, and production of new long-lived heavy particles~\cite{Coleman:2022abf}.  The  already-established anomalous muon production in UHECR air showers (which has eluded explanation in models tuned to LHC data)~\cite{PierreAuger:2016nfk} guarantees that discoveries will be made.  

In addition to studying hadron physics at ultra-high energies in UHECR air showers, experiments such as AugerPrime~\cite{PierreAuger:2016qzd}, GCOS~\cite{Horandel:2021prj} and POEMMA~\cite{POEMMA:2020ykm} can discriminate between nucleus-, photon- and neutrino-induced showers.  Observations of photons and neutrinos at such energies enable unique probes of new physics including instanton-induced decay of super-heavy relics from the Big Bang~\cite{PierreAuger:2022wzk,PierreAuger:2022ibr}, cosmic strings~\cite{Berezinsky:2011cp}, Lorentz Invariance Violation~\cite{PierreAuger:2021tog}, and  axion-photon conversion in large-scale magnetic fields  fields~\cite{Csaki:2003ef,Gorbunov:2001gc}. 

Neutrinos and gamma rays produced when ultra-high-energy cosmic rays interact with ambient gas and thermal photon backgrounds in their source environment also provide invaluable tests of fundamental symmetries~\cite{Engel:2022bgx, Ackermann:2022rqc, Arguelles:2022xxa, Abraham:2022jse}. (The boost factor of a neutrino of energy $10^7~{\rm GeV}$ is five orders of magnitude higher than has ever been observed for a proton primary!) Furthermore, cosmic neutrinos and gamma rays provide a unique indirect probe of particle dark matter~\cite{Engel:2022bgx, Ackermann:2022rqc, Coleman:2022abf}.  Next-generation gamma-ray telescopes such as the Southern Wide-field Gamma-ray Observatory (SWGO)~\cite{Albert:2019afb, SWGOPBH}, the Cherenkov Telescope Array (CTA)~\cite{CTAConsortium:2017dvg}, and the All-sky Medium-Energy Gamma-ray Observatory (AMEGO)~\cite{2020SPIE11444E..31K} foresee significant improvements in sensitivity, effective area, and field of view. The
IceCube-Upgrade \cite{Ishihara:2019aao} and other upcoming cosmic neutrino experiments at GeV energies will provide neutrino oscillation sensitivity complementary to long baseline experiments. Future PeV-EeV neutrino experiments (as summarized in Figure~\ref{fig:scales}) will advance neutrino physics at energies beyond the reach of colliders by measuring the properties of standard model particles and their interactions, such as neutrino-nucleon cross sections. Observations of neutrino flavors and neutrino-antineutrino ratios have the potential to probe beyond-standard-model (BSM) neutrino physics, such as interactions with sterile neutrinos and unknown electrically neutral mediators.  

Cosmic particle physics has an important synergy with accelerator-based particle physics. For instance, forward particle production plays a crucial role in astroparticle physics~\cite{Adhikari:LoI}, which will thereby benefit enormously from the far-forward experiments at the high-luminosity LHC to be studied at the Forward Physics Facility~\cite{Anchordoqui:2021ghd,Feng:2022inv}. Measurements of forward  neutrinos will provide critical information to understand the anomalous muon production observed in UHECR air showers, while multi-faceted measurements of the properties of UHECR air showers -- as are now possible with state-of-the-art UHECR observatories -- will  reciprocally inform theories of forward hadron production~\cite{Anchordoqui:2022fpn}.  Another example is constraints on forward charm production using LHC neutrinos, being a key input for current and upcoming generations of large-scale neutrino telescopes.

Complementing very high energy particles as probes of new physics, we have Gravitational Waves.  These can be measured in several frequency bands with few-to-tens of kilometer scale interferometers (LIGO/Virgo and future terrestrian GW observatories), pulsar timing arrays (PTAs), and future space-based interferometers such as the European Space Agency's Laser Interferometer Space Antenna (LISA, NASA) to be launched around 2037~\cite{Ballmer:2022uxx}.   These different techniques allow observation of mergers between neutron stars and black holes in the mass range of less than 1 to $\sim$ 3,000 solar masses (terrestrial GW observatories), the merger of $10^4$--$10^7$ solar mass supermassive black holes (LISA) and  the stochastic gravitational-wave background produced by a population of supermassive black holes (PTAs). 

Future gravitational-wave observatories \cite{Evans:2021gyd, Punturo:2010zz} will measure the dense matter equation of state with exquisite precision \cite{Evans:2021gyd, Kashyap:2022wzr, Bogdanov:2022faf}, probe the QCD phase transition in neutron star cores \cite{Prakash:2021wpz}, explore dark matter in astrophysical environments \cite{Berti:2022wzk, Brito:2022lmd} and potentially discover primordial black holes in the dark ages \cite{Brito:2022lmd, Ng:2021sqn}, begin a new era in precision cosmology \cite{Abdalla:2022yfr}, open a new window for probing extreme gravitational phenomena in the early Universe~\cite{Caldwell:2022qsj, Berti:2022wzk, Foucart:2022iwu, Asadi:2022njl, Achucarro:2022qrl}, and have arguably unprecedented discovery potential \cite{Evans:2021gyd}. Already, LIGO and Virgo limits on the difference in the speed of light and gravitational waves from a single, well-measured, binary neutron star merger have dramatically reduced the model-space for theories of modified gravity.  The  measured tidal deformability in this same merger implies that neutron stars of $\sim  1.5$ solar masses have surprisingly similar radii to neutron stars of $\gtrsim 2 M_\odot$ measured by NICER, with important implications for quark matter~\cite{Raaijmakers:2019dks}.  The large number of binary neutron star and neutron star-black hole mergers to be measured by LIGO/Virgo and future GW observatories will enable the exploration of transition between baryon and quark degrees of freedom and the quark matter equation of state to be mapped out in detail, confronting theoretical physics with data on non-perturbative QCD phenomena in an entirely new regime.  

GW observatories will also strongly constrain cosmology, most immediately by offering a clean, largely systematic-free measurement of Hubble constant $H_0$ that should definitively settle the question of whether the expansion rate of the universe today is the same as or different from the value obtained from CMB measurements using $\Lambda$CDM.  Furthermore, Cosmic Explorer and Einstein Telescope will enable structure and the black hole mass spectrum to be measured out to $z\approx 50$ and beyond, opening a completely new regime of testing $\Lambda$CDM and determining if there are primordial black holes.   

The dawn of a new Astrophysical Multimessenger Era has been heralded by the recent co-detection of gamma rays and gravitational waves in a binary neutron star merger~\cite{LIGOScientific:2017ync}, the co-detection of gamma rays and neutrinos in a blazar flare~\cite{IceCube:2018dnn} and recent examples of neutrinos consistent with production in tidal disruption events~\cite{2021NatAs...5..510S,2022PhRvL.128v1101R}.  Over the next decade, simultaneous observations  with different techniques promise to reveal where these extreme-energy cosmic messengers come from, and how they came to be~\cite{Engel:2022yig}.  Maximal exploitation of our cosmic probes will require a level of programmatic planning for complementarity between facilities with distinct goals, not previously attempted. The United States is well poised to lead this endeavor, through investment in facilities as well as the communities of scientists and specialists that build, maintain, and utilize them, but coordination between agencies will be indispensable to realizing this potential. 

\clearpage

\tableofcontents

\clearpage  

\pagenumbering{arabic}


\section{The Big Questions and Goals for the Next Decade}
\label{s:questions}

\noindent
\begin{minipage}{0.42\linewidth}
$\qquad$The seventh Cosmic Frontier (CF7), named Cosmic Probes of Fundamental Physics, was asked to  summarize current knowledge and identify future opportunities (both experimental and theoretical) in the use of  astrophysical and cosmological probes of fundamental physics. As a result of the breadth of this area of research, CF7 has been subdivided into five main topical areas:  {\it (i)}~History of the Universe and Cosmology; {\it (ii)}~Cosmic Probes of Dark Matter; {\it (iii)}~Astroparticle Physics; {\it (iv)}~Multimessenger Synergies in Particle Astrophysics; and {\it (v)}~Architecture of Spacetime.  All of these areas are aligned with the primary goals of High Energy Physics in general, and the APS Division of Particles and Fields (DPF) in particular.~~~
\end{minipage}
\hfill
\begin{minipage}[r]{0.56\linewidth}
    \centering
    \captionsetup{type=figure}
    \includegraphics[width=0.97\columnwidth]{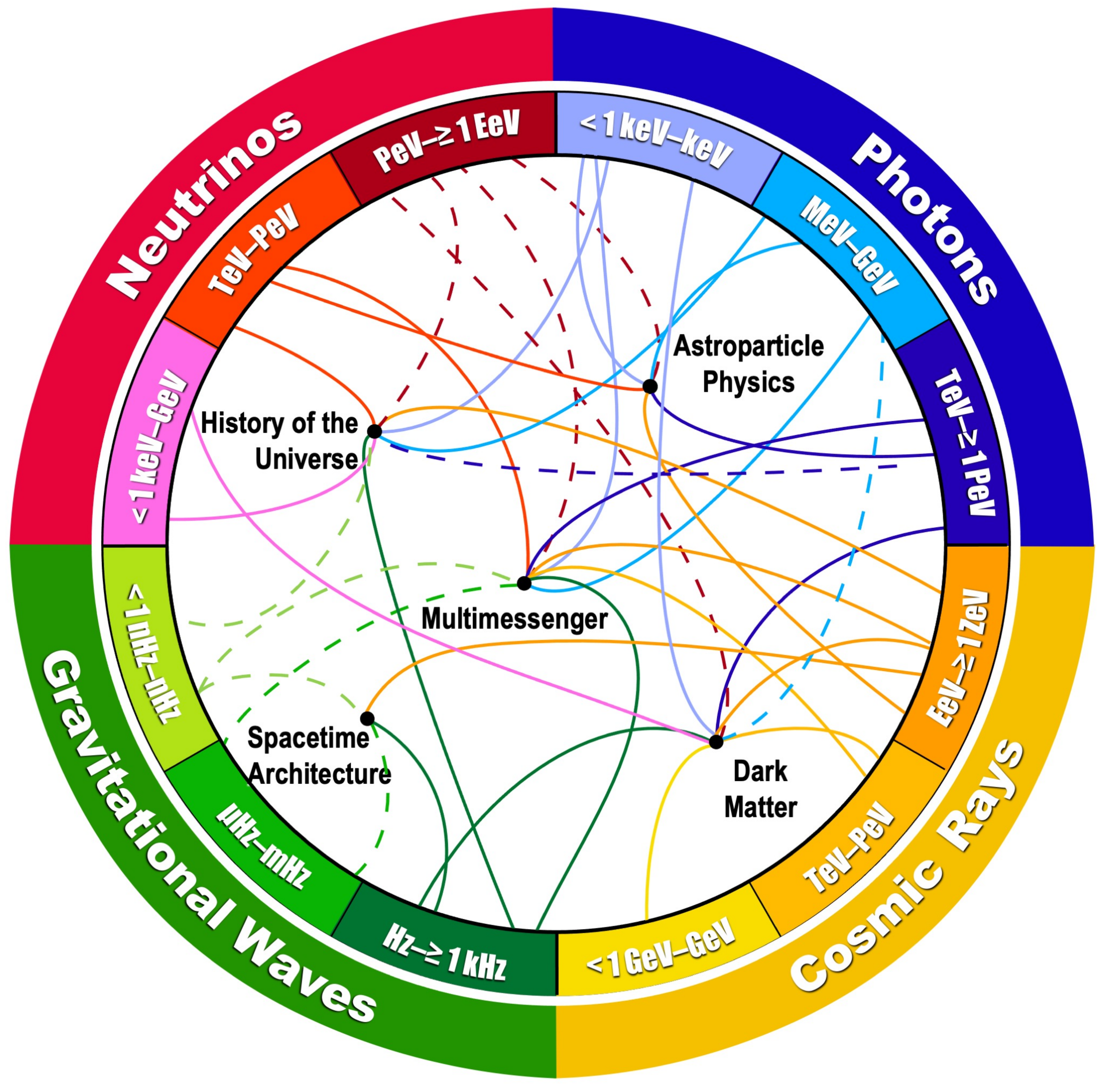}
    \vskip-5pt
  \captionof{figure}{Connections between messengers and fundamental physics topics. Current and future multimessenger landscapes are indicated by solid and dashed curves, respectively.}
    \vskip+5pt
\label{fig:summary}
\end{minipage}


We received 12 White Papers addressing the current challenges and future opportunities in each of these fields as they relate to the other frontiers of High Energy Physics. We have identified 27 big picture questions that will shape the course of discovery of CF7 in the coming decade.  The report is organized as follows. In 
this section, we lay out the big questions and goals for the next decade. In Secs.~\ref{s:cosmo}, \ref{s:DM}, \ref{s:astropart}, \ref{s:multimessenger}, and \ref{s:spacetime}, we introduce the theoretical inputs needed for addressing these questions, with each subsection corresponding to a question in Sec.~\ref{s:questions}. In Sec.~\ref{s:experiments}, we go through a description of existing and future experiments where a U.S. contribution is sought. In Sec.~\ref{s:opportunities}, we  identify important opportunities for complementarity with other frontiers. Finally, in Sec.~\ref{s:DEIA}  we explore some challenges and limitations that professionals experience in their daily working practice to identify strategies for expanding diversity, equity, inclusion, and accessibility. A summary of the topics and their relevance to various messengers is shown in Figure~\ref{fig:summary}.

Over the next decade, the physics community is poised to answer longstanding and profound questions about the nature of the Universe and its elementary components. This section highlights the big science questions in fundamental physics that can be addressed by \emph{cosmic probes} over the next 10 years.

\subsection{History of the Universe and Cosmology}

     \begin{itemize}
       
        \item Is the Hubble constant measured with low redshift probes different from the value inferred with $\Lambda$CDM normalized to the cosmic microwave background data? 
        
        \item Is the Hubble tension a footprint of physics beyond the Standard Model? 
        
        \item What is the absolute sum of neutrino mass? (given the lower limit of 0.06~eV from oscillations) Is the hierarchy  normal or inverted?
    
        \item What are the imprints of early Universe phase transitions and inflation in the stochastic gravitational-wave backgrounds?
   
        \item What role do ultra-high-energy cosmic rays and advances in constraint-based modelling of Grand Unified Theories play in  early Universe model building?

\end{itemize}

\subsection{Cosmic Probes of Dark Matter}

 \begin{itemize}       
        \item Is there a portal connecting the dark and visible sectors?
    
       \item What fraction of dark matter is held in primordial black holes? Are there currently evaporating primordial black holes? 

       \item Does the dark sector consist of a vast ensemble of particle species whose decay widths are balanced against their cosmological abundances?
       
       \item What is the gravitational-wave signature of dark matter?
       
       \item What are the gravitational-wave signatures of dilute dark matter distributions? 
       
\end{itemize}
\subsection{Astroparticle Physics} 

 \begin{itemize}        
        \item What are the properties of Standard Model particles and their interactions beyond the reach of terrestrial accelerators? \label{item:particleProperty} 
    
        \item How do neutrino flavors mix at high energies? Are neutrinos stable? Are there hidden neutrino interactions with cosmic backgrounds?
        
        \item Could an enhancement of strangeness production in hadronic collisions be the carrier of the observed muon deficit in air-shower simulations when compared to ultra-high-energy cosmic-ray data? Alternatively, do new particles and interactions exist at the highest energies?
    
        \item How does matter behave in the center of neutron stars? What are the physical properties of matter at ultra-high density, large proton/neutron number asymmetry, and low temperature?

        \item  Do the Lorentz and CPT symmetries that underpin the Standard Model break down in extreme cosmic environments?
        
         \item Does the QED domain (extreme magnetic fields) produce exotic particles or dark matter? 
        
\end{itemize}

 \subsection{Multimessenger Synergies in Particle Astrophysics}
      
      \begin{itemize} 
       \item How are particles accelerated in the cosmos to ultra-high energies?  Is the cosmic ray maximum energy a fingerprint of physics beyond the Standard Model?
       
       \item What role do hadrons play in the extreme-energy Universe? 
       
       \item How does diffuse emission from different messengers and energies contribute to cosmic evolution?

        \item How are Galactic TeVatrons and PeVatrons produced? Are gamma-ray halos a signal of physics beyond the Standard Model?

        \item How are heavy elements formed?
        
    \end{itemize}

\subsection{Architecture of Spacetime}

 \begin{itemize}
    \item What are the true degrees of freedom in gravitational-wave polarizations, how are gravitational waves produced and how do they propagate?
    
    \item Is there a modification of General Relativity that successfully takes into account the effects ascribed to dark matter and dark energy?
    
    \item Does the graviton have a mass, what is the speed of gravity, and is local Lorentz invariance a fundamental symmetry of nature? 

    \item Does General Relativity apply to electromagnetic and gravitational wave signals from dynamic black hole environments without modification?
    
    \item What are the ``ab initio'' models of nonsingular, horizonless alternatives to black holes, and self-consistent predictions of the ringdown spectra and echo signal they might produce?
    
    \item What is the space of low energy Effective Field Theories that admit an UV completion? What are the phenomenological implications of the Swampland conjectures for the topics discussed in this report?

\end{itemize}

\section{History of the Universe and Cosmology}
\label{s:cosmo}

The Universe is composed primarily of matter and energy we do not understand. Dark energy makes up about 70\% of the universe and we see its effects in the acceleration of the expansion of the universe over time, especially through measurements of high-redshift supernovae, anisotropies in the Cosmic Microwave Background, and the sub-critical density of large scale structure--- the distribution of galaxies and galaxy clusters. The past decade has seen renewed recognition for discoveries in dark energy's effects on the formation of the large-scale structure. Over the next decade, measurements of the Hubble constant, the cosmic microwave background, supernovae, and large-scale structure, especially in the ultraviolet band, will challenge the new gravity theories, the concordance model of cosmology, and even the Standard Model. These cosmic probes could provide complementary information about the unification of  fundamental forces.
    
\subsection{The Hubble Tension} 
\label{s:H0tension}

\begin{figure}[!ht]
\centering
  \includegraphics[width=1.00\columnwidth]{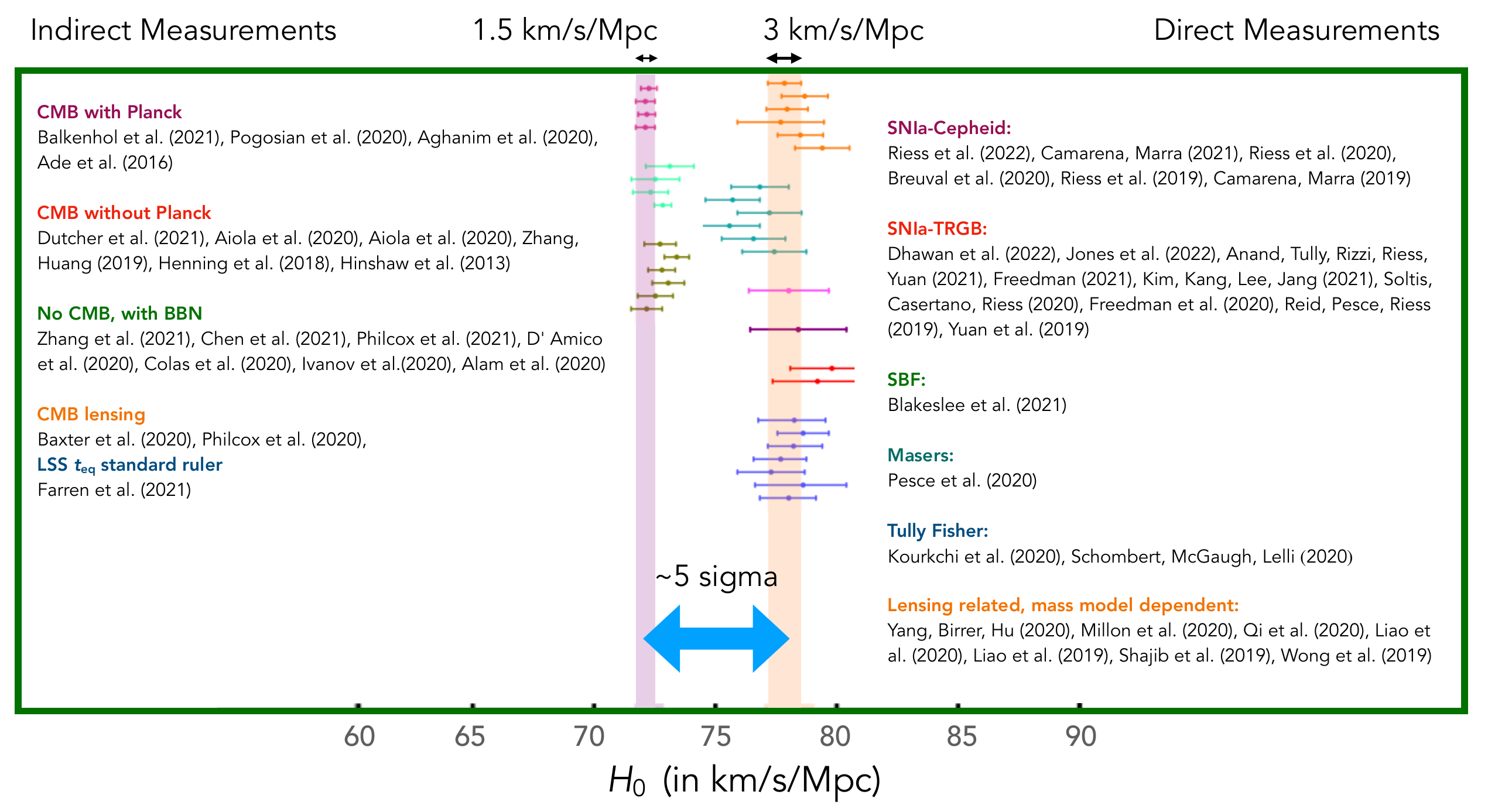}
   \caption{68\% CL constraints on $H_0$ from different cosmological probes. (Adapted from \cite{Abdalla:2022yfr}).}
\label{fig:whisker_H0}
\end{figure}

The $\Lambda$CDM model, in which the expansion of the Universe today is dominated by the cosmological constant $\Lambda$ and cold dark matter (CDM), is the simplest model that provides a reasonably good account of all  astrophysical and cosmological observations~\cite{ParticleDataGroup:2020ssz}. However, over the last decade, various discrepancies have emerged. In particular, local measurements of the Hubble constant $H_0 = 100 h {\rm \,km\,s^{-1}\,Mpc^{-1}}$ are increasingly in tension with the value inferred from a $\Lambda$CDM fit to the cosmic microwave background (CMB) and baryon acoustic oscillation (BAO) data~\cite{Abdalla:2022yfr}. Throughout, we refer to the {\it Hubble tension} as the $5.0\sigma$  disagreement  between the value inferred by the {\it Planck} Collaboration, $H_0=\left(67.27\pm0.60\right){\rm \,km\,s^{-1}\,Mpc^{-1}}$ at 68\% CL~\cite{Planck:2018vyg}, and the latest 2021 SH0ES Collaboration distance ladder constraint based on Type Ia supernovae (SNIa) calibrated by Cepheids, $H_0=(73.04 \pm 1.04){\rm \,km\,s^{-1}\,Mpc^{-1}}$ at 68\% CL~\cite{Riess:2021jrx}. However, these are not the only $H_0$ measurements--- there are actually two sets of measurements. Remarkably, all of the indirect model-dependent estimates at early times agree between themselves (such as those inferred from CMB and BAO experiments) and a similar agreement is reached by all of the direct late-time $\Lambda$CDM-independent measurements (such as distance ladders and strong lensing). Besides, an independent determination of $H_0$, based on the calibration of SNIa using the Tip of the Red Giant Branch, leads to $H_0=(72.4\pm2.0){\rm \,km\,s^{-1}\,Mpc^{-1}}$~\cite{Yuan:2019npk} and  $H_0=(69.6 \pm 0.8\,({\rm stat}) \pm 1.7\,({\rm sys})){\rm \,km\,s^{-1}\,Mpc^{-1}}$~\cite{Freedman:2020dne}, both at 68\%~CL. A measurement that is independent of SNIa, based geometric distance measurements to megamaser-hosting galaxies gives $H_0 = (73.9 \pm 3.0)~{\rm \,km\,s^{-1}\,Mpc^{-1}}$~\cite{Pesce:2020xfe}. A collection of $H_0$ measurements is shown in Fig.~\ref{fig:whisker_H0}.

\begin{figure*}
\centering
\includegraphics[width=0.8\textwidth]{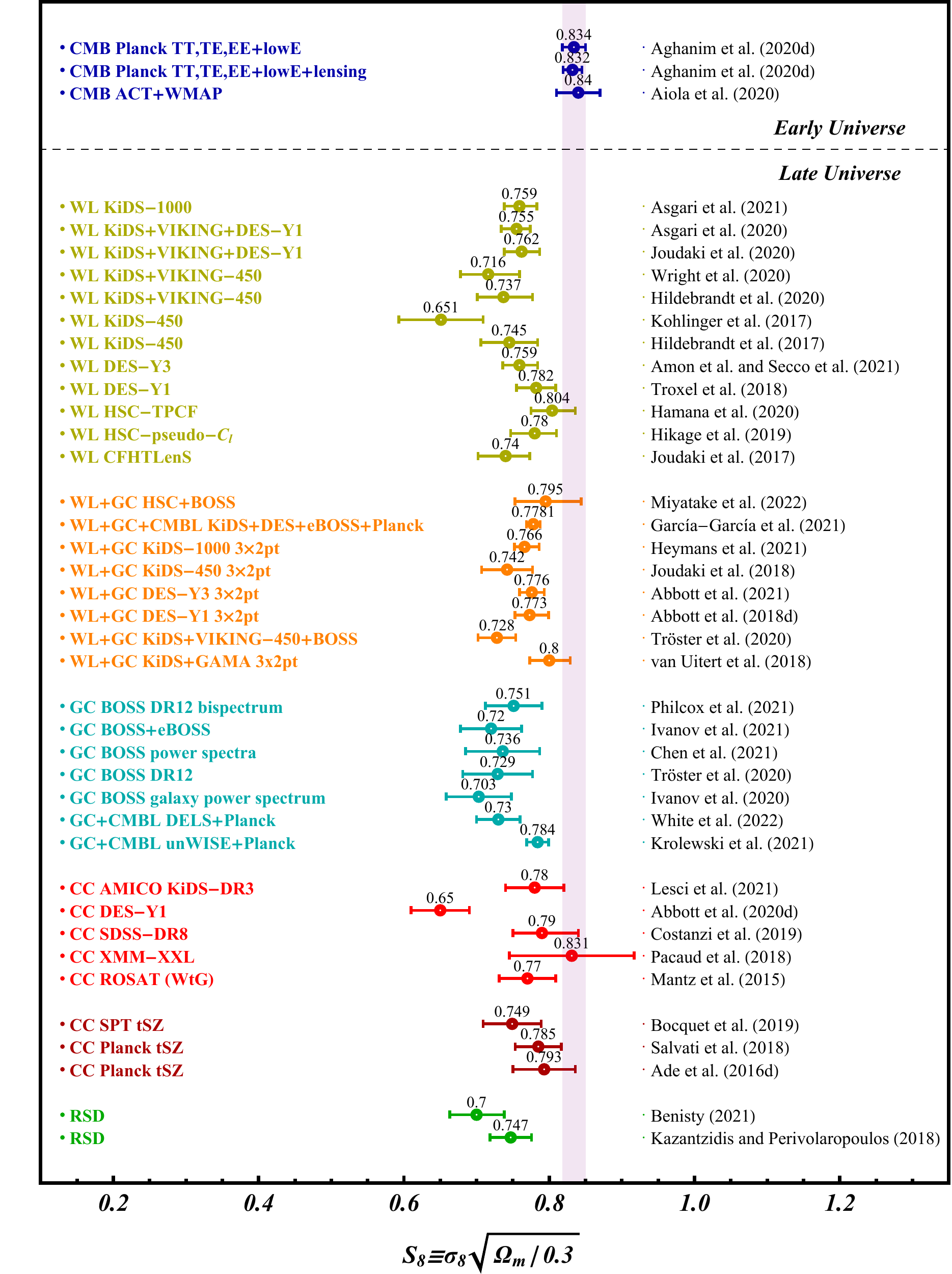}
\caption{Constraints on $S_8$ and its corresponding 68\% error. We show the nominal reported values by each study, which may differ in their definition of the constraints. 
The definition $S_8= \sigma_8 (\Omega_{\rm m}/0.3)^\alpha$ with $\alpha=1/2$ has been uniformly used for all points. In those cases where $\alpha \neq 1/2$ has been used in some references, the value of $S_8$ with $\alpha =1/2$ was recalculated (along with the uncertainties) using the constraints on $\sigma_8$ and $\Omega_{\rm m}$ shown in those references, assuming their errors are Gaussian. This concerns only 5 CC points where the published value of $\alpha$ was different from $1/2$ and the difference from the published $S_8$ (with different $\alpha$) is very small. The rest of the points are taken directly from the published values. Taken from Ref.~\cite{Abdalla:2022yfr}.}
\label{fig:whisker_S8}
\end{figure*}

Another seemingly different, but perhaps closely related, subject is the evidence of a growing tension between the Planck-preferred value and the local determination of $\sigma_8$, which gauges the amplitude of mass-density fluctuations when smoothed with a top-hat filter of radius $8h^{-1}~{\rm Mpc}$. More concretely, it is the combination $S_8 = \sigma_8 (\Omega_m/0.3)^{1/2}$ that is constrained by large-scale structure data, where $\Omega_m$ is the present-day value of the non-relativistic matter density parameter. On the assumption of $\Lambda$CDM, the Planck Collaboration reported $S_8 = 0.830 \pm 0.013$~\cite{Planck:2018vyg}, which is in $3\sigma$ tension with the result reported by KiDS-1000: $S_8 = 0.766^{+0.020}_{-0.014}$~\cite{KiDS:2020suj}. A collection of $S_8$ measurements is shown in Fig.~\ref{fig:whisker_S8}.

The discrepancy in the value of $H_0$ inferred from model-independent and -dependent experiments (each sensitive to different physics and systematic errors) might be a hint that the standard $\Lambda$CDM model needs to be modified. However,  more data are needed before we have a final verdict. An important role in reducing systematics of $H_0$ measurements will be played by gravitational-wave (GW) standard sirens (GWSS), the GW analog of astronomical standard candles~\cite{Abdalla:2022yfr, Berti:2022wzk, Engel:2022yig}. The amplitude of GWs is inversely proportional to the luminosity distance from the source, hence they can be used in conjunction with redshift information of the source location to probe the distance-redshift relation~\cite{Schutz:1986gp, Holz:2005df, Dalal:2006qt, Sathyaprakash:2009xt}. Observations of the binary neutron star merger GW170817, along with the redshift from its host galaxy (identified from the observation of an electromagnetic counterpart~\cite{LIGOScientific:2017ync}), yield $H_0=70_{-\,8}^{+12}{\rm km\,s^{-1}\,Mpc^{-1}}$ at 68\% CL~\cite{LIGOScientific:2017adf}. Despite the fact that the measurement has large uncertainties, it does not require any cosmic ``distance ladder'' and it is model-independent--- the absolute luminosity distance is directly calibrated by the theory of general relativity~\cite{Schutz:1986gp}. In other words, these GWSS are an ideal independent probe to weigh in on the Hubble tension.  Around 50 additional observations of GWSS with electromagnetic counterparts would be needed to measure $H_0$ with a precision of $1$--$2\%$~\cite{Nissanke:2013fka, Chen:2017rfc, Mortlock:2018azx}. Complementary dark GWSS (GW sources without EM counterparts)~\cite{LIGOScientific:2018gmd, DES:2019ccw, Palmese:2019ehe, DES:2020nay, LIGOScientific:2021aug} are expected to provide percent-level uncertainty on $H_0$ after combining a few hundreds to thousands of events using the statistical host identification technique~\cite{Gray:2019ksv} or by identifying the host galaxy of the nearby sources~\cite{Borhanian:2020vyr}. Improved measurements of the Hubble constant will also come from future CMB experiments~\cite{Chang:2022tzj} (including the Simon Observatory~\cite{SimonsObservatory:2018koc} and CMB-S4~\cite{Abazajian:2019eic, Abazajian:2022nyh}) which, combined with gigantic cosmic surveys (such as Euclid~\cite{EUCLID:2020syl} and Rubin~\cite{Blum:2022dxi}), are expected to measure $H_0$ with an uncertainty of about 0.15\%~\cite{DiValentino:2020zio}. A thorough discussion of probes that will help reducing uncertainties in $H_0$, $\sigma_8$, and $S_8$, as well as other anomalies of lower statistical significance (see e.g.~\cite{Webb:2022mrw}) is given in Ref.~\cite{Abdalla:2022yfr}.

\subsection{Model Building a Breakout from the Hubble Tension}

Models addressing the $H_0$ tension are extremely difficult to concoct~\cite{Abdalla:2022yfr}. One promising class of models involves a boost in the expansion rate close to the epoch of matter-radiation equality to reduce the size of the baryon-photon sound horizon $r_d$ at recombination and increase the Hubble rate inferred from the CMB. Extra relativistic degrees of freedom at recombination (scalars, Weyl fermions, and/or vector particles~\cite{Anchordoqui:2011nh, Weinberg:2013kea, Baumann:2016wac, Brust:2013ova}) parametrized by the number of equivalent light neutrino species $N_{\rm eff}$~\cite{Steigman:1977kc} is one such possibility. For three families of massless (SM) neutrinos, $N_{\rm eff}^{\rm SM} \simeq 3.044$~\cite{Mangano:2005cc, Bennett:2020zkv}, and so the contribution of extra light relics to the cosmological energy density is usually expressed as $\Delta N_{\rm eff} = N_{\rm eff} - N_{\rm eff}^{\rm MS}$. Current data are only sensitive enough to detect additional relics that froze out after the quark-hadron transition, hence  CMB-S4's ability to probe times well before that transition is a major advance~\cite{Dvorkin:2022bsc}. More concretely, CMB-S4 will constrain $\Delta N_{\rm eff} <0.06$ at 95\% C.L., achieving sensitivity to Weyl fermion and vector particles that froze out at temperatures a few hundred times higher than that of the QCD phase transition~\cite{Abazajian:2022nyh}. Another promising way to decrease $r_d$ is to include what is traditionally called ``early dark energy''~\cite{Poulin:2018cxd}. This type of models posit an additional energy density that briefly bumps up the expansion rate between the epoch of matter-radiation equality and recombination.

Deviations from $\Lambda$CDM that only affect pre-recombination physics have become more tightly constrained as the CMB data improve~\cite{Jedamzik:2020zmd, Lin:2021sfs}. Modifications of the late-time Universe have also been proposed~\cite{Salvatelli:2014zta, Berezhiani:2015yta, DiValentino:2017iww, DiValentino:2017rcr, Vattis:2019efj, Agrawal:2019dlm}. The basic idea behind this class of models is also simple: the matter-dark energy equality is shifted to earlier times than it otherwise would in $\Lambda$CDM to obtain a larger value of $H_0$. The challenge for this class of models is to increase $H(z)$ as $z \to 0$ while keeping a redshift-distance relation that is compatible with that inferred from the distance ladder~\cite{Benevento:2020fev, Alestas:2021luu}. This is because calibrated type Ia supernovae fundamentally tell us about their luminosity distances from us, which depends on the integrated expansion history and not just on $H_0$. 

All in all, a plethora of new ideas have been put forward to ameliorate the $H_0$ tension~\cite{Abdalla:2022yfr, DiValentino:2021izs}, but as yet, none of the extant new physics models on this front have done so to a satisfactory degree~\cite{Schoneberg:2021qvd}. The resolution of this conundrum will likely require a coordinated effort from the side of theory and phenomenology  (to construct model-independent consistency tests~\cite{Bernal:2021yli}), as well as data analysis and observation (to improve computational methods that could disentangle systematics). 

\subsection{Inferring the Neutrino Mass from Cosmological Probes}

The total neutrino mass can be measured with the cosmological data thanks to the cosmic neutrino background created at early times and the growth of structures at late times (see~\cite{Lesgourgues:2012uu,Lattanzi:2017ubx,Green:2021gdc}).
The main cosmological probes that we can use for this purpose are the CMB and the large scale structure (LSS) data.

The cosmic neutrino background (CNB) is formed
when neutrinos decouple, that is when the rate of the weak interaction reactions, which keep neutrinos in equilibrium with the primordial plasma, becomes less than the expansion rate of the Universe, at a temperature of about 1~MeV. After neutrino decoupling, photons are heated by electron-positron annihilation. After the end of this process, the ratio between the temperatures of photons and neutrinos will be frozen, although they cool as the Universe expands. Therefore, we expect today a CNB at a temperature of $T_\nu=(4/11)^{1/3}T_\gamma \approx 1.95$~K.
The CNB has not been directly detected and it will be the goal of the PTOLEMY project~\cite{PTOLEMY:2019hkd}. However, in the meantime, we have an indirect detection with the measurement of the effective neutrino number $N_{\rm eff} = 2.92^{+0.36}_{-0.37}$ at 95\% CL from Planck~\cite{Planck:2018vyg}, at many $\sigma$ different from zero.

Neutrinos are the only particles in the standard model to have the transition between the relativistic and the non-relativistic regime. 
When neutrinos are relativistic, they will contribute to the radiation content of the universe.
When they become non-relativistic, they will behave like matter contributing to the expansion of the Universe like baryons and cold dark matter. Neutrinos will only cluster on scales larger than their free streaming scale, thereby suppressing the structure formation at small scales, and affecting the large scale structures.

Since the CMB is formed at recombination, the effect of the neutrino mass can only manifest itself by changing the evolution of the background and introducing some secondary anisotropy corrections~\cite{Lesgourgues:2012uu,Lattanzi:2017ubx}.
Indeed, by varying their total mass we are changing the redshift of the matter-to-radiation equality $z_{eq}$, and the amount of matter density today $\omega_m = \omega_b + \omega_{cdm} + (\Sigma m_\nu)/93.14$~eV.
Therefore, the impact on the CMB will be the shift in the position of the peaks, the slope and the amplitude of the low-$\ell$ multipoles of the spectrum and the first peak, due to the ISW effect, and the damping of the high-$\ell$ tail, due to the lensing effect.

From Planck temperature and polarization spectra we have a very important upper limit on the total neutrino mass $\Sigma m_\nu<0.26$~eV~\cite{Planck:2018vyg}. This strong limit is completely due to the CMB gravitational lensing, indicating that we have a clear detection of this signal in the CMB spectra~\cite{Kaplinghat:2003bh}. In fact, the more massive the neutrino, the fewer structures we have, the less the CMB gravitational lensing should be. So a larger signal of the CMB lensing means a smaller neutrino mass. Given the neutrino effect on the structure formation, important observables are the LSS data, in particular the power spectrum of the non-relativistic matter fluctuations in Fourier space $P(k,z) = \langle | \delta_m (k,z)|^2 \rangle $, where $\delta_m = \delta \rho_m / \bar \rho_m$, and the two-point correlation function in the configuration space.

The shape of the matter power spectrum is a key observable for constraining neutrino masses with cosmological methods, and can be obtained with measurements of the CMB gravitational lensing, the clustering and the weak lensing of galaxies, and the galaxy cluster abundance~\cite{Hu:1997mj,Cooray:1999rv,Abazajian:2011dt,Chabanier:2019eai,TopicalConvenersKNAbazajianJECarlstromATLee:2013bxd,Green:2021gdc}. 
Unfortunately, this is really difficult to derive in non-linear and mildly non-linear scales, and needs the help of perturbation theory or N-body simulations. On the other hand, the BAO peak of galaxy correlation function, corresponding to the acoustic scale at decoupling, is one of the prominent observables in today's cosmology and easier to obtain, and is very sensitive to massive neutrinos. 
Combining Planck + BAO we get $\Sigma m_\nu<0.13$~eV~\cite{Planck:2018vyg}, because BAO data is directly sensitive to the free-streaming nature of neutrinos and helps in breaking the degeneracies among cosmological parameters.
Another important cosmological probe is the Redshift Space Distortions (RSD), which is obtained by analysing the clustering in redshift space, and it is the result of an anisotropic clustering along the radial direction because of the peculiar velocities~\cite{Hamilton:1997zq}. This RSD measures f$\sigma_8$, that is the product of the growth rate of structure (f) and the clustering amplitude of the matter power spectrum ($\sigma_8$). Massive neutrinos prefer a lower value for the f$\sigma_8$ data, so the inclusion of the latest RSD from eBOSS DR16~\cite{eBOSS:2020yzd} gives $\Sigma m_\nu < 0.087$~eV at 95\% CL~\cite{DiValentino:2021hoh}, disfavouring the minimal value allowed for Inverted Ordering (IO, $\Sigma m_\nu \gtrsim 0.1$~eV) at more than 2$\sigma$, but also the Normal Ordering (NO, $\Sigma m_\nu \gtrsim 0.06$~eV) at more than 68\% CL ($\Sigma m_\nu < 0.037$~eV). Current cosmological data do not allow to distinguish the ordering of the neutrino masses, but may give a preference for the NO when combined with oscillation and not oscillation data~\cite{Gerbino:2016ehw,Gariazzo:2018pei,Capozzi:2021fjo,Jimenez:2022dkn,Gariazzo:2022ahe}. It is worth underlining that in fact the total neutrino mass preferred by the cosmological data is zero or negative~\cite{eBOSS:2020yzd}, and
although this is not yet statistically significant, it shows a first hint of tension between cosmology and neutrino oscillation experiments.

These constraints could be drastically improved in the future.
Terrestrial CMB telescopes are currently the proposals with the highest probability of being realised. However, they need large angular scale measurements (such as Planck or future experiments) to measure the optical depth, that is strongly correlated with the neutrino masses, and a perfect a priori knowledge of the foregrounds.
The Simons Observatory~\cite{SimonsObservatory:2019qwx} aims to measure the total neutrino mass with an uncertainty $\sigma (\Sigma m_\nu)$ = 0.04~eV when combined with DESI BAO~\cite{DESI:2016fyo} and Rubin LSST~\cite{LSSTDarkEnergyScience:2018jkl} weak lensing data. The replacement of Planck with LiteBIRD’s future cosmic variance-limited measurements of the optical depth to reionisation SO can instead reach $\sigma (\Sigma m_\nu)$ = 0.02~eV. 
CMB-S4 measurements~\cite{Chang:2022tzj} of the lensing power spectrum (or cluster abundances), when combined with BAO from DESI and the current measurement of the optical depth from Planck, will provide a constraint on the sum of neutrino masses with a $\sigma (\Sigma m_\nu)$ = 0.024~eV, improving to $\sigma (\Sigma m_\nu)$ = 0.014~eV with better measurements of the optical depth.
PICO~\cite{NASAPICO:2019thw}, a proposal for a future CMB satellite experiment, plus BAO from DESI (or Euclid) should reach an uncertainty $\sigma (\Sigma m_\nu)$ = 0.014~eV, i.e. a 4$\sigma$ detection of the minimum sum for the NO. A satellite experiment is the only instrument that can measure very precisely all these neutrino properties together with the optical depth with the same single dataset without calibration problems.
Finally, CMB-HD~\cite{CMB-HD:2022bsz}, a futuristic millimetre-wave survey, could get an uncertainty on $\sigma (\Sigma m_\nu)$ = 0.013~eV (at least 5$\sigma$ detection for the sum of the neutrino masses), measuring the gravitational lensing of the CMB and the thermal and kinetic SZ effect on small scales.

All of these constraints have been obtained by assuming the $\Lambda$CDM cosmological model, which is the mathematically simplest model among those introduced in the literature. However, it cannot yet explain key pillars in our understanding of the structure and evolution of the Universe, namely, dark energy, dark matter and inflation. For this reason, the anomalies and tensions we see between some parameters coming from different cosmological probes may indicate the need for a paradigm shift~\cite{Abdalla:2022yfr}. We have, in fact, the $H_0$ tension at 5$\sigma$ and the $S_8$ tension at the level of $2-3\sigma$, and both these parameters are very important for the determination of the total neutrino mass because they are strongly correlated with it.
Furthermore, we have the $A_{lens}$ consistency check which fails in the Planck data and, due to their correlation, the upper limits on the total neutrino mass are strongly weakened, up to a factor of 2 when $A_{lens}$ is free to vary.
Finally, the global tensions between CMB datasets, at the level of 2.6$\sigma$ assuming the $\Lambda$CDM model and the Suspiciousness statistics~\cite{Handley:2020hdp}, is translated in $1\sigma$ preference from the terrestrial CMB telescopes ACT-DR4 and SPT-3G data for a neutrino mass different from zero, which is very similar to the value obtained from a combination of Planck + CMB Lensing, when $A_{lens}$ is free to vary~\cite{DiValentino:2021imh}. Furthermore, if the neutrino limits are obtained in a 10 parameter model, ACT-DR4 and SPT-3G can host even larger neutrino masses~\cite{DiValentino:2021imh}, and when CMB and BAO constraints are considered in these extended cosmologies, they provide constraints on the $\Sigma m_\nu$ vs $H_0$ plane that clearly show a correlation between these two parameters, that is exactly the opposite of what is obtained with the standard $\Lambda$CDM model, allowing this combination to solve the $H_0$ tension with massive neutrinos.

To conclude, the indication for anomalies and tensions present in the cosmological data could significantly influence the current cosmological constraints on the neutrino properties, presenting a serious limitation to precision cosmology. 
Until the nature of these anomalies (whether new physics or systematic errors) is clear, we should be very cautious when considering cosmological constraints.

\subsection{Imprints from the Early Universe on the Gravitational Wave Background}

\paragraph{Cosmic (super)strings}
Cosmic strings are topological defects that can form during phase transitions in the early Universe~\citep{Kibble:1976sj, Vilenkin:2000jqa}, and cosmic superstrings are the fundamental strings of string theory stretched to cosmological scales due to the expansion of the Universe~\citep{Jones:2002cv, Sarangi:2002yt, Dvali:2003zj, Jones:2003da, Copeland:2003bj, Jackson:2004zg}. In a cosmological setting, and for the simplest superstring models, cosmic string and superstring networks evolve in the same way. For a detailed review of cosmic (super)string network evolution and observational signatures, see, e.g.,~\citep{Copeland:2009ga}. Cosmic (super)strings can exchange partners when they meet and produce loops when they self-intersect. These loops then oscillate and lose energy to GWs generating bursts and a stochastic background~\citep{Berezinsky:2001cp,Damour:2000wa,Damour:2001bk,Damour:2004kw,Siemens:2006vk,Siemens:2006yp}--- signals that can be potentially detected by space-based  and terrestrial GW detectors and pulsar timing arrays~\citep{Pastorello:2019akb, Blanco-Pillado:2017rnf}. Strings are characterized by their mass per unit length $\mu$, which is normally given in terms of the dimensionless parameter $G\mu/c^2$, the ratio of the string energy scale to the Planck scale squared.

The cosmic string GW spectrum is broad-band, spanning many orders of magnitude in frequency, and hence accessible to a number of GW experiments including the Laser Interferometer Gravitational-wave Observatory (LIGO)~\cite{Harry:2010zz} and Virgo~\cite{VIRGO:2014yos}, Laser Interferometer Space Antenna (LISA)~\cite{LISA:2017pwj}, and the Pulsar Timing Arrays (PTAs)~\cite{Hobbs:2009yy}. PTAs are currently the most sensitive experiment for the detection of cosmic (super)strings and have set the most stringent bounds on the energy scale and other model parameters. The best limit on the string tension, $G\mu/c^2 < 5.3(2) \times 10^{-11}$, is several orders of magnitude better than constraints from CMB data, and comes from the NANOGrav Collaboration~\citep{NANOGRAV:2018hou}. In fact, PTAs might have already seen the first hints of a stochastic background~\cite{NANOGrav:2020bcs}. A definitive detection of GWs from cosmic (super)strings would be transformative for fundamental physics which could be enabled by PTAs over the next decade.

\paragraph{Primordial gravitational waves from inflation}
The inflationary paradigm that the very early Universe saw a period of exponential expansion accounts for the observed homogeneity, isotropy, and flatness~\citep{Brout:1977ix, Starobinsky:1980te, Kazanas:1980tx, Sato:1980yn, Guth:1980zm, Linde:1981mu, Albrecht:1982wi}. Additionally, by expanding quantum fluctuations present in the pre-inflationary epoch, inflation seeds the density fluctuations that evolve into the large-scale structures we see in the Universe today~\citep{Mukhanov:1981xt, Hawking:1982cz, Guth:1982ec, Starobinsky:1982ee, Bardeen:1983qw} and produces a stochastic background of GWs~\citep{Starobinsky:1979ty, Rubakov:1982df, Abbott:1984fp}, which have so far eluded detection. This GW background is broad-band, like the one produced by cosmic strings, and potentially detectable by multiple experiments.

Detecting primordial GWs from inflation has been a critical objective of CMB experiments for some time~\citep{Kamionkowski:2015yta}. The CMB is sensitive to the lowest frequency portion of the GW spectrum from inflation, and CMB data can be used to constrain the tensor-to-scalar ratio, which is the ratio of the size of GWs produced to that of scalar perturbations that seed density fluctuations, as described above.  For standard inflation models, the GW background in the PTA band is likely to be fainter than that of supermassive black hole binaries depending on the nature of the latter at lowest frequencies~\citep{Sampson:2015ada}.  In addition, some inflationary models have a spectrum that rises with frequency.  Thus, GW detectors operating at higher frequencies than CMB experiments, like PTAs and space-based  and terrestrial interferometers, can be used to constrain the shape of the inflationary GW spectrum. Indeed, PTA, CMB, and GW interferometer data across 29 decades in frequency have already begun to place stringent limits on such models~\citep{Lasky:2015lej}, though future observations are necessary to detect the background or tighten the constraints on model parameters.

\paragraph{Gravitational waves from phase transitions}
The early Universe may have experienced multiple phase transitions as it expanded and cooled. Depending on the detailed physical processes that occur during a phase transition, GWs can be generated with wavelengths of order the Hubble length at the time of the phase transition. That length scale, suitably redshifted, translates into a GW frequency today. Thus, GW experiments at different frequencies today probe horizon-sized physical processes that occurred at different times in the early Universe, with higher frequency experiments probing earlier and earlier times.

For example, the nanohertz frequency band accessible to PTAs maps onto the era in the early Universe when the quantum chromodynamics (QCD) phase transition took place, about $10^{-5}$~s after the Big Bang. The horizon at that time was on the order of 10~km, and any GWs generated at that length scale at that time would today be stretched to about 1~pc (or 3~light-years), which corresponds to GW frequencies of about 10~nHz, and lie within the PTA sensitivity band. The possibility that interesting QCD physics can result in a GW signal detectable by PTAs was first pointed out by Witten in the 1980s~\citep{PhysRevD.30.272}. More recently, Caprini et al.~\citep{Caprini:2010xv} considered the possibility of a first-order phase transition at the QCD scale. In standard cosmology, the QCD phase transition is only a cross-over and we do not expect it to generate GWs. However, if the neutrino chemical potential is sufficiently large, it can become first order (it is worth pointing out that if sterile neutrinos form dark matter, we expect a large neutrino chemical potential). There is also the possibility that the fluctuations of gluon fields could generate scalar GWs from the conformal anomaly in the quark-gluon plasma phase~\cite{Mottola:2016mpl}. Thus, PTAs provide a window onto physical processes occurring in the Universe at the time of the QCD phase transition or before and could detect GWs from a first order phase transition at that time.

\subsection{Ultra-High-Energy Cosmic Rays as Probes of the Early Universe} 

\label{s:SHDM}

The motivations for super-heavy dark matter (SHDM) particles were recently revived by the possibility that new physics could only manifest at the Planck scale or at the scale of Grand Unified Theories (GUTs)~\cite{Coleman:2022abf, Anchordoqui:2021crl}. This possibility is motivated not only by the absence of any sign of new physics at the TeV scale, but also by the precise measurements of the mass of the Higgs boson and the Yukawa coupling of the top quark that make it possible to extrapolate the SM all the way to the Planck mass without encountering any inconsistency that would make the electroweak vacuum of the SM unstable. This vacuum lies, in fact, close to the boundary between stability and metastability~\cite{Degrassi:2012ry}.

Super-heavy dark matter particles that are only gravitationally coupled could have been produced at the end of inflation via the freeze-in mechanism, which relies on annihilations of SM fields to populate the dark sector. An interesting consequence is that, so as to produce enough such very weakly coupled heavy particles, the reheating temperature must be relatively high, which implies a tensor/scalar ratio of the primordial modes possibly detectable in the power spectrum of the CMB. The limits inferred from the Planck satellite on this ratio thus constrain the possible phase space for the mass of the particles and the value of the Hubble rate at the end of inflation~\cite{Garny:2015sjg}.

Another possibility to constrain these models is to look for the secondary products produced via particle decay. In the minimalist benchmark described above, dark matter (DM) particles are protected in the perturbative domain by the conservation of quantum numbers, and so would only decay through non-perturbative effects. One of these effects is due to the non-trivial vacuum structure of non-commutative gauge theories and the possibility of the generation of one quantum number for the benefit of another through the change of configuration of gauge fields by tunnel effect (instantons)~\cite{Kuzmin:1997jua}. This mechanism offers the possibility of providing metastable particles, which can produce detectable secondaries.

If SHDM particles decay into SM fields, then a flux of ultra-high-energy photons could be observed preferentially from regions of denser DM density, such as the center of our Galaxy~\cite{Aharonian:1992qf, Berezinsky:1997hy, Birkel:1998nx, Evans:2001rv, Aloisio:2006yi}. AugerPrime is in a prime position to collect a large exposure in the direction of the Galactic center~\cite{PierreAuger:2016qzd}. Indeed the non-observation of photons in Auger data has allowed limits to be set on the gauge coupling in the dark sector~\cite{PierreAuger:2022wzk,PierreAuger:2022ibr}, which are complementary to those obtained via the tensor/scalar ratio of the primordial modes. With increased sensitivity to the tensor-to-scalar ratio on the one hand and to ultra-high-energy photons thanks to the planned extreme-energy cosmic-ray observatories in the next decade~\cite{POEMMA:2020ykm, Horandel:2021prj} on the other, the GUT parameter space will continue to shrink towards the low-mass particle range and/or small gauge coupling values. 

While the observation of ultra-high-energy photons could open a window to explore high-energy gauge interactions and possibly GUTs in the early Universe, the observation of extreme-energy neutrinos could provide a method of searching for strongly coupled string moduli, which complements searches based on gravitational effects of cosmic strings (including structure formation, CMB data, gravitational radiation, and
gravitational lensing)~\cite{Anchordoqui:2018qom}. In particular, the future Probe of Extreme Multi-Messenger Astrophysics (POEMMA)~\cite{POEMMA:2020ykm} will be able to detect extreme energy neutrinos from cosmic strings with $G \mu/c^2 \sim 10^{-20}$~\cite{Berezinsky:2011cp}. Thus, POEMMA will be sensitive to dimensionless string tensions down to 9 orders of magnitude below the current upper limit from the NANOGrav Collaboration~\citep{NANOGRAV:2018hou}.

\section{Cosmic Probes of Dark Matter}
\label{s:DM}

A second component of the dark, or hidden, sector is dark matter. Dark matter is distinct from dark energy in that it is evidenced by its gravitational pull on celestial objects, but its nature has otherwise eluded searches during decades. However, over the past decade, new tools and techniques to search for dark matter have come to the fore to probe new parameter spaces. Searches for primordial black holes, which should radiate both baryonic and dark matter as they decay, are prominent in high-energy gamma rays, and gravitational-wave facilities now look for evidence of dilute distributions and coalescence of dark matter. Connections between Standard Model particles and the hidden sector may now be revealed through a variety of messengers, including key measurements of cosmic rays, neutrinos, gravitational waves, gamma rays, and other photon wavelengths.
        
\subsection{Connection between Visible and Hidden Sectors} 

\label{s:portals}

The nature of dark matter (DM) is one of the great fundamental puzzles of particle physics and cosmology~\cite{Boddy:2022knd,Ando:2022kzd,Carney:2022gse,Aramaki:2022zpw,Leane:2022bfm,Baryakhtar:2022hbu}. The DM distribution in galaxies and other virialized systems is a powerful indicator of its nature and a portal towards understanding the DM phenomenon~\cite{Salucci:2018hqu}. The annihilations and decays of DM could produce visible particles over a wide range of energy scales, which subsequently decay producing a range of visible secondary particles. Long-standing efforts have been dedicated to searches for such signals in photons, cosmic rays, and neutrinos, and future experiments offer the prospect of significantly improved sensitivity.

Indirect searches for DM based on gamma-ray, cosmic-ray, and neutrino signals are highly complementary~\cite{PerezdelosHeros:2020qyt}. For example, the production of high-mass quarks and gluons leads to copious production of gamma rays, neutrinos, antiprotons, and antinuclei, while the production of electrons or muons leads to strong signals in searches for cosmic-ray positrons. Dark matter decaying or annihilating into neutrinos can be well constrained by high-energy neutrino telescopes. In scotogenic models where the neutrinos mass is achieved via interaction with DM, neutrinos might be the principal portal to the dark sector. Combining constraints from all these channels allows us to avoid blind spots in sensitivity, and probe the lifetime or annihilation rate of DM in broadest possible range of scenarios. UHECR experiments could be also sensitive to interactions induced by macroscopic dark quark nuggets~\cite{Bai:2018dxf} in the atmosphere, offering further windows to identify the nature of DM~\cite{Coleman:2022abf, Anchordoqui:2021xhu}.

Searches for DM often rely critically on an understanding of astrophysical backgrounds or systems, including diffuse astrophysical backgrounds, as discussed in Sec.~\ref{subsec:diffuseBG}, and emission by individual astrophysical sources, as discussed in Sec.~\ref{subsec:TeVatronPeVatron}. Poorly understood systematic errors associated with multimessenger astrophysics can be the major limiting factor for sensitivity to dark matter signals. In the event of a possible detection of DM in an astrophysical data set, searches for multimessenger counterpart signals will be crucial in determining whether the apparent detection is truly associated with DM, and if so, determining the properties of that DM.

\subsection{Primordial Black Holes as Dark Matter}\label{s:PBHs}

Primordial Black Holes (PBHs) have long been considered as plausible cold DM (CDM) candidates~\cite{Carr:1975qj, Brito:2022lmd}, potentially forming a significant fraction of the DM~\cite{Bird:2016dcv, Ali-Haimoud:2017rtz, Raidal:2017mfl, Raidal:2018bbj, Vaskonen:2019jpv, Atal:2020igj, DeLuca:2020qqa, Wong:2020yig, Franciolini:2021tla}. The detection of binary black holes of masses in excess of 30 $M_\odot$ has brought renewed attention to this possibility by positing that progenitor black holes in these systems could be primordial in origin~\cite{Bird:2016dcv, Clesse:2016vqa, Sasaki:2016jop}. It is not possible to obtain conclusive evidence for PBHs in these detections as astrophysical models are able to explain their existence~\cite{Clesse:2020ghq}. The primordial origin of binary black holes would be compelling if either of their masses are below one solar mass or if they arise in the dark ages when stars could not have produced black hole binaries~\cite{Ng:2021sqn, Ng:2022agi, Raidal:2017mfl, Raidal:2018bbj, Vaskonen:2019jpv, Atal:2020igj, Mukherjee:2021ags, Mukherjee:2021itf, Ng:2021sqn, Franciolini:2021xbq}. PBHs can also be distinguished on the basis of their source properties such as mass, and eccentricity spin, the redshift evolution of BBH merger rates, and their spatial distribution, though a firm detection is required.

PBHs have been predicted as a generic outcome of density perturbations in the early Universe~\cite{1967SvA....10..602Z, 1971MNRAS.152...75H, 1975ApJ...201....1C, 1980PhLB...97..383K, 1985MNRAS.215..575K, Carr:2005zd, Clesse:2016vqa, Sasaki:2018dmp, Sasaki:2016jop, Raidal:2017mfl, Raidal:2018bbj, Vaskonen:2019jpv, Gow:2019pok, Jedamzik:2020ypm, Jedamzik:2020omx, DeLuca:2020jug, Atal:2020igj, DeLuca:2020qqa, Clesse:2020ghq}. The formation of black holes in the early Universe appears to be quite generic~\cite{Garcia-Bellido:1996mdl} and does not require special conditions such as large density fluctuations of matter. Large, non-Gaussian exponential tails in the density fluctuations arising from quantum processes during inflation could produce them~\cite{Ezquiaga:2019ftu}. Moreover, even the known thermal history of the Universe may play an important role by providing the required lack of pressure to allow gravitational collapse at certain well-defined epochs in the evolution of the Universe~\cite{Carr:2019kxo}; these are the Electroweak and quantum chromodynamics (QCD) epochs at the time of $e^+ e^-$ annihilation, which generate a multimodal PBH mass function with peaks at $10^{-5},\ 1,\ 10^{2},\ \mathrm{and}\ 10^{6}\ M_\odot$. These black holes come with different fractional abundances depending on the underlying inflationary potential and this may be used as a window into the early Universe and fundamental physics. For example, if an excess of $10^{-10}\ M_\odot$ is found in microlensing events, or in the induced Stochastic Gravitational Wave Background (SGWB) at LISA frequencies, we may infer the existence of new fundamental particles at scales above those reached by present particle physics accelerators, which become non-relativistic and momentarily decrease the radiation pressure at a time when the mass within the horizon is precisely that mass scale.

The possibility that PBHs may be lurking in the dark Universe as building blocks of the CDM fluid is extremely attractive~\cite{Clesse:2017bsw}. In fact, the non-Gaussian exponential tails mentioned above may give rise to enhanced clustering, which leads, after recombination, to a population of PBH clusters with intermediate masses of order $10^{6}\ M_\odot$ that could be searched for with the microlensing of quasars around clusters or the perturbations they induce on stellar tidal streams around our galaxy and Andromeda~\cite{Montanari:2020gcr}. These clusters could explain where most of the mass in the halo of galaxies is, thus evading the microlensing limits coming from stars in the Large Magellanic Cloud and the Galactic bulge~\cite{Calcino:2018mwh}, which had been used in the past to rule out PBHs as the main component of CDM~\cite{Wyrzykowski:2019jyg}. Moreover, such dense objects may help explain many unexpected correlations in the radio and X-ray backgrounds at high redshift~\cite{Kashlinsky:2016sdv, Kashlinsky:2019kac}, as well as the unusually high number of massive galaxies and quasars at high redshift, unaccounted for by the standard $\Lambda$CDM scenario.

There is, nowadays, a great opportunity for testing all these ideas with new astrophysical and cosmological observations. For example, if PBHs existed before recombination, they should have left their imprint in an excess of injected energy in the plasma in the form of spectral distortions at high frequencies that a CMB experiment dedicated to it could detect~\cite{Chluba:2019kpb}. One could further use the James Webb Space Telescope (JWST) to look for the first stars and galaxies at redshifts bigger than 10 or 20, confirming the role of black holes in early star formation \cite{Hasinger:2020ptw,Cappelluti:2021usg}.

Moreover, GWs may also allow us to detect PBHs over a wide range of masses, being complementary to other proposed probes and able to distinguish between BHs of astrophysical origin and PBHs using either resolved events~\cite{Hall:2020daa, Wong:2020yig, DeLuca:2021wjr, Hutsi:2020sol, Franciolini:2021tla} or the stochastic GW background~\cite{Mandic:2016lcn, Mukherjee:2021ags, Mukherjee:2021itf}. With the advent of the next generation of GW antennas, like the Cosmic Explorer~\cite{Evans:2021gyd} and Einstein Telescope~\cite{Punturo:2010zz, Maggiore:2019uih} on the ground and LISA in space~\cite{Barausse:2020rsu}, we should be able to reach black hole fusions at redshifts $z\sim 100,$ where no plausible stellar evolution could have generated such a population, and thereby convincingly proving their primordial nature. A further hint at their primordial origin, which would link their formation with the cosmic history, would be the discovery of the induced SGWB from second-order perturbations of large-amplitude fluctuations entering at the same time as the formation of PBHs~\cite{Garcia-Bellido:2017aan}, when the size of the horizon redshifted today gives LISA frequencies (mHz), or perhaps in the PTA range (nHz). Such a discovery would open a new window into the early Universe, where we could explore independent constraints on the existence of PBHs~\cite{Kohri:2018awv, Espinosa:2018eve, Wang:2019kaf} such as the non-Gaussian character of the fluctuations giving rise to the PBH mass spectrum, as well as the number of relativistic species present at that time~\cite{Carr:2019kxo}, well beyond the present reach with particle accelerators. In addition, terrestrial GW detectors, LISA and PTAs, could  give us conclusive evidence of whether or not PBHs form a significant fraction of DM in a wide range of masses~\cite{Singh:2020wiq,LIGOScientific:2021job,LIGOScientific:2019kan}.

The detection of PBHs of any size would acutely constrain our understanding of the physics of the early Universe. Such a monumental reward motivates the search for signs of PBHs across all facets of the multimessenger landscape, including the hunt for gamma-ray and neutrino signatures of PBH evaporation. The prediction that a black hole will thermally radiate (evaporate) with a blackbody temperature inversely proportional to its mass was first calculated by Hawking~\cite{Hawking1974}--- the emitted radiation consisting of all fundamental particles with masses less than $\sim$T$_\mathrm{BH}$~\cite{MacGibbon1990}. While Hawking radiation for black holes in the stellar mass range and above is nearly negligible, this process dominates the evolution of lower-mass PBHs over time. PBHs with initial masses of $\sim$10$^{14}$--10$^{15}$~g should be expiring today, producing short bursts (lasting a few seconds) of high-energy radiation in the GeV--TeV energy range~\cite{MacGibbon2008, Ukwatta:2015tza}. Their final moments would thus be an ideal phenomenon to observe with current space-based  and terrestrial gamma-ray telescopes, as well as neutrino observatories~\cite{FermiPBH1, FermiPBH2, HAWCPBH, Tavernier2021HESSPBH, Archambault:2017asc, Dave:2019epr}. Improvements in sensitivity, effective area, and field of view seen with the proposed next generation of gamma-ray telescopes, such as the Southern Wide-field Gamma-ray Observatory (SWGO)~\cite{Albert:2019afb, SWGOPBH}, the Cherenkov Telescope Array (CTA)~\cite{CTAConsortium:2017dvg}, and the All-sky Medium-Energy Gamma-ray Observatory (AMEGO)~\cite{2020SPIE11444E..31K}, present a boundless new frontier for discovery beyond the Standard Model and characterization of early Universe conditions with PBHs. While this mass regime is not currently a candidate for PBHs as dark matter, confirmation of an evaporation signal from a PBH of any size would lend significant credence to that dark matter model.

For reference, it is worth pointing out the Snowmass paper on Primordial Black Holes, Ref.~\cite{Bird:2022wvk}, which specifically focuses on these natural candidates for dark matter, describes the science cases (the origin of PBH dark matter in the early Universe) and the existing observational constraints, and expands on the theoretical work and data analysis required to improve the constraints and/or enable a possible detection in the future.

\subsection{Properties of Non-Minimal Dark Sectors} 
\label{sec:DDM}
       
For several decades, it has been suspected that the dark sector consists of one stable weakly interacting massive particle. However, some critical thinking was recently adopted to build up a more generic view of the hidden sector in which a given dark matter particle need not be stable if its abundance at the time of its decay is sufficiently small. Dynamical Dark Matter (DDM) is a framework for non-minimal dark sectors which posits that the dark matter in the Universe comprises a vast ensemble of interacting fields $\chi_\ell$ with a variety of different masses, lifetimes, and cosmological abundances~\cite{Dienes:2011ja}. In general, the mass spectra and corresponding lifetimes and abundances of the individual states within the DDM ensemble turn out to be tied together through scaling relations involving only a few scaling exponents. As a result, the DDM ensemble is described by only a few free parameters, rendering the DDM framework every bit as constrained and predictive as more traditional dark matter scenarios.

The DDM framework might be experimentally tested and constrained through dark matter direct- and indirect-detection~\cite{Dienes:2012cf, Dienes:2013lxa, Boddy:2016fds, Boddy:2016hbp} experiments, and at colliders~\cite{Curtin:2018ees, Dienes:2019krh, Dienes:2021cxr}. Since there may be a large number of transitions between the ensemble of DDM states, there may be a variety of lifetimes and long-lived particles which, on decay, can produce spectacular signals at the Forward Physics Facility (FPF)~\cite{Feng:2022inv}. DDM scenarios can also leave observable imprints across the cosmological timeline, stretching from structure formation~\cite{Dienes:2020bmn, Dienes:2021itb} all the way to late-time supernova recession data~\cite{Desai:2019pvs} and unexpected implications for evaluating Ly-$\alpha$ constraints~\cite{Dienes:2021cxp}. Such dark sectors also give rise to new theoretical possibilities for stable mixed-component cosmological eras~\cite{Dienes:2021woi}. DDM scenarios also bring about enhanced complementarity relations~\cite{Dienes:2014via, Dienes:2017ylr} between different types of experimental probes.

DDM scenarios in which the constituents decay entirely within the dark sector---i.e., to final states comprising other, lighter ensemble constituents and/or dark radiation---are particularly challenging to test. Nevertheless, there exist observational handles that can be used to probe and constrain DDM ensembles that decay primarily via ``dark-to-dark'' decay processes of this sort, and thus potentially permit us to distinguish them from traditional DM candidates.

Dark-to-dark decays of this sort modify the way in which the expansion rate of the Universe, as described by the Hubble parameter $H(z)$, evolves with redshift.  These modifications, in turn, affect the functional relationship between the redshifts $z$ and luminosity distances $D_L(z)$ of Type-Ia supernovae~\cite{Desai:2019pvs}. Since the dark-to-dark decays of a DDM ensemble alter the dependence of $H(z)$ on $z$, the DDM framework can potentially also provide a way of addressing the $H_0$ tension~\cite{Abdalla:2022yfr}.  The advantage of a DDM ensemble relative to a single decaying dark matter species is that the timescale across which the decays have a significant impact on the expansion rate can be far broader. Nevertheless, DDM models in which the $\chi_\ell$ decays into dark radiation via a two-body process of the form  $\chi_\ell \to  \psi \bar \psi$, where $\psi$ is a massless dark-radiation field, cannot ameliorate the $H_0$ tension~\cite{Anchordoqui:2022gmw}. Models in which the $\chi_\ell$ decays primarily via intra-ensemble processes---e.g., of the form $\chi_\ell \rightarrow \chi_m \bar{\psi} \psi$, where $\psi$ once again denotes a dark-radiation field---could be more promising~\cite{Anchordoqui:2020djl}. Such decays endow the final-state $\chi_m$ with non-negligible velocities, thereby modifying the equation of state $w_m(z)$ for each ensemble constituent and modifying the DM velocity distribution of the ensemble as a whole. Moreover, complementary scattering processes of the form $\chi_\ell \psi \rightarrow \chi_m \psi$, through which the different ensemble constituents interact with the dark radiation, could potentially also help to ameliorate the $\sigma_8$ tension in the same way that they do in partially acoustic DM scenarios~\cite{Chacko:2016kgg}.

A concise description (put together in 13 ``take-away lessons'' for Snowmass 2021~\cite{Dienes:2022zbh}) of collective phenomena that can arise in dark sectors, which contain a large number of states, underscores the need to maintain a broad perspective when contemplating the possible signals and theoretical possibilities associated with non-minimal hidden sectors.

\subsection{Gravitational-Wave Probes of Coalescing Dark Matter} 

Some of the DM may have clustered gravitationally in the early Universe, forming compact dark objects. These structures may cause a transient magnification of light from distant stars via microlensing, which remains one of the most powerful techniques to constrain compact dark objects in a wide range of masses~\cite{1986ApJ...304....1P}.

DM clumps near (or within) the Earth can alter the planet's tidal field---which is well monitored for decades and therefore well constrained---or cause sudden accelerations, leading to interesting constraints on asteroid-like clumps~\cite{Seto:2007kj, Namigata:2022vry, Kashiyama:2018gsh}. Albeit small, the interaction cross section of DM with Standard Model fields can lead to the deposition of small DM cores at the center of stars~\cite{Press:1985ug}, with capture rates that can be enhanced by the large density of white dwarfs and neutron stars. For fermionic fields, the accumulation of DM could eventually lead to cores more massive than the Chandrasekhar limit, collapse of the DM core to a black hole (BH), and eventually to the disruption of the star by accretion onto the newly formed BH~\cite{Goldman:1989nd}. For bosonic DM, this fate may be eluded via gravitational cooling~\cite{Brito:2015yga}.

Another possibility is that standard CDM models could produce small-scale clumps. A CDM clump moving near the Earth or a pulsar produces an acceleration that could be measurable in PTA data, providing an opportunity to test the CDM paradigm~\citep{Siegel:2007fz, Kashiyama:2018gsh}.

The possibility of compact objects harboring DM cores is intriguing. If these cores are sufficiently massive, the star is effectively described by a different equation of state and its properties change. The coalescence of DM stars will differ from the prediction of standard GR, leading to peculiar signatures in the GW signal close to merger~\cite{Ellis:2017jgp, Bezares:2019jcb}. In fact, DM clumps can also form in isolation and bind to compact stars in their vicinity. Compact DM cores orbiting neutron stars (either in their exterior or in their interior) may give rise to detectable signals in our Galaxy~\cite{Horowitz:2019aim}.

The general GW signatures of the coalescence of DM clumps or ``blobs'' have been explored by various authors~\cite[see, e.g.,][]{Giudice:2016zpa, Diamond:2021dth}, but precise calculations of the signal from the coalescence of two DM clumps require an underlying theory with a well-posed initial value problem. One example are compact configurations made of self-gravitating scalar fields, also known as boson stars~\cite{Palenzuela:2017kcg, Cardoso:2017cfl, Sennett:2017etc, Bustillo:2020syj}.


\subsection{The Gravitational-Wave Signatures of Dilute Dark Matter}  

One of the most solid experimental pillars of modern physics is the equivalence principle, which ensures that all forms of matter couple universally to gravity. Even if DM does not form compact objects, dilute DM configurations must still interact gravitationally; dense DM spikes can then develop in the vicinity of isolated compact bodies such as BHs~\cite{Gondolo:1999ef, Sadeghian:2013laa}. Massive BHs are expected to be present at the center of many galaxies. In these environments the DM density should therefore be substantially higher than in the Solar System. Compact objects (BHs or neutron stars) moving in these dense DM environments will be subject to accretion and dynamical friction, leading to small changes in their dynamics that require a detailed understanding of the physics involved in these processes. Preliminary studies indicate that DM-induced changes in the GW phase of compact objects could be detectable by next-generation GW interferometers~\cite{Barausse:2014tra, Cardoso:2019rou, Kavanagh:2020cfn, Annulli:2020lyc, Annulli:2020ilw, Traykova:2021dua, Vicente:2022ivh}.

If DM has a very large Compton wavelength, as in the case of ``fuzzy'' ultralight DM fields of mass $10^{-23}$--$10^{-22}$~eV, it may give rise to small pressure oscillations at low frequency (e.g., of the order of nHz), that could affect the motion of stars and binary systems~\cite{Khmelnitsky:2013lxt, Porayko:2018sfa}. These minute changes can be tracked with PTA experiments. These oscillations can also affect the GW detectors themselves: the direct couplings to the beam splitter of GW detectors can be used to set stringent constraints on the abundance and coupling strength of DM~\cite{Vermeulen:2021epa, Pierce:2018xmy}.

\paragraph{Nonperturbative effects: ultralight bosonic fields}
The simplest possibility for new matter sectors are bosonic or fermionic degrees of freedom minimally coupled to gravity. These fields could form all or part of the DM. Their scale is set by their mass $\mu$, which could range from cosmological scales to very heavy particles~\cite{Dine:1982ah, Preskill:1982cy, Arvanitaki:2009fg}. Bosonic fields with Compton wavelengths comparable to the Schwarzschild radius of astrophysical BHs of mass $M$, i.e.,  $GM\mu/(c\hbar)\sim 1$, can trigger a new fascinating phenomenon caused by the existence of {\it ergoregions} around spinning BHs~\cite{Penrose:1969pc, Brito:2015oca}. Spinning BHs can spontaneously transfer their rotational energy to a boson ``condensate'' or ``cloud'' co-rotating with the BH and carrying a significant fraction of its angular momentum. The bosonic cloud is a classical object of size much larger than the BH itself, and it can contain up to $10\%$ of its mass~\cite{Brito:2015oca}. The BH/cloud system is similar to a huge gravitational ``lighthouse'' which extracts energy from the BH by emitting a nearly monochromatic GW signal.

Proposed ways to rule out or constrain light bosons as DM candidates include~\cite{Brito:2015oca}:

\begin{itemize}
    \item[(i)] Monitoring the spin and mass distribution of astrophysical BHs. Measurements of highly spinning BHs will immediately rule out fields with Compton wavelengths comparable to the horizon radius, as these BHs should have been spun down on relatively short timescales~\cite{Brito:2015oca,Baryakhtar:2017ngi,Davoudiasl:2019nlo}.

        \item[(ii)] Direct searches for the resolvable or stochastic  monochromatic GW signals produced by the boson cloud~\cite{Arvanitaki:2010sy, Arvanitaki:2014wva, Brito:2014wla, Arvanitaki:2016qwi, Brito:2017wnc, Brito:2017zvb, Ghosh:2018gaw}, which are now routinely carried out by the LIGO/Virgo collaboration~\cite{LIGOScientific:2021jlr}.

            \item[(iii)] Searches for electromagnetic (EM) emission from BH/boson cloud systems. Axion-like particles have been proposed in many theoretical scenarios, including variations of the original solution to the strong CP problem of QCD. Self-interactions and couplings with Standard Model fields can lead to periodic bursts of light, ``bosenovas,'' and other interesting phenomenology~\cite{Yoshino:2012kn, Ikeda:2018nhb}. In addition, axion-like particles should couple to photons and produce preferentially polarized light~\cite{Chen:2019fsq}.

                \item[(iv)] Observations of peculiar stellar distributions around massive BHs. The nonaxisymmetric boson cloud can cause a periodic forcing of other orbiting bodies, possibly leading to Lindblad or corotation resonances  where stars can cluster~\cite{Ferreira:2017pth, Boskovic:2018rub}.
                \end{itemize}

These are only some of the possible strategies. Superradiance does not require any ``seed" boson abundance: any vacuum fluctuation will lead to energy extraction and exponential growth in time. In this sense, BHs are natural particle detectors, complementary to terrestrial colliders~\cite{Brito:2015oca, Barack:2018yly}. It is important to remark that the complementary role of the different GW and EM instruments necessary to probe the large range of mass/energy scales--- astrophysical BHs span about ten orders of magnitude in mass, thus allowing us to constrain ultralight bosonic fields across ten orders of magnitude in mass (or energy).

Most of the discussion above was focused on a neutral DM environment and gravitational dynamics. Another possibility is that beyond-the-Standard-Model fermions may carry a fractional electric charge or be charged under a hidden $U(1)$ symmetry~\cite{DeRujula:1989fe, Perl:1997nd}. Modified theories of gravity can also lead to compact stars or BHs carrying nonzero scalar charges~\cite{Damour:1993hw, Doneva:2017bvd, Silva:2017uqg}.  In all of these theoretical scenarios, BHs and compact stars can carry non-negligible charges that would lead to different inspiral and merger signals~\cite{Zilhao:2012gp, Cardoso:2016olt, Alexander:2018qzg, Kopp:2018jom, Dror:2019uea, Bozzola:2020mjx, Maselli:2021men}; GW observations can be used to reveal or constrain these charges and the underlying theories.

\section{Astroparticle Physics}
 \label{s:astropart}
  
In order to understand the fundamental physical forces at play in sources that human hands did not create, there is a need to collaborate with astronomers and astrophysicists to discover the nature of matter and emission from cosmic sources. This is akin to understanding the machinery along a particle accelerator to put the detection in context. While the specialties of building detectors and interpreting the results go hand in hand for particle and astroparticle physics, specialities in particle instrumentation and astrophysical progenitors require more intentional collaboration for a common goal. Nevertheless, the inclusion of each area of expertise is vital to broadening our understanding of physics as a species since each piece of the Standard Model puzzle is necessary for its completion. 
    
\subsection{Properties of Standard Model Particles and their Interactions} 
\label{s:astropart_1}

\begin{figure}[t]
  \centering
  \includegraphics[width=\textwidth]{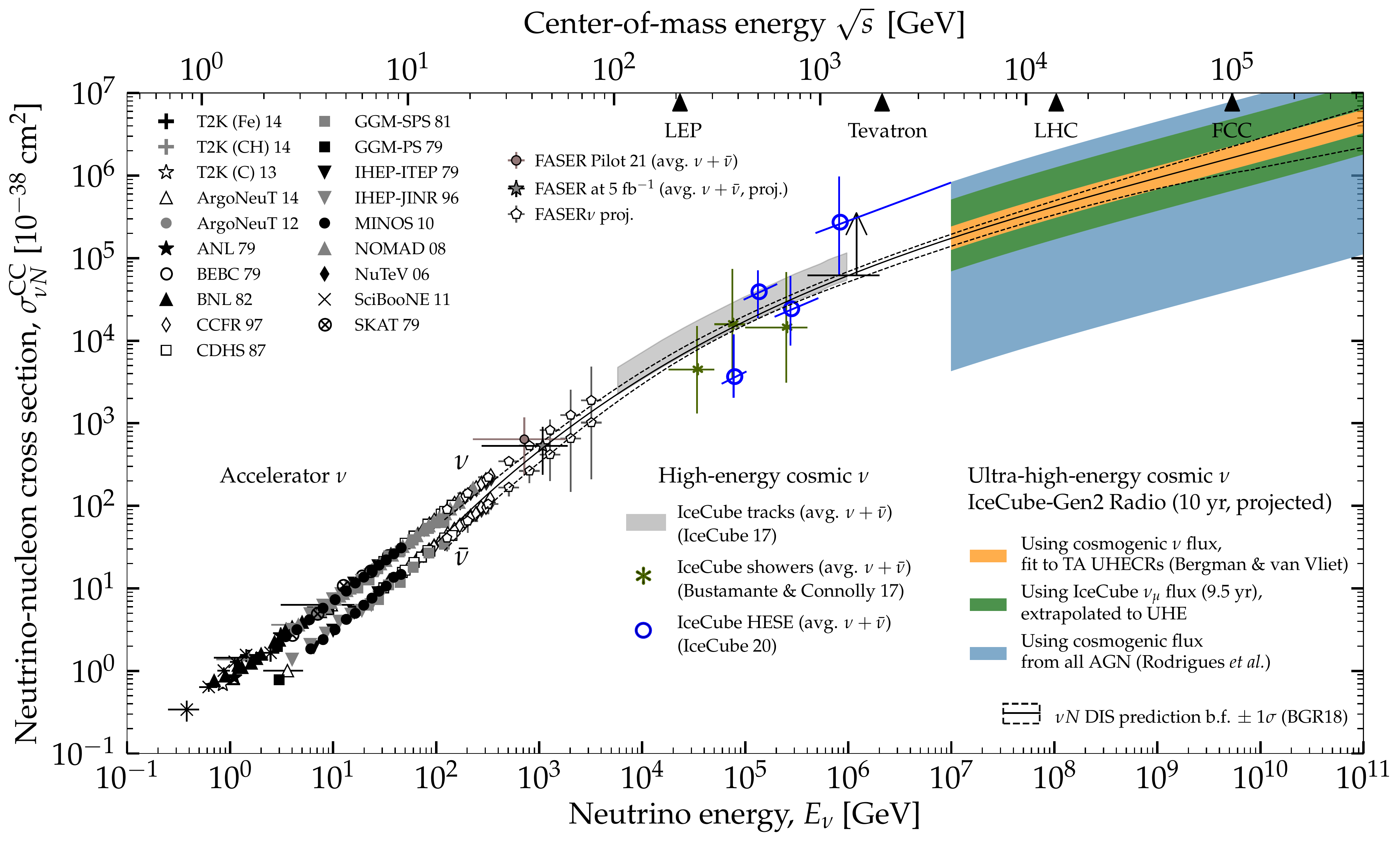}
  \caption{Neutrino-nucleon cross section measurements, compared to deep-inelastic-scattering (DIS) cross section predictions from Ref.~\cite{Bertone:2018dse} (BGR18).  In the TeV range, FASER and FASER$\nu$ have started measurements~\cite{Arakawa:2022rmp}. Measurements in the TeV--PeV range are based on IceCube showers~\cite{Bustamante:2017xuy, IceCube:2020rnc} and tracks~\cite{IceCube:2017roe}.  Projected measurements at energies above 100~PeV~\cite{Valera:2022ylt} are based on 10~years of operation of the radio component of IceCube-Gen2, assuming a resolution in energy of 10\% and a resolution in zenith angle of 2°.  Since the flux at these energies remains undiscovered, projections for the measurement of the cross section are for different flux predictions.  From Ref.~\cite{Ackermann:2022rqc}, adapted from Ref.~\cite{Valera:2022ylt}.}
  \label{fig:cross_section_uhe}
\end{figure}

\begin{figure}[t]
  \centering
  \includegraphics[width=0.7\textwidth]{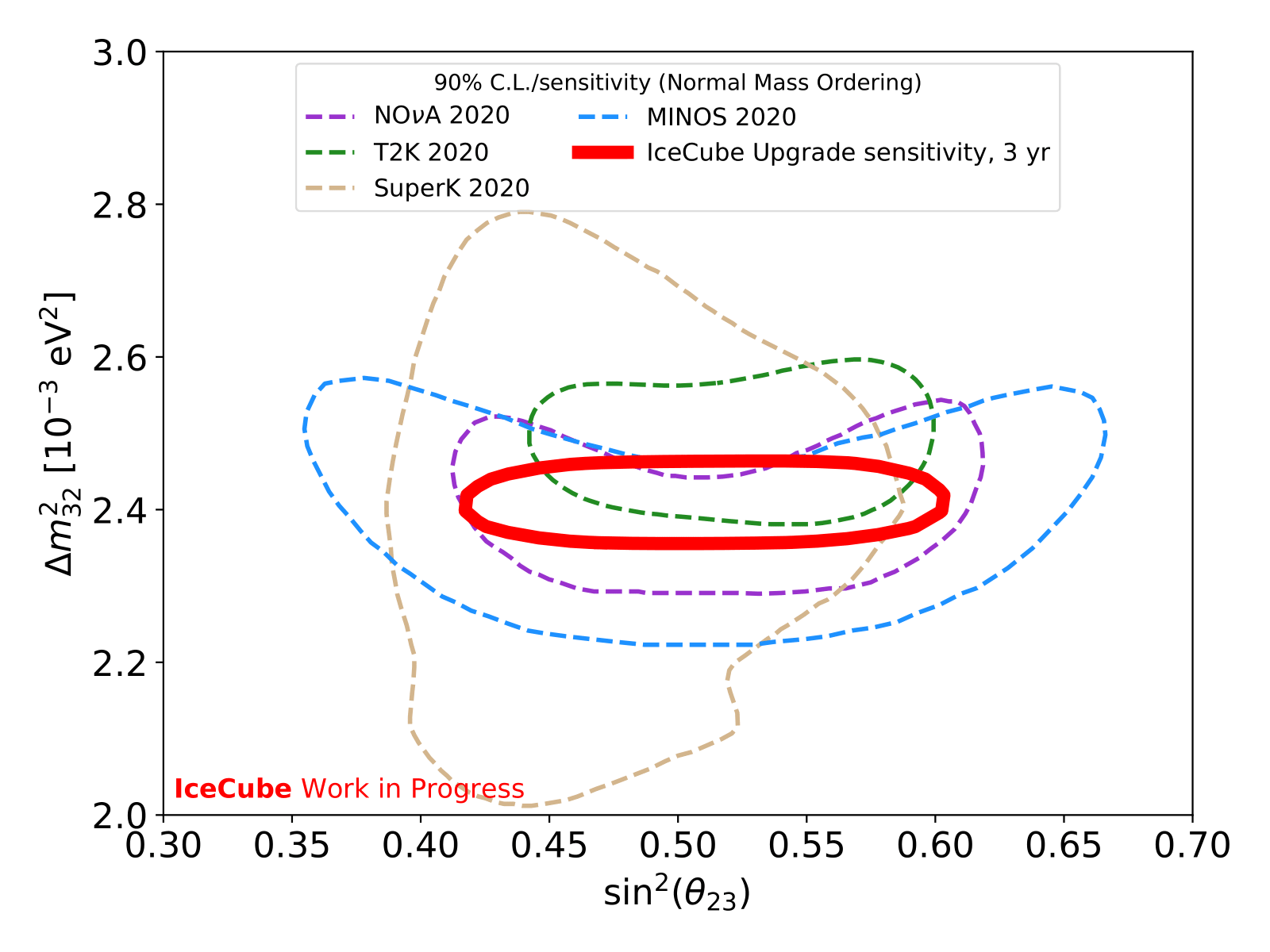}
  \caption{
  Measurements of $\sin^2 (\theta_{23})$ and $\Delta m^2_{32}$ with the IceCube Upgrade (inner fiducial volume) with 3 years of data, comparing with long-baseline neutrino oscillation facilities and other Cherenkov detectors \cite{Super-Kamiokande:2017edb, OPERA:2018nar, Haegel:2017Ck, Whitehead:2016xud}. From Ref.~\cite{Ishihara:2019aao, Stuttard22}.  
  }
  \label{fig:oscillation}
\end{figure}

The history of cosmic-ray and neutrino studies has witnessed many discoveries central to the progress of High Energy Physics, from the watershed identification of new elementary particles in the early days, to the confirmation of long-suspected neutrino oscillations, to measuring cross sections and accessing particle interactions far above accelerator energies. There have recently been two major achievements towards this progress: {\it (i)}~the measurement of the proton-proton cross section at $\sqrt{s} \sim 75~{\rm TeV}$~\cite{PierreAuger:2012egl, TelescopeArray:2015oxb, Abbasi:2020chd}, which provides evidence that the proton behaves as a black disk at asymptotically high energies~\cite{Block:2012nj, Block:2015mjw}, and {\it (ii)}~the measurements of both the charged-current neutrino-nucleon cross section~\cite{IceCube:2017roe, Bustamante:2017xuy} and the neutral-to-charged-current cross-section ratio~\cite{Anchordoqui:2019ufu} at $\sqrt{s} \sim 1~{\rm TeV}$, which provide restrictive constraints on fundamental physics at sub-fermi distances. Moreover, at ultra-high energies, neutrino interactions probe the structure of the proton in kinematic regions that cannot be explored by accelerator experiments. In the coming decade, neutrino-nucleon cross sections will be probed well above the energy scale of colliders, testing many allowed novel-physics scenarios; see Fig.~\ref{fig:cross_section_uhe} and Refs.~\cite{Ackermann:2022rqc,Denton:2020jft,Esteban:2022uuw,Valera:2022ylt} for details. Additionally, the inelasticity distribution of events detected at neutrino telescopes has also been envisioned as an important tool for revealing new physics processes~\cite{Anchordoqui:2006wc}. Current IceCube data in the TeV--PeV range are in good agreement with the SM predictions~\cite{IceCube:2018pgc}. Finally, the cosmic neutrino observatories such as the upcoming IceCube Upgrade will provide complimentary sensitivity to neutrino oscillation measurements by long baseline oscillation facilities, as shown in Fig.~\ref{fig:oscillation} and discussed in the white papers \citep{Ackermann:2022rqc, Abraham:2022jse}. 

Above $\sim 1$~PeV, $W$-boson production becomes relevant from two processes: electron anti-neutrino scattering on atomic electrons and neutrino-nucleus interactions in which the hadronic coupling is via a virtual photon. The former produces the distinct Glashow resonance at 6.3~PeV~\cite{Glashow:1960zz}. The latter can reach up to 5--10\% of the deep-inelastic-scattering cross section in the PeV range~\cite{Seckel:1997kk, Alikhanov:2015kla, Gauld:2019pgt, Zhou:2019vxt, Zhou:2019frk}. IceCube recently reported the detection of a particle shower compatible with a Glashow resonance event~\cite{IceCube:2021rpz}. $W$-boson production~\cite{Seckel:1997kk, Alikhanov:2015kla, Zhou:2019vxt, Zhou:2019frk} can play a significant role in the detection of tau neutrinos from cosmic origin~\cite{Soto:2021vdc}. Future analyses with more neutrino data above PeV energies will better probe these effects.

LHC experiments provide a laboratory for measurements relevant to understand the subtleties of astroparticle physics. Atmospheric neutrinos, produced in the interaction of cosmic rays with nuclei in the Earth’s atmosphere and the subsequent decay of mesons, are an irreducible background to searches for cosmic  neutrinos (see e.g.~\cite{IceCube:2021uhz}). An accurate understanding of the physics of cosmic sources therefore requires an in-depth understanding of the atmospheric neutrino flux. The CERN's Forward Physics Facility (FPF) will provide key information to reduce the uncertainties for cosmic neutrino searches in the context of multimessenger astrophysics~\cite{Feng:2022inv, Anchordoqui:2021ghd,Bai:2022jcs,Jeong:2021vqp}. More concretely, LHC neutrinos to be measured at the FPF experiments could give critical information on charm production at Feynman-$x$ close to 1. This process could potentially become a source of background for cosmic neutrinos above $10^{7.3}~{\rm GeV}$~\cite{Anchordoqui:2022ivb} and, at the moment, we have no supportive data and no theory for the process.

\subsection{Beyond-Standard-Model Neutrino Physics} 
\label{s:nuosc}

Neutrinos could have beyond-Standard-Model (BSM) interactions and, if the coupling strengths are weak or if heavy particles mediate the interactions, these interactions may only manifest themselves in the high-energy (HE) and ultra-high-energy (UHE) neutrino sector. Possible scenarios include BSM neutrino interactions with dark matter---including heavy dark matter---and with sterile neutrinos.

While the SM allows for interactions among neutrinos, these interactions are all highly suppressed by the electroweak scale. It is still unknown whether there are additional, BSM secret interactions solely among neutrinos that are stronger. In a UV-complete BSM model, this implies the existence of some new electrically neutral mediator, significantly lighter than the Z boson, that couples to neutrinos.

New interactions of this type can significantly modify the character of the HE and UHE neutrino flux arriving at the Earth by the scattering of the HE/UHE neutrinos off of nearly at-rest cosmic neutrino background (C$\nu$B) neutrinos. Such spectral distortion features may appear in the PeV--EeV regime, depending on the absolute mass of neutrinos~\cite{Hooper:2007jr, Lykken:2007kp, Ioka:2014kca, Ng:2014pca, Ibe:2014pja, Blum:2014ewa, DiFranzo:2015qea, Cherry:2016jol, Kelly:2018tyg, Barenboim:2019tux, Murase:2019xqi, Bustamante:2020mep, Creque-Sarbinowski:2020qhz}. The secret neutrino interactions are independently motivated by the neutrino mass generation mechanism~\cite{Blum:2014ewa}, muon $g-2$ anomaly~\cite{Araki:2015mya}, small-scale problems in dark matter substructures~\cite{vandenAarssen:2012vpm, Tulin:2017ara}, and apparent Hubble tension~\cite{Cyr-Racine:2013jua, Kreisch:2019yzn, Blinov:2019gcj,Carpio:2021jhu}.

In addition, flavor and $\nu/\bar{\nu}$ ratios provide complementary probes of new neutrino physics and neutrino production mechanisms~\cite{Ackermann:2022rqc, Engel:2022bgx, Arguelles:2022xxa, Abraham:2022jse}. Due to the fact that neutrinos are predominately expected from the decays of muons and charged pions, the nominal expectation is that only electron and muon neutrinos are generated at the sources and that $\nu$ and $\bar{\nu}$ are produced in comparable numbers. After leaving the sources, oscillations over cosmological distances are expected to distribute the flux nearly evenly among all flavors by the time the neutrinos reach Earth. In reality, however, different neutrino production channels become accessible at different energies and, as a result, the flavor and $\nu/\bar{\nu}$ ratios should vary with energy~\cite[see, e.g.,][]{Anchordoqui:2003vc, Anchordoqui:2004eb, Lipari:2007su,  Bustamante:2015waa}. Following this, the expected flavor ratios at Earth might deviate from a democratic flavor composition, and may do so as a function of energy.  Hence, the flavor ratios measured at Earth~\cite{IceCube:2015rro} combined with information about the values of the neutrino mixing parameters~\cite{Esteban:2020cvm} can be used to infer the flavor ratios at the sources~\cite{Palomares-Ruiz:2015mka, Bustamante:2019sdb, Song:2020nfh}. However, as discussed in Sec.~\ref{s:portals}, large deviations are possible in some BSM scenarios (e.g., neutrino decay, pseudo-Dirac states, new neutrino interactions with dark matter or sterile neutrinos, violation of Lorentz and CPT symmetries) which can alter the oscillation parameters~\cite{Beacom:2002vi, Beacom:2003eu, Barenboim:2003jm, Beacom:2003nh, Anchordoqui:2005ey, Bustamante:2010nq, Arguelles:2015dca, Shoemaker:2015qul, Rasmussen:2017ert, Ahlers:2018yom, Denton:2018aml, Ahlers:2020miq,Abdullahi:2020rge}. Large event statistics and complementary flavor-specific detection techniques are needed to identify flavor-specific signals and to measure the flavor composition statistically in a sample of collected events.

Constraints on BSM neutrino interactions using current IceCube data have been derived in Refs.~\cite{Bustamante:2020mep, Esteban:2021tub}. Future detectors, with improvements in particular in detector energy resolution and capability to identify neutrino flavor, are crucial to probing the BSM neutrino physics.   

 The ANtarctic Impulsive Transient Antenna (ANITA) has observed two anomalous events, which qualitatively look like air showers initiated by energetic ($\sim 500~{\rm PeV}$) particles that emerge from the ice along trajectories with large elevation angles ($\sim 30^\circ$ above the horizon)~\cite{ANITA:2016vrp,ANITA:2018sgj}. As was immediately noted by the ANITA Collaboration, these events may originate in the atmospheric decay of an upgoing tau-lepton produced through a charged current interaction of a tau-neutrino inside the Earth. However, for the angles inferred from ANITA observations, the ice would be well screened from up-going high-energy neutrinos by the underlying layers of Earth, challenging SM explanations~\cite{Romero-Wolf:2018zxt}. As of today, the origin of these anomalous events remains unclear; follow-up observations of these unusual events by EUSO-SPB2~\cite{Eser:2021mbp} and PUEO~\cite{PUEO:2020bnn} are well-motivated~\cite{Ackermann:2022rqc}.

 \subsection{The Muon Puzzle of Ultra-High-Energy Cosmic Rays} 
    \label{s:muonpuzzle}

The muonic component of cosmic-ray air showers is generally used as a probe of the hadronic interactions during the cascade development~\cite{Coleman:2022abf}. Various measurements of atmospheric muons with energies  $1 \lesssim E_\mu/{\rm GeV} \lesssim 10$ have revealed a discrepancy between simulated and observed muon production in air showers. The highest energy cosmic rays currently observed by the Pierre Auger Observatory (Auger)  show a significant discrepancy in the shower muon content when compared to predictions of LHC-tuned hadronic event generators~\cite{PierreAuger:2014ucz, PierreAuger:2016nfk}. More concretely, the analysis of Auger data suggests that the hadronic component of showers (with primary energy $10^{9.8} < E/{\rm GeV} < 10^{10.2}$) contains about $30\%$ to $60\%$ more muons than expected with a significance somewhat above $2.1\sigma$. The discrepancy between experiment and simulations has also been observed in the Telescope Array data analysis at the same energy range~\cite{TelescopeArray:2018eph}. Auger findings have also been recently confirmed by studying air-shower measurements over a wide range of energies. The muon deficit between simulation and data, dubbed the {\it muon puzzle}, seems to start at $E \sim 10^8~{\rm GeV}$, increasing noticeably as primary energy grows, with a slope that was found to be significant at $\sim8\sigma$~\cite{Albrecht:2021cxw}. However, the muon deficit of simulated events has not been observed in IceTop data with $10^{6.4} < E/{\rm GeV} < 10^{8.1}$~\cite{IceCube:2022yap}. It is noteworthy that, in this energy range, the cosmic-ray spectrum has a significant contribution of nulcei~\cite{Coleman:2022abf}, and so the center-of-mass energy per nucleon pair of the showers in the IceTop sample is well below those of LHC collisions. In contrast, the center-of-mass energy per nucleon of a $10^{10.3}~{\rm  GeV}$ helium nucleus incident upon a nucleon in the atmosphere is 100~TeV, and so many secondary interactions would be above the LHC center-of-mass energy~\cite{Allen:2013hfa}. Within this decade, ongoing detector upgrades of existing facilities---such as AugerPrime~\cite{PierreAuger:2016qzd} and IceCube-Gen2~\cite{IceCube-Gen2:2020qha}---will enhance the precision of air-shower measurements and reduce uncertainties in the interpretation of muon data. In particular, as a part of the upcoming AugerPrime upgrade, each surface station will have additional detectors that will provide complementary measurements of the incoming shower particles, consequently leading to improved reconstruction of muons and electromagnetic particles~\cite{PierreAuger:2016qzd}. This will allow for the measurement of the properties of extensive air showers initiated by the highest energy cosmic rays with unprecedented precision, providing a unique probe of hadronic collisions at center-of-mass energies that surpass the LHC energy.

In solving the muon puzzle, one has to simultaneously get good agreement with the measurements of the distribution of the depth of shower maximum $X_{\rm max}$~\cite{PierreAuger:2014sui} and the fluctuations in the number of muons~\cite{PierreAuger:2021qsd}. A thorough phenomenological study has shown that an unrivaled solution to the muon deficit, compatible with the observed $X_{\rm max}$ distributions, is to reduce the transfer of energy from the hadronic shower into the electromagnetic shower by reducing the production or decay of neutral pions~\cite{Allen:2013hfa}. Hence, the amount of forward strangeness production could be of particular relevance in addressing the muon puzzle~\cite{Allen:2013hfa, Anchordoqui:2016oxy, Baur:2019cpv, Anchordoqui:2022fpn}. Strangeness production is traced by the ratio of charged kaons to pions, for which the ratio of electron and muon neutrino fluxes is a proxy that will be measured by the FPF  experiments~\cite{Feng:2022inv, Anchordoqui:2021ghd}. Electron neutrino fluxes are a measurement of kaons, whereas both muon and electron neutrinos are produced via pion decay. However, $\nu_\mu$ and $\nu_e$ populate different energy regions, which can help to disentangle them. In addition, neutrinos from pion decay are more concentrated around the line-of-sight  than those of kaonic origin, given that $m_\pi < m_K$, and thus neutrinos from pions obtain less additional transverse momentum than those from kaon decays. Thereby, the closeness of the neutrinos to the line-of-sight or, equivalently, their rapidity distribution, can be used to disentangle different neutrino origins to get an estimate of the pion-to-kaon ratio. This implies that measurements at the FPF will improve the modeling of high-energy hadronic interactions in the atmosphere, reduce the associated uncertainties of air-shower measurements, and thereby help to understand the properties of cosmic rays, such as their energy and baryonic structure, which is crucial to discover their origin.

There is also the possibility that the muon puzzle does not originate from an incomplete understanding of the forward particle physics. If this were the case, future ultra-high-energy cosmic-ray (UHECR) measurements would provide a unique probe of BSM physics with a high potential for discovery~\cite{Coleman:2022abf}.

\subsection{The Nature of Matter in Neutron Star Interiors} 
\label{s:NS_EOS}

The nature of matter at ultra-high densities ($\rho_s > 2.8 \times 10^{14}$\,g\,cm$^{-3}$), large proton/neutron number asymmetry, and low temperatures ($\lesssim 10^{10}~{\rm K}$) is, at present, one of the major outstanding problems in modern physics, owing to a number of challenges both in the experimental and theoretical realms~\cite[see, e.g.,][for a review]{Bogdanov:2022faf, Watts:2016uzu}. A plethora of well-motivated theoretical predictions for the state of matter in this temperature-density regime have been proposed, ranging from normal nucleonic matter, to particle exotica such as hyperons, deconfined quarks, color superconducting phases, and Bose-Einstein condensates (for a review see~\cite{Watts:2016uzu}. Matter in this extreme regime is known to only exist stably in the cores of neutron stars (NSs).
\vskip5pt

\noindent
\begin{minipage}{0.48\textwidth}
Neutron stars are host to the densest matter in the Universe. The density increases toward the center of the star, reaching densities of 5--10\,$n_s$, that is, several times the nuclear saturation density of $n_s = 0.16/$fm$^3$. At these densities, we currently do not know how matter behaves, what the phase structure is, or what the dynamic degrees of freedoms are. Neutron stars offer a unique laboratory to study strongly interacting matter and the underlying theory of QCD in the most extreme conditions. 
\end{minipage}
\hfill
\begin{minipage}{0.48\textwidth}
    \centering
    \includegraphics[width=0.98\textwidth]{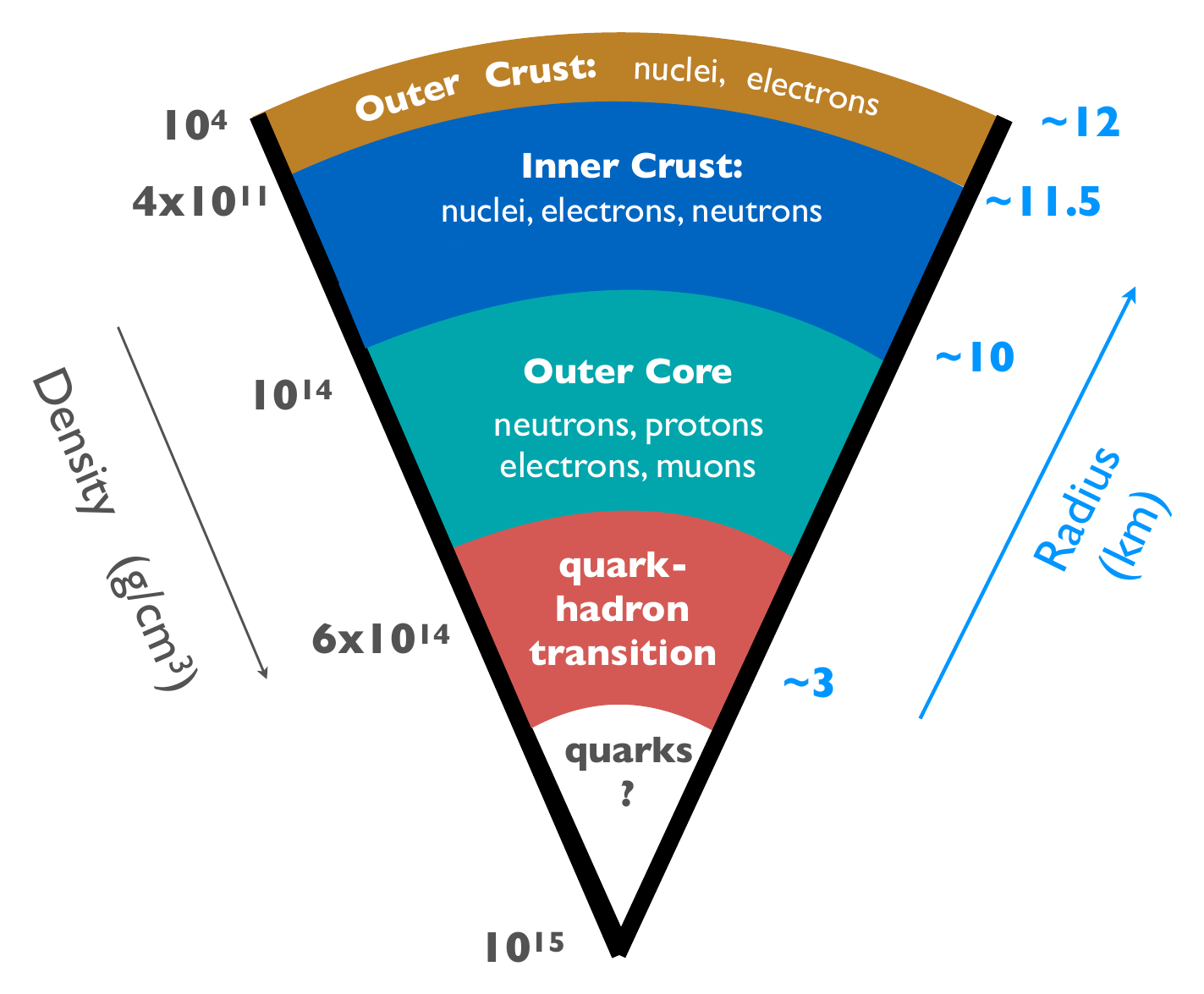}
    \vskip-0.5cm
    \captionsetup{type=figure}
    \captionof{figure}{The structure of a neutron star as predicted by theory.}
    \label{fig:structure} 
\end{minipage}
\vskip5pt
They have the potential to facilitate the discovery of novel exotic phases of matter in their cores, including the appearance of strangeness in the form of hyperonic matter and ultimately the melting of nucleonic structure, giving rise to novel forms of cold quark matter.

The structure of NSs is determined by the competition between self gravity and pressure of strong nuclear interactions keeping the star in a hydrostatic equilibrium. This interaction is described in the simplest case of non-rotating NSs by the Tolman-Oppenheimer-Volkoff (TOV) equations~\cite{Oppenheimer:1939ne, Tolman:1939jz}, which map the equation of state (EOS) of dense nuclear matter to the macroscopic properties of NSs, making the EOS the primary object of interest for the nuclear physics of NSs. A large community effort has been put to investigating the EOS using different \emph{ab-initio} calculations as well as various models. 

\begin{figure*}[t!]
\centering
\includegraphics[width=0.8\textwidth]{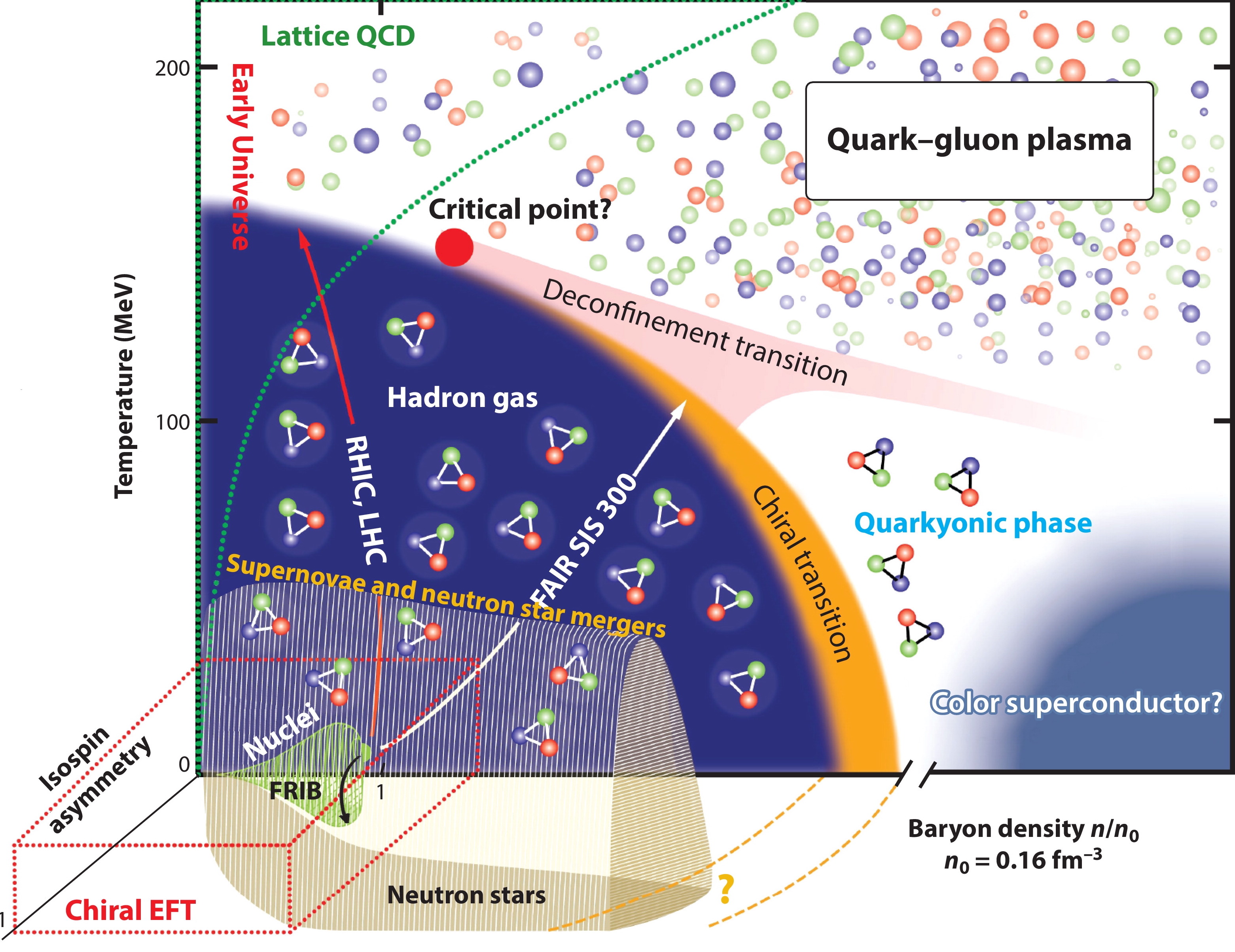}
\caption{Schematic view of the QCD phase diagram. The figure highlights regions probed by the RHIC, LHC, FAIR, and FRIB experiments, regions of validity for lattice QCD and chiral EFT, and environments reached in neutron stars, supernovae, and neutron star mergers. Abbreviations: EFT=effective field theory, QCD=quantum chromodynamics.[Borrowed from \url{https://www.annualreviews.org/doi/full/10.1146/annurev-nucl-102419-041903}]}
\vspace{-0.3cm}
\label{fig:eosphysics}
\end{figure*}

The evidence from observations, in particular with the advent of multimessenger astronomy, is more complicated than the simple static systems described by the TOV equations and a significant push has been made in the past years to numerically solve the combined Einstein and relativistic-fluid-dynamic equations~\cite{Romatschke:2017ejr}.  Fig.~\ref{fig:structure} displays a schematic figure of the NS structure. The crust and the outer core down to a depth of roughly $0.5$~km, where densities are of the order of nuclear density, is under good theoretical control~\cite{Baym:1971pw, Negele:1971vb}. Beyond that, our understanding of the structure relies on theoretical extrapolations. In particular, the phase of the inner core is currently unknown. Fig.~\ref{fig:eosphysics} is a schematic view of the hypothesized phase diagram of QCD. It is a firm prediction of QCD wherein, at sufficiently high temperatures and/or densities, ordinary hadronic matter melts to a partonic form of matter--- Quark Matter. 
In the regime of high temperatures and low baryon densities, the deconfiement transition to Quark Matter is well studied using lattice field theory~\cite{Fodor:2004nz,Aoki:2006we, Karsch:2001cy} and its existence is confirmed in two decades of experimentation with ultra-relativistic heavy-ion collisions~\cite{Margetis:2000sv} at the RHIC and the LHC. Further experimental program runs at the LHC aim to quantify the transport properties and the conditions of the onset of Quark Matter.

The deconfinement transition is a cross-over at low baryon densities but it is been long hypothesized that the transition becomes stronger with increasing baryon density. New theoretical arguments based on topological features of QCD have been recently put forward supporting the first-order-nature of the transition~\cite{Cherman:2018jir, Fukushima:2010bq}. The beam energy scan (BES) program at the RHIC~\cite{STAR:2017sal} and the future FAIR facility~\cite{CBM:2016kpk} are geared to discover the critical point separating the crossover transition from the first-order transition. The discovery of the critical point would have a profound impact on physics of NSs.

At very-high densities, owing to an attractive interaction between quarks in QCD, it is expected that Quark Matter is in the form of a color superconductor, fundamentally affecting the transport properties of Quark Matter~\cite{Alford:2007xm, Alford:1997zt}. Based on large-$N_c$ arguments, it has also been speculated~\cite{McLerran:2018hbz} that, at low temperatures, there may be a further intermediate phase that is still confined but where the chiral symmetry is restored. Owing to the similarities with both the hadronic and Quark Matter phases, this hypothetical phase is dubbed the quarkyonic phase.

Gravitational-wave observations of binary neutron star mergers can constrain the cold state of ultra-high density matter in neutron stars from tidal effects during the inspiral phase of the binary made possible with GW170817 \cite{LIGOScientific:2018cki, De:2018uhw, Bauswein:2017vtn}, as well as observe the dynamics of the hot, dense matter after merger, which will become possible with the next generation of gravitational-wave detectors.

Constraints on the EOS from mass and radius observations of NSs can be complemented with constraints on the nuclear-symmetry energy obtained from nuclear experiments and \textit{ab-initio} neutron matter theory.  It has been known for some time that there is a high degree of correlation between neutron star radii and the pressure of neutron star matter slightly above the nuclear saturation density ($n_s=0.16$ fm$^{-3}$ or $\rho_s=2.7\times10^{14}$ g cm$^{-3}$). The NS matter pressure at $n_s$ is nearly completely determined by the slope $L$ of the symmetry energy at the same density~\cite{2001ApJ...550..426L}, which is experimentally probed by nuclear binding energies, dipole polarizabilities and neutron skin thicknesses of neutron-rich nuclei, and, theoretically, from neutron matter studies.

An important set of measurements in nuclear astrophysics comes from studying the timing properties of radio pulsars--- rotating NSs that emit beamed radiation along their magnetic poles. An early example is the discovery and analysis of the ``Hulse-Taylor" radio pulsar-binary system which (indirectly) confirmed of the existence of gravitational radiation~\cite{Weisberg:1981bh, Taylor:1982zz} and produced mass estimates with high precision. The recent discoveries of additional relativistic pulsar orbits and high-mass NSs, as well as the eventual detection of nanohertz-frequency gravitational radiation through pulsar timing, show that radio pulsars continue to serve as ideal laboratories for fundamental physics.

Space-borne observations at X-ray energies offer various means for obtaining strong constraints on the allowed dense-matter EOS, providing unique insight into the high-density, low-temperature region of the QCD phase diagram. While current telescopes have made important headway, they lack the required capabilities to fully exploit the information about the dense-matter EOS encoded in the observed X-ray emission from NSs. This important undertaking requires a new generation of X-ray facilities with at least an order-of-magnitude improvement in sensitivity relative to current observatories, while also offering the high time resolution required for effective studies of rapidly spinning NSs.

Collectively, electromagnetic (radio and X-ray) and gravitational-wave astrophysical measurements, combined with terrestrial laboratory constraints, hold the promise to provide definitive empirical constraints on the true nature of the densest matter in the Universe~\cite{Miller:2019cac, Raaijmakers:2019qny, 2020ApJ...892...55J, 2020PhRvD.101l3007L}.

\subsection{Tests of Lorentz  and  CPT invariance} 
\label{s:LIV}

Both Lorentz and CPT symmetries are fundamental to our understanding of the SM and General Relativity~\cite{Coleman:1998ti, Stecker:2017gdy, Altschul2011}. Lorentz invariance (LI)---one of the main symmetries that govern the SM of elementary particles---requires the structure of spacetime to be the same for all observers. However, proposed Grand Unified Theories suggest that our understanding of spacetime symmetries may be incomplete and that fundamental modifications to the Lorentz symmetry could be made to account for quantum effects, thereby potentially violating this symmetry when approaching the Planck scale~\cite{Addazi:2021xuf}. Lorentz invariance violation (LIV) at a high-enough energy scale could actually arise in loop quantum gravity or string theory~\cite{Alfaro:2004aa, AmelinoCamelia2001, Bluhm:2013mu, Calcagni:2016zqv, Colladay:1998fq, EllisMavromatosNanopoulos1999, GambiniPullin1999, Kostelecky:1988zi, Nambu1968, Potting2013}. Closely intertwined with Lorentz symmetry is the CPT symmetry--- the established discrete spacetime symmetry of charge, parity, and time reversal. In a local quantum theory, it is impossible to violate CPT invariance without also breaking LI~\cite{Greenberg:2002uu}. Thus, many tests of LI can also be interpreted as tests of CPT. 

Even a small violation of LI could easily affect the propagation of particles on a cosmological scale~\cite{Coleman:1998ti, Aloisio:2000cm}.  Moreover, at extreme energies, like those available in the collision of ultra-high-energy cosmic rays in the Earth's atmosphere, one could also expect a change in the interactions driving the air-shower development due to LIV~\cite{Coleman:2022abf, Tomar:2017mgc}.  

As there are several signatures of the violation of these fundamental symmetries, there are a variety of tests that may be performed to search for them. The Pierre Auger Collaboration has derived limits on LIV by comparing the energy spectrum and cosmic-ray composition with upper limits on the photon flux~\cite{PierreAuger:2021tog} and by comparing Monte Carlo expectations to muon fluctuation measurements~\cite{PierreAuger:2021mve}. In the years ahead, the most restrictive bounds on LIV could be coming from UHECR experiments~\cite{PierreAuger:2021tog, Coleman:2022abf}. Additionally, as discussed in Sec.~\ref{s:nuosc}, the flavor ratio of cosmic neutrinos provides a powerful test of Lorentz and CPT symmetries. 

Precise measurements of very-high-energy photons can also be used to test LIV~\cite[see, e.g.,][]{Vasileiou:2013vra, HAWC:2019gui, LHAASO:2021opi}. One consequence of LIV is that photons of sufficient energy are unstable and decay over short timescales~\cite{Martinez-Huerta:2017ulw}. This means that high-energy photons from astrophysical objects may decay well before they can arrive at Earth. Constraints to the LIV energy scale have been established by looking at the highest-energy photons from the Crab nebula, eHWC J1825-134, and LHAASO J2032+4102~\cite{LHAASO:2021opi, HAWC:2019gui}. However, higher limits are expected from continued observations of even more high-energy sources, such as RXJ1713.7-3946, with upcoming observatories including the Southern Wide-field Gamma-ray Observatory (SWGO)~\cite{Albert:2019afb, Hinton:2021rvp, Schoorlemmer:2019gee} and the Cherenkov Telescope Array (CTA)~\cite{CTAConsortium:2017dvg}. The higher the energy of a detected gamma ray and the narrower its energy uncertainty, the more stringent the constraints would be. Thus, instruments optimized at the highest energies, such as SWGO, LHAASO~\cite{LHAASO:2019qtb}, and CTA, would be optimal instruments to search for LIV signatures.

\subsection{Production of Exotic Particles in the QED Domain}

Exotic quantum electrodynamics (QED) processes may operate in extremely strong magnetic fields with $B> B_{\rm cr} = 4.4\times 10^{13}$~G, when $h\nu_B \sim m_e c^2$ is achieved.   
Magnetars, a topical subclass of neutron stars with surface fields exceeding $10^{14}$~G  \cite{Harding:2006qn, Mereghetti:2008je, Turolla:2015mwa, Kaspi:2017fwg}, provide a cosmic lab to test QED in this domain. The potential action of exotic QED mechanisms of photon splitting and magnetic pair creation yields distinctive imprints on magnetar polarization and Comptonization, which may be observed in the sub-MeV waveband \cite{Wadiasingh:2017rcq, Hu:2019nyw} by future MeV telescopes like the All-sky Medium-Energy Gamma-ray Observatory (AMEGO)~\cite{AMEGO:2019gny}.

\section{Multimessenger Synergies in Particle Astrophysics}
\label{s:multimessenger}     

Multimessenger astrophysics encompasses the measurement of any cosmic event with more than one type of signal--- photons, gravitational waves, neutrinos, or cosmic rays. The dawn of the modern Multimessenger Era was heralded by the co-detection of gamma rays and gravitational waves in a binary neutron star merger~\cite{LIGOScientific:2017ync} and by the co-detection of gamma rays and neutrinos in a blazar flare~\cite{IceCube:2018dnn}. We learned more from each of those singular co-detection events than a decade of astrophysical observations could have told us with photons alone. Over the next decade, multimessenger detections will become more important to accelerating the rate of discoveries in cosmic particle physics by constraining coincident event types from different messengers simultaneously. The United States currently leads efforts in multimessenger astrophysics through the investments DOE, NSF-Physics, and NASA have made over the past several decades. Maintaining U.S. primacy in this field will require the support of a well-balanced program of facilities across all messengers in complementarity with our collaborators around the world and leadership in the rigorous task of coordinating between them. The following section highlights several compelling astroparticle physics areas that are best addressed with multimessenger methods. 
     
\subsection{Pinpointing the Sources of the Highest-Energy Cosmic Rays} 
      \label{UHECRsources}

It is well known that the cosmic microwave background (CMB) makes the Universe opaque to the propagation of ultra-high-energy cosmic rays (UHECRs). The so-called GZK interactions of cosmic rays above the photopion production threshold (or nucleus photodisintegration) and the relic photons lead to a sharp cutoff in the UHECR spectrum above about $10^{10.6}~{\rm GeV}$~\cite{Greisen:1966jv, Zatsepin:1966jv}. It was recently noted that the characteristic cosmic-ray energy of the GZK cutoff could coincide with the species scale ($\hat M \sim 10^{10}~{\rm GeV}$), where 
physics becomes strongly coupled to gravity~\cite{Montero:2022prj}. This suggests that the cosmic-ray maximum energy may be driven by the species scale. Hence, aside from its astrophysical motivations, understanding the origin of the abrupt cutoff observed in the UHECR flux around $10^{10.6}~{\rm GeV}$~\cite{PierreAuger:2008rol, HiRes:2007lra} could have direct applications to probe BSM physics and posses a huge challenge for UHECR experiments within the next decade. In particular, a precise characterization of the source spectra of the highest-energy cosmic rays has the potential for breakthrough results in fundamental physics~\cite{Anchordoqui:2022ejw}. 

Rapid progress in computational high-energy astrophysics is dramatically advancing the study of acceleration mechanisms. Some of the current contenders for acceleration mechanisms and source types are shock acceleration in systems ranging from the large-scale shocks surrounding galaxy clusters~\cite{Kang:1996rp,Ryu:2003cd} to internal or external shocks of starburst-superwinds~\cite{Anchordoqui:1999cu, Anchordoqui:2018vji, Anchordoqui:2020otc}, active galactic nuclei (AGN)~\cite{Biermann:1987ep, Takahara:1990he, Rachen:1992pg, Blandford:2018iot, Matthews:2018laz, Matthews:2018rpe, Eichmann:2022ias} or gamma-ray burst~\cite{Waxman:1995vg, Vietri:1995hs, Wang:2007xj, Murase:2008mr, Baerwald:2013pu, Globus:2014fka, Zhang:2017moz} jets, and the jets of tidal disruption events (the transient cousins of AGN jets)~\cite{Farrar:2008ex,Farrar:2014yla,Pfeffer:2015idq}. Other contenders are shear acceleration~\cite{Rieger:2004jz,Kimura:2017ubz} and one-shot mechanisms such as ``espresso"~\cite{Caprioli:2015zka}, in which an AGN or other jet boosts a galactic cosmic ray of the host galaxy; electromotive force acceleration as in fast-spinning pulsars~\cite{Blasi:2000xm, Fang:2012rx, Fang:2013cba} and magnetars~\cite{Arons:2002yj}, black holes~\cite{Blandford:1977ds, Neronov:2009zz}, and potentially reconnection, explosive reconnection, gap and/or wakefield acceleration~\cite{Chen:2002nd, Murase:2009pg, Ebisuzaki:2013lya}. The abundance of possibilities suggests there may well be multiple sources of UHECRs---some of which may be transient---making the identification of sources even more challenging and essential~\cite{Coleman:2022abf, Engel:2022yig}. The Study of particle acceleration in astrophysical plasmas is a near-term application of the accelerator physics as pointed out by the Snowmass 2021  Accelerator Frontier White Paper: ``Near Term Applications driven by Advanced Accelerator Concepts" \citep{Emma:2022zdv}. 

On the experimental side, the Pierre Auger Collaboration (Auger) has discovered a large-scale dipole anisotropy above $10^{9.9}~{\rm GeV}$ with a significance $> 6\sigma$~\cite{PierreAuger:2017pzq}. Given that the dipole direction is $\sim 115^\circ$ away from the Galactic center, this is evidence of the extragalactic origin of cosmic rays above this energy threshold. Intriguingly, the dipole direction is not aligned with the CMB dipole, the local matter over-density, or any obvious individual source. A further analysis finds a $4\sigma$ significance for correlation of cosmic rays above $10^{10.6}~{\rm GeV}$ with a model based on a catalog of bright starburst galaxies and a $3.1\sigma$ correlation with a model based on a \textit{Fermi}-LAT catalog of jetted AGNs~\cite{PierreAuger:2018qvk,PierreAuger:2022axr}. The best-fit Gaussian angular scales correspond to a top-hat radii of $25^\circ$ and the signal fractions range from 5--10\%. Most of the anisotropy signal comes from the so-called Centaurus region (which contains the jetted AGN Centaurus A as well as the starburst galaxies NGC4945 and M83). The starburst model also benefits from one prominent source candidate, NGC253, being close to the southern Galactic pole where a warm spot of Auger events is found. When data from the Telescope Array are included in the analysis, the correlation with starburst galaxies is mildly stronger than the Auger-only result with a post-trial significance of $4.2\sigma$~\cite{TelescopeArray:2021gxg}. Continuing operation of Auger should yield a significance level of $5\sigma$ for the Centaurus region excess by the end of 2025 ($\pm 2$ calendar years), possibly preceded by a similar significance milestone in the correlation with the starburst catalog if the warm spot continues to grow~\cite{Coleman:2022abf}.

Interaction between UHECRs and the CMB leads to the production of ultra-high-energy (UHE) neutrinos~\cite{Berezinsky:1969erk}. The so-called GZK process is effective when the energies of UHECR nucleons are higher than $\sim 5\times 10^{10}$~GeV and the corresponding cosmogenic neutrinos see their main flux around and below $\sim 10^{9}$~GeV~\cite{Stecker:1978ah,Yoshida:1993pt, Takami:2007pp, Anchordoqui:2007fi, Kotera:2010yn, Ahlers:2010fw, AlvesBatista:2018zui}. Upper limits on the cosmogenic neutrino fluxes have been obtained by IceCube~\cite{IceCube:2018fhm}, the Pierre Auger Observatory~\cite{PierreAuger:2019ens}, and ANITA~\cite{Gorham:2019guw}. Cosmogenic neutrinos when combined with UHECR observations could provide a unique multimessenger signature of GZK interactions, but as of today no neutrino has been observed with energy above $10^{7}~{\rm GeV}$.

Sources of Galactic UHECR neutrons, when combined with the antineutrino flux resulting from neutrons decaying on flight at lower energies provide a unique beam to test neutrino oscillations, as the expected Earthly neutrino flavor ratio differs from the nearly even distribution among electron, muon, and tau flavors (1:1:1) of astrophysical neutrinos originating via charged pion decay~\cite{Anchordoqui:2003vc}.

\subsection{Probing Extreme-Energy Hadron Acceleration and Interaction with Neutrinos} 

While gamma rays may be produced by both leptonic process, such as inverse Compton scattering of background photons, and hadronic process, such as pion decay, high-energy neutrinos may only be produced when hadronic cosmic rays interact with surrounding matter ($pp$) and light ($p\gamma$).  Thus, high-energy neutrinos provide a unique probe of hadron acceleration and interaction in astrophysical environments. 

The origin of the bulk of the high-energy neutrinos remains unknown \cite{IceCube:2019cia}, though hints to the first sources have been found. The coincident observations of a high-energy neutrino event, IceCube-170922A, with X-rays and gamma rays from the blazar TXS 0506 +056~\cite{IceCube:2018cha, IceCube:2018dnn} make this blazar the first candidate high-energy neutrino source. In addition, the ten-year point-source searches with IceCube indicated that NGC~1068 is the most significant steady source of neutrinos at a significance of $\sim3\sigma$~\cite{IceCube:2019cia}. 

Neutrinos are an important  probe of dense environments that are not visible with photons. Interestingly, gamma-ray limits and observations of these early sources indicate that the energy carried by hadrons must be significantly higher than that carried by leptons. Models that may explain the observed neutrinos require a large baryonic loading, i.e., a large fraction of the available energy imparted to cosmic rays, which may be theoretically challenging. 

The flux and spectral index of the TeV--PeV diffuse neutrino background are comparable to that of the GeV--TeV diffuse gamma-ray background, and the latter tightly constraining the flux of the electromagnetic cascades of the gamma-ray counterparts of high-energy neutrinos. The current IceCube measurements already indicate that unless new physics processes are at play~\cite{Anchordoqui:2021dls}, the bulk of the neutrino sources are likely opaque to gamma rays \cite{Murase:2015xka, Fang:2022trf}. Future observation of TeV and sub-TeV neutrinos may confirm the present indications that neutrinos originate in cosmic environments that are optically thick to GeV--TeV gamma rays. Such gamma-ray-obscured sources may be bright in 1--100~MeV energies and be observed by future MeV gamma-ray facilities like the All-sky Medium-Energy Gamma-ray Observatory (AMEGO)~\cite{AMEGO:2019gny}. This is a clear example of the predictive power of multimessenger science, which will be testable within this decade.

Future firm detections of high-energy neutrino sources and characterization of their spectra are crucial to the understanding of hadron acceleration and interaction in the cosmos. The Snowmass 2021 whitepaper ``Snowmass2021 Cosmic Frontier: Advancing the Landscape of Multimessenger Science in the Next Decade"~\cite{Engel:2022yig} discusses the current and future multimessenger network and the collaboration and infrastructure needed for successful multimessenger observations of neutrino sources.  

With the improved statistics, sensitivity, and sky coverage offered by upcoming neutrino experiments, we can expect to expand our view of the neutrino sky, including firmly establishing neutrino sources. Next-generation telescopes currently in th planning stage or under construction, as discussed in the Snowmass 2021 whitepaper ``High-Energy and Ultra-High-Energy Neutrinos"~\cite{Ackermann:2022rqc}, will allow detailed studies of high-energy neutrinos, including their energy spectrum, flavor composition, and the identity of their sources.

\subsection{Diffuse Backgrounds}

\label{subsec:diffuseBG}

Diffuse astrophysical backgrounds arise in all of the astrophysical messengers, not just due to the limitations of the resolutions of current detectors, but as an indication of large-scale and diffuse structure in the Universe. These diffuse backgrounds are studied extensively for individual messengers, but future insights to the origin of the cosmos may arise from considering their similarities and collaboration across diffuse working groups for each messenger~\cite{Engel:2022yig}.

{\bf Diffuse gamma-ray background} (DGRB). The DGRB is defined as a smooth residual component of the measured gamma-ray emission emerging after the subtraction of known sources of gamma rays, including both point-like and extended sources. The unresolved gamma-ray background (UGRB) can be explained by the cumulative emission of randomly distributed gamma-ray sources whose flux is below the sensitivity of the observing instrument. The UGRB between 100~MeV and 800~GeV is measured by the {\it Fermi} Large-Area Telescope (LAT)~\cite{Fermi-LAT:2010pat, Fermi-LAT:2014ryh}. The UGRB is expected to be contributed to largely by the faint subgroups of the bright gamma-ray source populations, including blazars~\cite{Ando:2006mt, Fermi-LAT:2015otn, DiMauro:2017ing} and star-forming galaxies~\cite{2010ApJ...722L.199F, Linden:2016fdd, Roth:2021lvk}. More exotic scenarios may contribute, as well-annihilating or decaying particles of dark matter in extragalactic halos may explain the diffuse backgrounds~\cite{Camera:2012cj, Camera:2014rja, Shirasaki:2014noa, Lisanti:2017qlb}. 

{\bf Diffuse Supernova Neutrino Background} (DSNB). The detection of 25~MeV neutrinos from SN1987A in the Large Magellanic Cloud marked the first time neutrinos were detected from a massive star undergoing core collapse~\cite{Kamiokande-II:1987idp, Bionta:1987qt}. While the low galactic supernova rate requires much larger neutrino detectors to detect more supernovae neutrinos from nearby galaxies (1--10~Mpc), another avenue to study these explosions is available through the detection of the DSNB, which consists of MeV neutrinos from all past core-collapse supernovae. The discovery prospects of the DSNB in the next decade are promising with the gadolinium-enhanced Super-Kamiokande detector~\cite{Beacom:2003nk, Super-Kamiokande:2021the}, Jiangmen Underground Neutrino Observatory (JUNO~\cite{JUNO:2015zny}), and Hyper-Kamiokande~\cite{Abe:2011ts}.  

{\bf Astrophysical Diffuse Neutrino Background}. The flux of diffuse neutrinos at TeV--PeV energies of astrophysical origin has been measured by IceCube with a significance well above $5\sigma$~\cite{IceCube:2013low, IceCube:2014stg}. Specifically, the flux has been measured using a sample of high-energy neutrinos, which includes both tracks and cascades with interaction vertices within the instrumented volume~\cite{IceCube:2020wum}, a sample of up-going tracks (mostly muon neutrinos)~\cite{IceCube:2021uhz}, a sample of cascade-like events (mostly electron and tau neutrinos)~\cite{IceCube:2020acn}, and a sample of tracks that start within the instrumented volume~\cite{IceCube:2018pgc}. An apparent slight tension between the different measurements could be due to differences in flavor composition, energy range, the accounting of atmospheric backgrounds, and the spectral model used. Future statistics and analyses with improved calibration and simulations will lead to improvements of the accuracy of the measurement and a reduction of systematic uncertainties. 

{\bf Cosmogentic Neutrino Background}. Whether the diffuse astrophysical neutrino flux extends to higher energies is unknown. Determining if this flux has or does not have a cutoff in the 10--100~PeV range is crucial for understanding the physics underlying UHECR accelerators and identifying source classes. Studying the diffuse neutrino flux in this energy regime also opens an avenue to probe fundamental neutrino physics and BSM physics at an energy scale that would be otherwise unreachable.

{\bf Galactic Diffuse Emission.} In addition to the extragalactic diffuse emission, Galactic diffuse emission of gamma rays and neutrinos is produced by energetic cosmic rays interacting with the interstellar medium and radiation fields in our Galaxy. The Galactic diffuse gamma-ray emission has been measured by {\it Fermi}-LAT between 0.1~GeV and 1~TeV and by H.E.S.S.~\cite{HESS:2014ree} and HAWC~\cite{HAWC:2021bvb} above 1~TeV. The Galactic diffuse neutrino emission has been constrained by IceCube and ANTARES~\cite{IceCube:2017trr, ANTARES:2018nyb}, but is expected to be detected in the near future~\cite{Fang:2021ylv}.   

The perspective of an {\bf indirect detection of dark matter} with the diffuse backgrounds depends on the level of the  understanding of the astrophysical sources of astroparticles including, but not limited to, the fraction of astrophysical contribution, the faint end of the luminosity function of the astrophysical contributors, and the cosmological evolution of the source classes.  
 
\subsection{Galactic TeVatrons and PeVatrons}
\label{subsec:TeVatronPeVatron}

The recent launch and operation of wide-field air-shower observatories, including the Tibet AS$\gamma$, HAWC, and LHAASO experiments, has opened up the view of the Universe in the very-high-energy (0.1--100~TeV) and ultra-high-energy ($> 100$~TeV) (note that these definitions of the energy ranges are adopted in gamma-ray astrophysics) regimes with unprecedented sensitivities. Ultra-high-energy gamma rays are produced by cosmic-ray protons and electrons at PeV energies. Detecting PeV proton accelerators, a.k.a. PeVatrons, are crucial to solving the long-standing puzzle of the ``knee" feature in the Galactic cosmic-ray spectrum. Several candidates have been identified so far~\cite{HAWC:2018gwz, Abeysekara:2021yum, TibetASg:2021kgt}, though more discoveries of sources and differentiation between leptonic and hadronic scenarios are needed to identify the highest-energy Galactic accelerators. Future VHE and UHE detectors with improved sensitivities, like the Southern Wide-field Gamma-ray Observatory (SWGO)~\cite{Albert:2019afb} and the Cherenkov Telescope Array (CTA)~\cite{CTAConsortium:2017dvg}, and neutrino experiments at TeV--PeV like  IceCube-Gen2~\cite{IceCube-Gen2:2020qha}, KM3NeT~\cite{KM3Net:2016zxf}, P-ONE~\cite{P-ONE:2020ljt}, and Baikal-GVD~\cite{Baikal-GVD:2019kwy}, have the potential to unveil the nature of PeVatrons.

Dozens of VHE and UHE sources have been discovered by HAWC~\cite{HAWC:2020hrt} and LHAASO \cite{2021Natur.594...33C}, including many new ones that were not seen in other wavelengths. In particular, detection of few-degrees-extended gamma-ray emission, called halos, were first reported by HAWC~\cite{HAWC:2017kbo} around Geminga and Monogem, the two closest middle-aged pulsars, that could contribute to the positron excess measured by PAMELA~\cite{PAMELA:2013vxg} and AMS-02~\cite{PhysRevLett.122.041102}. More TeV halos have then been found by HAWC~\cite{HAWC:2020hrt} and LHAASO~\cite{LHAASO:2021crt}. The small angular size of the gamma-ray halos challenges traditional views of particle diffusion in the interstellar medium~\cite{Hooper:2017gtd, Tang:2018wyr, Fang:2018qco, DiMauro:2019yvh, DiMauro:2019hwn} and, so far, no convincing theoretical explanation of this effect has been proposed~\cite{Giacinti:2019nbu, Lopez-Coto:2017pbk, Evoli:2018aza, Liu:2019zyj, Recchia:2021kty, DeLaTorreLuque:2022chz}. The unexpectedly efficient confinement of electrons and positrons by pulsars could limit the astrophysical interpretation to the positron flux and hint at the necessity of exotic physics~\cite{HAWC:2017kbo}. Better understanding of the TeV halo population and their forming mechanism with future wide-field gamma-ray experiments is thus needed for the indirect detection of dark matter~\cite{Engel:2022yig}.

\subsection{Production of the Heavy Elements}
\label{s:r-process}
The synthesis of the elements~\cite{burbidge1957synthesis, Frebel:2018slj} in the periodic table are part of the overall Hot Big Bang theory. After baryogenesis, the neutrinos decouple in the first two seconds, nearly freezing out the neutron/proton ratio. The light elements (hydrogen, helium, deuterium) are produced within the first several minutes and, after $\sim$300,000 years, the electrons and protons in the plasma recombine into neutral atoms, allowing the CMB to stream freely, enabling a host of high-precision cosmological measurements.

Elements in the periodic table up to iron are made in the hot cores of massive stars. Heavier elements are made in the slow neutron capture process (the \emph{s-process}), but this only accounts for around half of the heavy isotopes. The rest must be created by a high-density, rapid neutron capture process (the \emph{r-process}) either in the explosions of suernovae or through the merger of neutron star binaries.

After the spectacular binary neutron star merger GW170817~\cite{Monitor:2017mdv}, there has been a renewed interest in the so-called \emph{r-process}~\cite{Cowan:2019pkx}, by which $\approx$50\% of the heavy elements in the Universe are produced. In particular, primordial black holes could leave direct observational imprints of \emph{r-process} nucleosynthesis~\cite{Fuller:2017uyd}. In order to more precisely determine the contribution of various processes to the isotopic abundances, inputs from the EM/GW observations of neutron star mergers will have to be combined with precision measurements at accelerator facilities (e.g., the Facility for Rare Isotope Beams~\cite{Balantekin:2014opa}). 

\begin{figure}[ht]
\centering
\includegraphics[width=0.8\columnwidth]{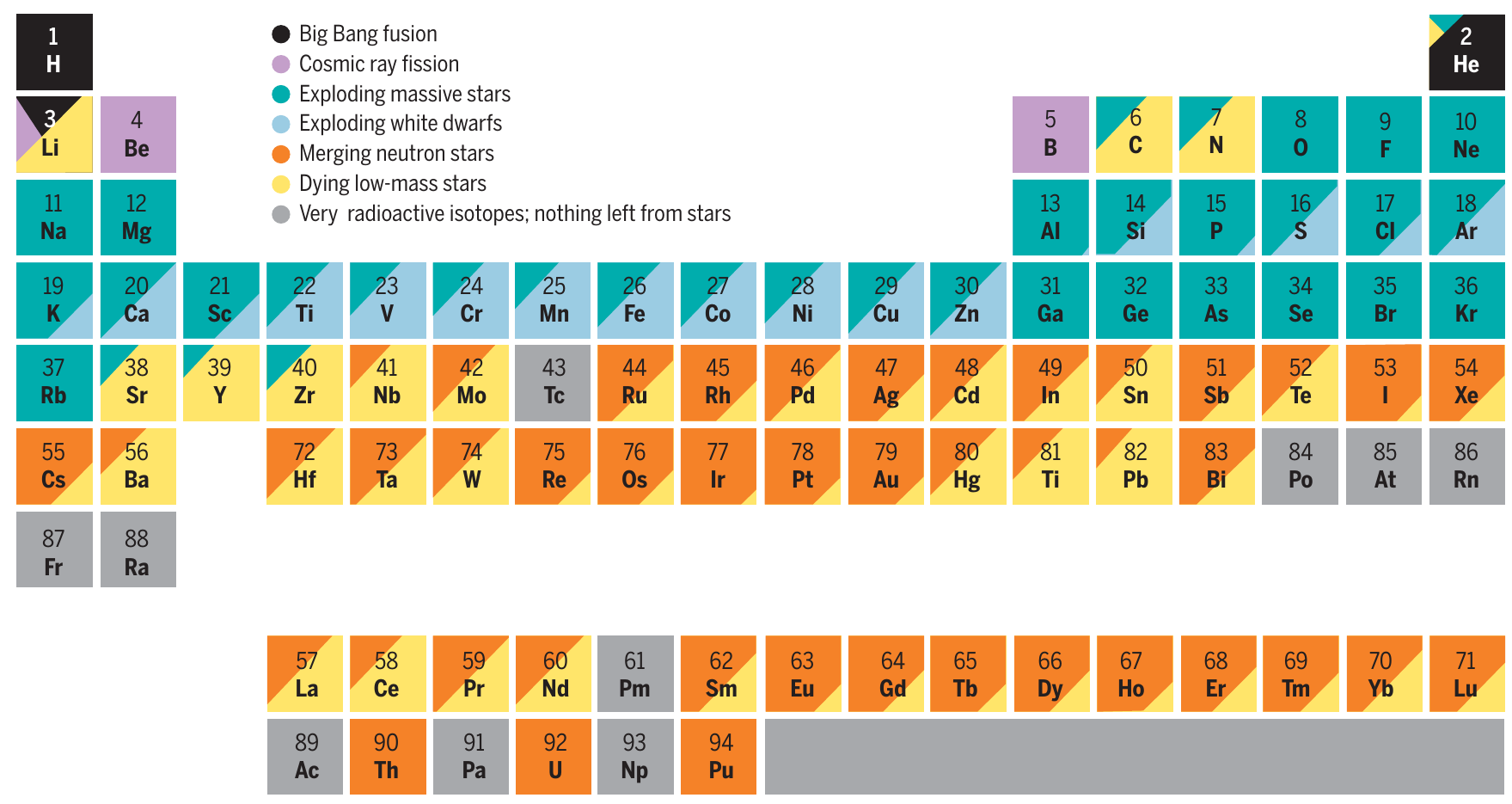}
\caption[Origin of the elements]{Nucleosynthetic sources of elements in the Solar System. Each element in this periodic table is color-coded by the relative contribution of nucleosynthesis sources, scaled to the time of the Solar System formation. Only elements that occur naturally in the Solar System are shown; artificially made elements and elements produced only through radioactive decay of long-lived nuclei are shown in grey. Taken from Ref.~\cite{Johnson:2019}.}
\label{fig:periodic_table}
\end{figure}

The periodic table shown in Fig.~\ref{fig:periodic_table} summarizes the origin of the elements in the Solar System that we see today. Cosmic nucleosynthesis is one of the the challenges ahead for the multimessenger program~\cite{Diehl:2022jnq}.

\section{Architecture of Spacetime}
\label{s:spacetime}

General Relativity (GR) is an incredibly successful theory describing the relationship between mass-energy and spacetime curvature. With the recent explosion of gravitational-wave (GW) detections, the prospects for testing the fundamental structure of spacetime are now looming closer~\cite{Berti:2022wzk}. In the sections below, we describe some of the most prominent examples.

\subsection{The Birefringence of Spacetime}

High-precision GW measurements coupled with multimessenger astronomy allow one to search for violations of GR in the propagation of waves~\cite{Berti:2022wzk}. These effects can largely be parameterized into the basis of graviton mass, dispersion in the GW propagation (to be discussed in the following sections), and birefringence of spacetime. These effects are all absent in Einsteinian gravity. However, non-zero gravition mass and dispersion or birefringence of gravitational waves can be linked to violations of Lorentz and CPT symmetry~\cite{ONeal-Ault:2021uwu}. Thus, testing whether spacetime is birefringent may amount indirectly to testing local Lorentz invariance and CPT symmetry.

While parity symmetry is conserved in GR, GW birefringence arises in effective-field-theory extensions of GR when  parity symmetry is broken. This causes the left- and right-handed polarizations to propagate differently from the source to the detector. Chern-Simons gravity~\cite{Alexander:2007kv, Yoshida:2017cjl}, Ho\v{r}ava-Lifshitz gravity~\cite{Horava:2009uw}, certain scalar-tensor theories of gravity~\cite{Crisostomi:2017ugk}, and the symmetric teleparallel equivalent of GR~\cite{Conroy:2019ibo} that have been proposed to account for dark matter and dark energy typically lead to parity violation. Moreover, such violations can also arise at large enough energy scales in quantum gravity theories such as the Loop Quantum Gravity and String Theory~\cite{Alexander:2007kv}.

Future GW detectors such as the pulsar timing arrays (PTAs), space-borne interferometers, and terrestrial laser interferometers will be able to fully constrain birefringence of the spacetime structure. To do so, it is necessary to observe GWs either for a long-enough duration or with enough number of non-collocated detectors so as to resolve their polarization states. LISA will track GWs from compact binary systems for years. As of 2022, the terrestrial detector network consists of the two LIGO detectors in the U.S. and the Virgo detector in Italy. Since the two LIGO detectors are nearly co-aligned, tests for non-GR polarizations have been limited.  By the end of the decade, the KAGRA detector in Japan and the third LIGO detector in India~\cite{Saleem:2021iwi} should be coming on-line, allowing for the full measurement of all polarization modes. With more sensitive detectors such as LIGO Voyager~\cite{LIGO:2020xsf}, Einstein Telescope~\cite{Punturo:2010zz}, and Cosmic Explorer~\cite{Evans:2021gyd}, it would be possible to place exceedingly tight constraints on a host of alternative theories of gravity; even more so using multiband analyses~\cite{Cutler:2019krq, Muttoni:2021veo, Gupta:2020lxa} with space-borne interferometers.

\subsection{Modified Gravity as an Alternative to Dark Energy \& Dark Matter}

The $\Lambda$CDM model, based on the theory of GR, has been very successful in explaining the observable properties of big bang nucleosynthesis, cosmic microwave background (CMB) observations, and large-scale structure. This success is achieved at the price of assuming that the energy content of the Universe today is dominated by dark energy and dark matter. However, only the large-scale gravitational interaction of the dark components has been detected so far and their fundamental properties remain largely unknown. As of today, we do not even know if the dark components are associated with new elementary particles or represent a mirage produced by modifications of the laws of gravity. Over the past decade, various discrepancies have emerged between $\Lambda$CDM predictions and cosmological observations, e.g., the tensions in the Hubble expansion rate and the clustering of matter discussed in Sec.~\ref{s:H0tension}. Several modified gravity models have been constructed to resolve the $H_0$ and $S_8$ tensions, but there seems to be no consensus on a  satisfactory solution to this problem yet~\cite{Abdalla:2022yfr}.

In the next decade, GW standard sirens are expected to provide strong constraints on dark energy, modified gravity, and dark matter and shed light on several other important aspects in cosmology (see Sec.\,IXA7 of Abdalla et al.~\cite{Abdalla:2022yfr} and references therein). Imprinted in the observed GWs is the nature of gravity. Thus, any modification of gravity beyond GR will leave a fingerprint in the GW signal.

Firstly, modified gravity theories are proposed mainly to explain the late-time acceleration of the Universe (dark-energy-dominated era), but they can also induce amplitude and phase corrections on the GW signal over cosmological volumes. The time variation of the gravitational constant could be inferred using a multimessenger~\cite{Engel:2022yig} approach, exploiting the unique relation between the GW luminosity distance, BAO angular scale, and the sound horizon at decoupling~\cite{Mukherjee:2020mha}. 

Secondly, by changing the gravitational interaction in a binary system, one induces a change in the generation mechanism of the gravitational radiation. Such changes can be quantified through the parameterized post-Newtonian~\cite{Arun:2006hn, Mishra:2010tp} or post-Einsteinian framework~\cite{Yunes:2009ke}. Future terrestrial and LISA observations can lead to improvements of 2--4 orders of magnitude with respect to present constraints, while multiband observations can yield improvements of 1--6 orders of magnitude~\cite{Perkins:2020tra}.

Finally, an interesting possibility is the detection of stochastic gravitational waves. The existence of a stochastic background is a robust prediction of several well-motivated cosmological and astrophysical scenarios operating at both the early and late Universe~\cite{Maggiore:1999vm, Caprini:2018mtu, Giovannini:2019oii}. As previously described, the existence of such backgrounds can be probed with GW observatories on ground and in space as well as PTAs.

\subsection{The Graviton Mass}

Regardless of the specifics of the theory one considers, there are general properties of the graviton  (understood as a gauge boson that mediates the gravitational interaction) that one may wish to measure or test to ensure our description is as prescribed by Einstein's theory. One such property is the graviton's mass which, according to GR, is exactly zero. Theories such as massive gravity~\cite{deRham:2014zqa} and bi-gravity~\cite{Crisostomi:2015xia} predict a non-zero value. In fact, many modified theories created to explain the present-day cosmic acceleration also predict deviations in the propagation of GWs~\cite{Cardoso:2002pa, Saltas:2014dha, Lombriser:2015sxa, Lombriser:2016yzn,  Belgacem:2017ihm, Nishizawa:2017nef, Belgacem:2018lbp} and in the gravitational lensing of GWs~\cite{Congedo:2018wfn, Mukherjee:2019wcg, Mukherjee:2019wfw, Ezquiaga:2020spg}. Gravitational waves thus have the potential to place stringent bounds on the graviton mass because a non-zero value leads to a modified dispersion relation~\cite{Will:1997bb, Kostelecky:2016kfm}. On very general grounds that rely only on special relativity, a non-zero graviton mass implies that the GW frequency does not just depend on its wave-vector, but rather also on the mass, leading to a compression of the GW train that accumulates with distance travelled~\cite{Will:1997bb}.

Current GW observations are already placing constraints on the mass of the graviton, but much more can be achieved in the next decade. Current LIGO/Virgo observations have constrained the graviton mass to be less than $4.7 \times 10^{-23} \; {\rm{eV}}/c^2$~\cite{LIGOScientific:2019fpa}. Constraints on the mass of the graviton, however, can be shown to scale as $[f_{\rm low}/(D_L \rho)]^{1/2}$, where $D_L$ is the luminosity distance, $\rho$ is the SNR, and $f_{\rm low}$ is the lowest frequency detected~\cite{Perkins:2020tra}--- this is because the larger the distance, the longer the GW train compression can accumulate for, leading to a stronger constraint. As a result, in the next few years and then in the next decade, future observations with LIGO/Virgo/KAGRA/LIGO-India and XG ground-based detectors can place constraints better than $10^{-25} \; {\rm{eV}}/c^2$ and $10^{-26}$, respectively, while space-borne detectors like LISA can improve these constraints down to $3 \times 10^{-27} \; {\rm{eV}}/c^2$~\cite{Chamberlain:2017fjl, Perkins:2020tra}. These numbers are interesting because if one associates the late-time acceleration of the Universe to a non-zero graviton mass, then the graviton would have to be of the scale of the Hubble constant, $10^{-33}$~eV. By stacking events from LISA and XG detectors we may begin to approach this scale and thus confirm or rule out a non-zero graviton mass as an explanation for the late-time acceleration of the Universe.

Another property of the graviton as a particle that one may wish to probe is its group velocity in the high-energy limit $E \gg m_g$. In Einstein's theory, this group velocity is equal to the speed of light, but in other theories of gravity, this need not be the case~\cite{Baker:2017hug, Ezquiaga:2017ekz, Creminelli2017, Sakstein:2017xjx, Boran:2017rdn, Akrami:2018yjz}. For example, in the Einstein-\AE{}ther theory, the graviton travels at a constant group speed that is faster than the speed of light, avoiding causality violations~\cite{Jacobson:2000xp, Jacobson:2007veq}. The measurement of the speed of the graviton, unfortunately, is rather difficult because it requires that we compare the time of arrival of a GW to some other baseline. This is where multimessenger events shine. If an event produces both GWs and electromagnetic (EM) waves simultaneously, then one can, in principle, compare the speed of the GWs to the speed of the EM waves (i.e., the speed of light) by comparing their times of arrival.

This is exactly what was done with the first LIGO/Virgo binary neutron star observation, GW170817, which was accompanied by a short gamma-ray burst emitted shortly after the merger~\cite{Monitor:2017mdv, Mukherjee:2019wcg, Mukherjee:2019wfw, Baker:2020apq, Ezquiaga:2020dao}. This single observation was sufficient to infer that the speed of the graviton is equal to that of the photon to better than one part in $10^{15}$. Such a measurement had the effect of severely constraining a variety of modified theories of gravity. Future terrestrial observations with LIGO/Virgo/KAGRA/LIGO-India or with XG detectors will allow for additional measurements of the speed of the graviton along other lines of sight, and thus allow us to test local position invariance~\cite{Will:2005va, Yunes:2013dva}.

With LISA, we may detect supermassive black hole (BH) binaries at mHz frequencies and measure time delays between the arrivals of photons and gravitons. This will present some advantages. First, the longer timescales of these massive mergers can facilitate triggered EM precursor observations. The inevitable periodic modulations of the EM signal due to Doppler and lensing effects during the inspiral stage arise from the same orbital motion as the GWs and can be phased in a robust way without the need to model the astrophysical source in detail~\cite{Haiman:2017szj, Tang:2018rfm}. The measurements will also provide tighter limits, due to the high SNRs and large horizon distances achievable with LISA. The frequency dependence of the time delay would further probe Lorentz-violating theories~\cite{Kocsis:2007yu, Haiman:2009te, Mirshekari:2011yq}.

Some modifications of GR, invoked to explain the present-day cosmic acceleration, predict deviations between the propagation properties of EM radiation and GWs~\cite{Saltas:2014dha, Lombriser:2015sxa, Lombriser:2016yzn, Belgacem:2017ihm, Nishizawa:2017nef, Belgacem:2018lbp, Mastrogiovanni:2020gua}. A multimessenger, data-driven measurement of the running of the effective Planck mass and its redshift dependence is possible by combining three length scales, namely the GW luminosity distance, baryon acoustic oscillations (BAO), and the sound horizon from the CMB~\cite{Mukherjee:2020mha}. Sources detectable at higher redshifts (such as supermassive BH binaries) are most useful to measure the redshift dependence and running of the effective Planck mass. Such measurements may be possible by cross-correlating binary BHs with galaxies~\cite{Mukherjee:2020mha}. General relativity propagation effects could also be probed using other techniques--- e.g., by using the mass distribution of binary neutron stars~\cite{Finke:2021eio} and BHs~\cite{Leyde:2022orh}.


\subsection{Tests of Black Hole Dynamics}

The dynamical content of the underlying theory of gravity can be probed in violent, dynamical situations giving rise to strong bursts of GW emission. After the violent merger of two compact objects leading to BH formation, GR predicts the formation of a Kerr BH, wherein the spacetime is described by only two parameters. The relaxation to this state is described by a set of exponentially damped sinusoids (``ringdown'') whose frequencies and damping times depend only on the mass and spin~\cite{Kokkotas:1999bd, Berti:2009kk}. Since GW observations provide a measurement of frequencies and damping times, the ``ground state'' quasi-normal mode (QNM) allows us to infer the mass and spin. Any measurement of additional QNM frequencies (``excited states'') can then be used as a null test of the Kerr nature of the remnant.

The idea of treating BHs as ``gravitational atoms,'' thus viewing their QNM spectrum as a unique fingerprint of spacetime dynamics (in analogy with atomic spectra), is usually referred to as ``BH spectroscopy''~\cite{Dreyer:2003bv, Berti:2005ys, Berti:2007zu, Gossan:2011ha, Meidam:2014jpa}.  The seeds of this idea were planted in the 1970s~\cite[see, e.g.,][for a detailed chronology]{Berti:2009kk}. Chandrasekhar and Detweiler developed various methods to compute the QNM spectrum, identifying and overcoming some of the main numerical challenges~\cite[see, e.g.,][]{Chandrasekhar:1975zza}. In particular, Detweiler concluded the first systematic calculation of the Kerr QNM spectrum~\cite{Detweiler:1980gk} with a prescient statement: {\em ``After the advent of gravitational wave astronomy, the observation of [the BH’s] resonant frequencies might finally provide direct evidence of BHs with the same certainty as, say, the 21 cm line identifies interstellar hydrogen.”}

 Early estimates~\cite{Berti:2005ys,Berti:2007zu} showed that the detection and extraction of information from ringdown signals requires events whose signal-to-noise ratio (SNR) in the ringdown {\em alone} is larger than those achievable now (for example, the first GW detection (GW150914) had a combined SNR of $24$, with an SNR $\sim7$ in the ringdown phase~\cite{TheLIGOScientific:2016wfe, TheLIGOScientific:2016src}). There are claims that overtones have been detected in GW150914~\cite{Isi:2019aib} and higher modes have been measured in GW190521~\cite{Capano:2021etf}, but the detection of modes other than the fundamental is debatable at current SNRs~\cite{Bustillo:2020buq, LIGOScientific:2020tif, Cotesta:2022pci}.
 Any deviation from the QNM spectrum of classical GR would indicate substructure of BH ``atoms'' inconsistent with the standard picture. In particular, a non-singular horizonless object would lead to different boundary conditions than the classical theory and departures from the BH QNM spectrum. In any case, conclusive tests should be achievable once LIGO and Virgo reach design sensitivity, and certainly with the next-generation (XG) observatories (Cosmic Explorer or the Einstein Telescope) or with space-based detectors such as LISA~\cite{Berti:2016lat}. If the frequencies turn out to be compatible with the predictions of GR, parametrized formalisms can be used to constrain theories of gravity that would predict different spectra~\cite{Cardoso:2019mqo, McManus:2019ulj, Maselli:2019mjd, Carullo:2021dui}.

The existence and properties of horizons can be inferred and quantified with a variety of observations~\cite{Cardoso:2019rvt}. It is believed that accreting horizonless objects would reach thermal equilibrium with the environment rather quickly, whereas accreting supermassive BHs do not--- the luminosity contrast between the central accreting object and its accretion disk imposes stringent constraints on the location and property of a putative surface~\cite{Broderick:2009ph, Cardoso:2019rvt}. However, constraints based on accretion models are model-dependent and have also been questioned~\cite{Carballo-Rubio:2018jzw}. They still leave open the possibility of a surface close to the would-be event horizon, as predicted in thin-shell gravastar models~\cite{Mazur:2004fk, Mazur:2015kia, Beltracchi:2021zkt, Beltracchi:2021lez}. The planned EHT and future surveys of tidal disruption events will improve current constraints on the location of a hypothetical surface by two orders of magnitude.

The EM observations above are done, essentially, in a fixed-background context in which the BH spacetime is an arena where photons propagate. One can also consider situations probing both the background {\it and} the field equations. A stellar-mass BH or a neutron star orbiting a supermassive BH will slowly inspiral due to emission of GWs, ``sweeping'' the near-horizon geometry and being sensitive to tiny near-horizon changes, such as tidal deformability or tidal heating, or to non-perturbative phenomena like resonances of the central object~\cite{Cardoso:2017cfl, Maselli:2017cmm, Cardoso:2019nis, Maggio:2021uge, Fang:2021iyf}. Accurate tracking of the GW phase by the future space-based detector LISA may constrain the location of a putative surface to Planckian levels~\cite{Cardoso:2019rvt}.


\subsection{Alternative Black Hole Models}

The absence of an horizon can also lead to smoking-gun effects in the GW signal. An ultracompact, horizonless vacuum object sufficiently close to the Kerr geometry outside the horizon behaves as a cavity for impinging GWs, which end up being trapped between the object's interior and its light ring~\cite{Cardoso:2016rao, Cardoso:2016oxy, Cardoso:2019rvt}. Thus, perturbations of such objects, and possibly mergers as well, lead to a GW signal which is---by causality principle---similar to that emitted by BHs on sufficiently small timescales. However, at late times, the signal trapped in the ``cavity'' leaks away as a series of ``echoes'' of the original burst, which may carry a significant amount of energy. LIGO/Virgo observations have so far shown no evidence for such echoes~\cite{LIGOScientific:2020tif, LIGOScientific:2021sio}. The absence of such structure in future observations by LIGO and LISA will allow the exclusion---or detection---of any significant structure a Planckian distance away from the Schwarzschild radius, with important implications for fundamental physics~\cite{Cardoso:2019rvt}.

Setting stringent constraints on the nature of compact objects---in particular quantifying the existence of horizons in the Universe---requires advanced detectors. It is also a challenging task from the modelling and computational point of view, as one needs: \textit{(i)} a physically motivated, well-posed theory solving, at least partially, the conceptual problems of GR; \textit{(ii)} the existence in such theories of ultracompact objects which arise naturally as the end-state of gravitational collapse; and \textit{(iii)} the solution of the relevant partial differential equations describing the mergers of such objects. There is pressing need for progress on all of these fronts to confront the increasingly precise data expected from a wide variety of new experimental facilities.

\subsection{Quantum Gravity Constraints on Low-Energy Dynamics}
\label{s:swampland}

Low-energy Effective Field Theories (EFTs) are the central tool in the theoretical description of low-energy particle physics, cosmology, and gravitational theories. Modern perception has it that the SM and Einstein gravity should both be understood as leading terms in an EFT expansion. It is thus of paramount importance to understand the space of allowed low-energy (IR) EFTs. Recently, there has been significant progress in understanding what is the space of low-energy EFTs that admit an UV completion.

\begin{figure}[ht]
\centering
\includegraphics[width=0.7\textwidth]{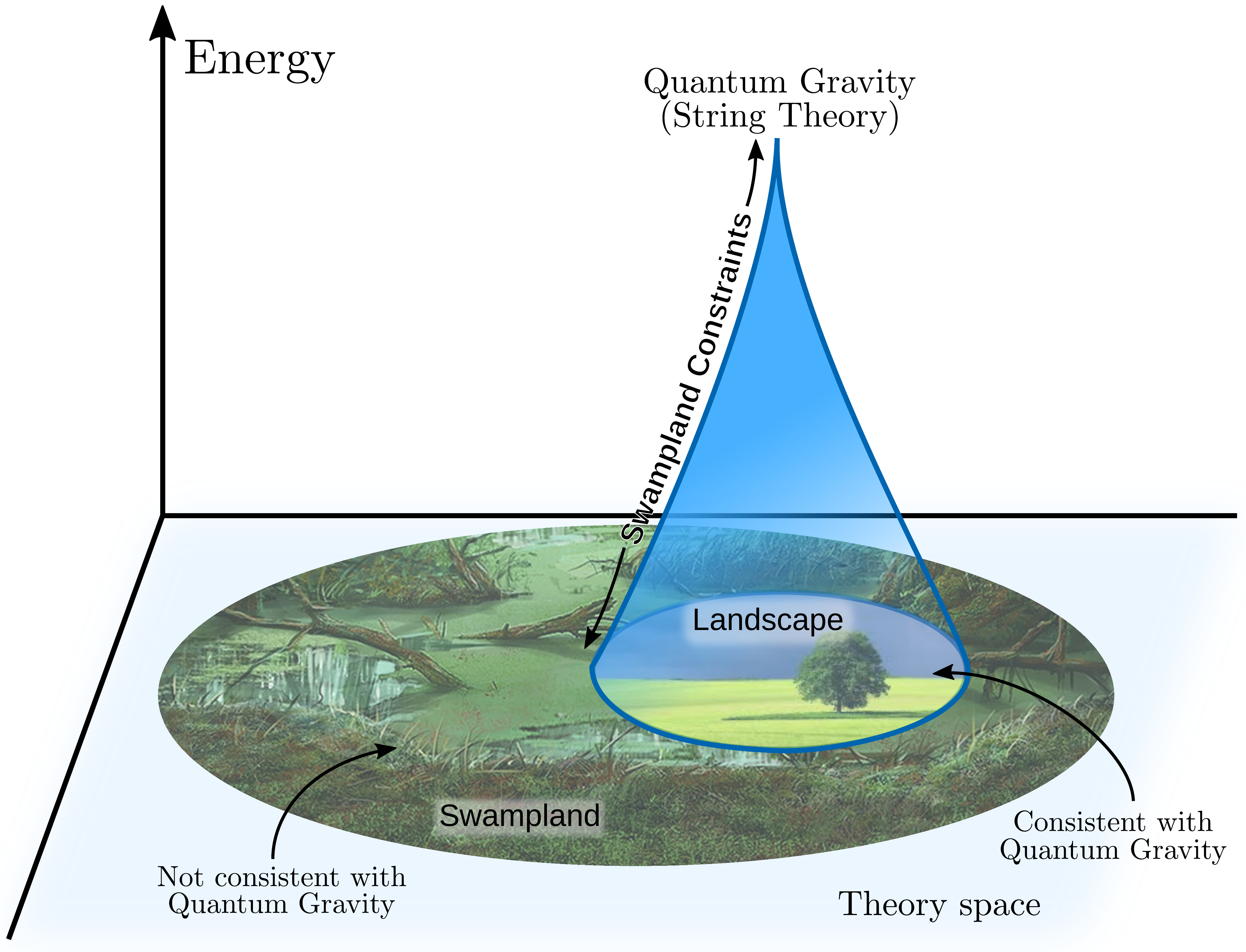}
\caption{The swampland and landscape of EFTs. The space of consistent EFTs forms a cone because swampland constraints become stronger at high energies. From Ref.~\cite{vanBeest:2021lhn}.}
\label{fig:cone}
\end{figure}

Despite decades of research, a full-fledged theory of quantum gravity (QG) remains elusive. Nonetheless, along the way we have learned some generic features that a QG theory should possess.  The Swampland Program seeks to delineate the boundary between the landscape of EFTs that are compatible with these features of QG and the swampland of EFTss that are not~\cite{Vafa:2005ui}. The set of QG features are sometimes referred as swampland conjectures~\cite{Palti:2019pca}. As illustrated in Fig.~\ref{fig:cone}, the swampland conjectures become more constraining as the energy at which the EFT should be valid increases, picking out, in the end, a (possibly unique) QG theory.

The accelerating expansion of the Universe is a phenomenon that is apparently IR but intrinsically UV. This cosmological hierarchy opens up the opportunity to probe physics beyond $\Lambda$CDM and the SM by analyzing some phenomenological implications of the swampland conjectures. For example, it has been conjectured that scalar field potentials $V$ that can be derived from putative QG theories obey the bound $V' \geq c V/M_{\rm Pl}$, where $c$ is a positive and dimensionless order one constant~\cite{Obied:2018sgi}. The most obvious consequence of this constraint is that de Sitter vacua are forbidden, ruling out the cosmological constant $\Lambda$ as a source of dark energy~\cite{Agrawal:2018own}. The model building of quintessence fields playing the role of dark energy has been featured extensively through the  swampland program~\cite{Abdalla:2022yfr,Vafa:2019evj,OColgain:2018czj,Colgain:2019joh}. Of particular interest here, quintessence models tend to exacerbate the $H_0$ tension~\cite{Banerjee:2020xcn}. More generally, the swampland conjectures make it difficult for fundamental theories based on compactification from extra dimensions to accommodate a period of accelerated cosmic  expansion~\cite{Montefalcone:2020vlu}. Such a restriction can be avoided in models whose internal space is not conformally Ricci flat~\cite{Anchordoqui:2020sqo}, e.g., the Salam-Sezgin model~\cite{Salam:1984cj}. Within this supergravity model, dark matter could acquire a mass term which depends on the value of the quintessence field~\cite{Anchordoqui:2019amx}, thus realizing an effective dark matter-dark energy coupling  which
could help to reduce (though not fully eliminate) the $H_0$
 tension~\cite{Agrawal:2019dlm}. Examined separately, the axion weak-gravity conjecture~\cite{Arkani-Hamed:2006emk} leads to a bound on early dark energy models proposed to resolve the $H_0$ tension~\cite{Rudelius:2022gyu}.

On a separate track, the distance conjecture~\cite{Ooguri:2006in,Lust:2019zwm}, combined with the cosmological hierarchy and bounds on deviations from Newton's law~\cite{Lee:2020zjt}, give rise to an exponentially light tower of states with two mass scales: {\it (i)}~the mass scale of states in the tower, $m \sim \Lambda^{1/4}/\lambda$, and {\it (ii)}~the scale at which the local EFT description breaks down, dubbed the species scale, $\hat M \sim \lambda^{-1/3} \ \Lambda^{1/12} \ M_{\rm Pl}^{2/3}$~\cite{Montero:2022prj}. For $\lambda \sim 10^{-3}$, $m \sim 1~{\rm eV}$ is of the order of the neutrino scale and $\hat M \sim 10^{10}~{\rm GeV}$ coincides with the sharp cutoff observed in UHECR data. This implies that the highest-energy cosmic rays could be an incisive probe of UV physics~\cite{Anchordoqui:2022ejw}. Moreover, this framework has interesting implications for the abundance of primordial black hole dark matter~\cite{Anchordoqui:2022txe}, while the excitations of the graviton in the bulk provide an alternative dark matter candidate~\cite{Gonzalo:2022jac} and  a particular realization of the DDM scenario~\cite{Dienes:2011ja} discussed in Sec.~\ref{sec:DDM}. 

The study of UV constraints on IR physics is a burgeoning field, with many new conceptual and technical developments. Promising future directions are summarized in Ref.~\cite{deRham:2022hpx}.

\section{Current and Future Experiments}
\label{s:experiments}

\begin{figure}[ht!]
    \includegraphics[width=1.0\textwidth]{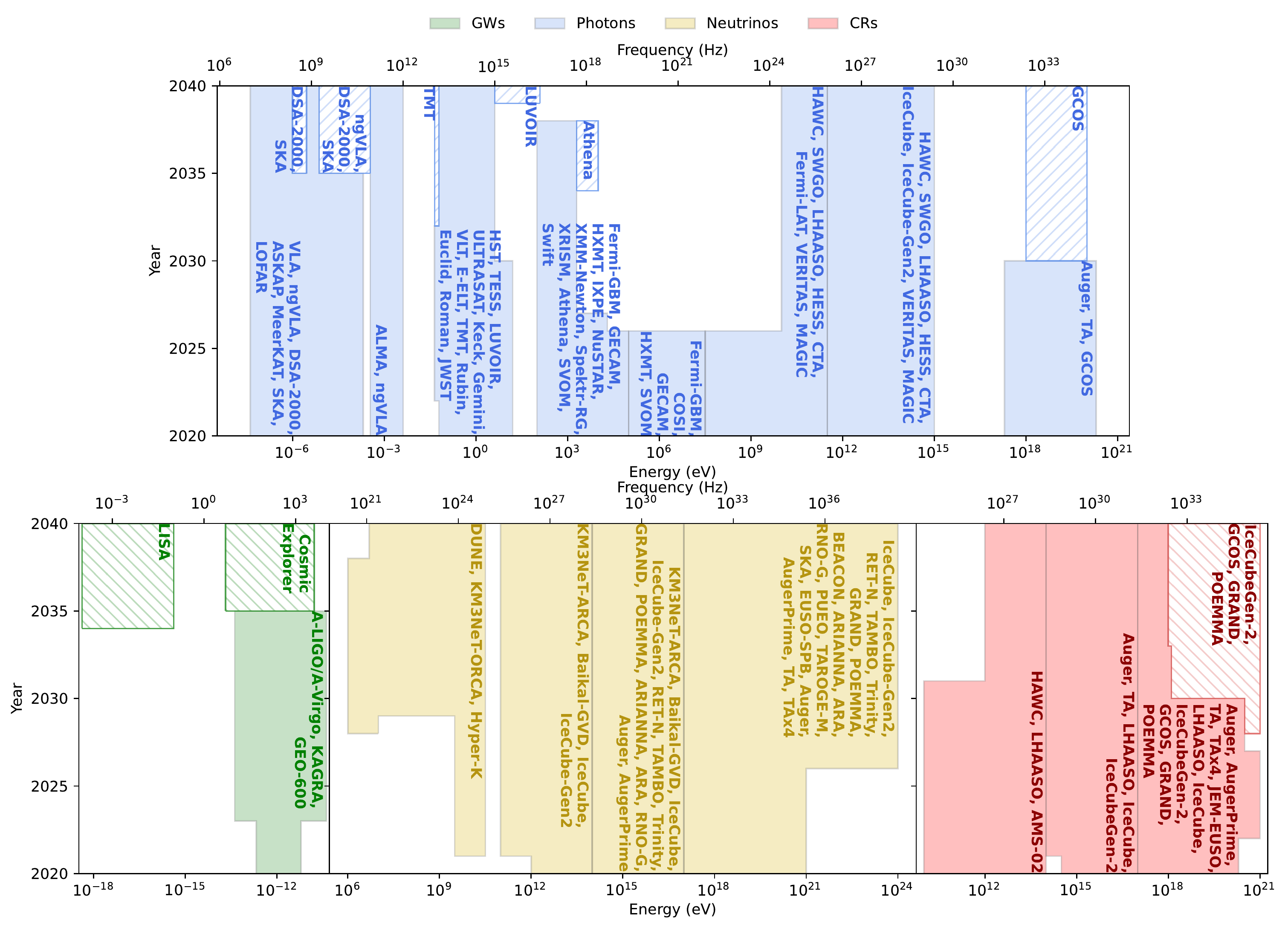}
    \caption{Timeline of current and proposed photon, gravitational-wave (GW), neutrino, and cosmic-ray (CR) facilities. Hatched regions indicate energies which proposed experiments would observe that would not be simultaneously observed by any current facilities. Over time, most messengers plan to increase their spectral coverage. The photon frame in blue illustrates continuous multi-wavelength coverage for the next two decades, with the glaring exception of MeV, GeV, and ultra-high-energy gamma rays. This impending gamma-ray gap is concerning to the broader multimessenger community. From Ref.~\cite{Engel:2022yig}.}
    \label{fig:gantt_chart}
\end{figure}

A new age of precision cosmology and elucidation beyond the Standard Model with astroparticle physics has just begun. In the coming decades, there will be a large set of new probes to determine the cosmological parameters with unprecedented rigor, as well as an array of multimessenger experiments to discover new and exciting physics at energies not achievable by terrestrial accelerators. In this context, programmatic balance is imperative (see Fig.~\ref{fig:gantt_chart}).

The individual instruments involved in the multimessenger program are some of the most finely tuned that human hands have developed. Current gamma-ray, neutrino, cosmic-ray, and gravitational-wave facilities plan generally to increase their spectral coverage over the next two decades, while proposed, but currently unfunded, future facilities go well beyond just picking up at the sunset of their predecessors--- they are poised to unravel the mysteries of Universe and serve as fertile grounds for the discovery of new and exciting physics.

The importance and great benefit of the involvement of DOE National Laboratories with those future experiments, both cosmological and astrophysical, cannot be overstated. The wealth of knowledge employed at these laboratories can be put to use by these experimental collaborations to achieve greater theoretical and technical progress than previously envisaged. This relationship is also incredibly symbiotic. As seen with the relationships between, e.g., Los Alamos National Laboratory and the HAWC Observatory, SLAC and \textit{Fermi}, Argonne National Laboratory and VERITAS, Fermilab and the Pierre Auger Observatory and Dark Energy missions, etc., this partnership enables the labs to work on smaller-scale experiments, in addition to their larger projects, and is a lucrative pathway for the recruitment and retention of highly skilled scientific minds to National Laboratories.

As elaborated in greater detail below and elsewhere in this Report,  each of the Cosmic Probes brings unique access to one or more aspects of physics of the Standard Model and BSM physics and merits support as part of the HEP mission.  Reflecting the maturity of the respective fields, the 4 types of Cosmic Probes (photons, neutrinos, UHECRs and GW) have different needs for development and increased US support in the next decades:
\begin{itemize}[noitemsep,topsep=0pt]
\item Next-stage gamma and neutrino investments should continue to be supported by US commitments including NASA and DOE, including infrastructure and financing. DOE investments in technology development for MeV gamma ray detection in colliders
and in next-generation air shower gamma-ray detectors will benefit this field as well. 
\item UHECRs give unique access to UHE phenomena, but facilities are currently mainly funded by Europe; it would benefit the US HEP community to maintain and grow US involvement in the next generation.  
	\item Cosmic Explorer is US-lead, with international participation.  Cosmic Explorer is probably the most dramatic new opportunity in the entire Cosmic Frontier portfolio.  It is at a critical moment when additional involvement of HEP physicists and support for infrastructure and R\&D will have disproportionate returns.  
\end{itemize}

\paragraph{Gamma-ray facilities} Gamma rays are vital messengers that carry information about an abundance of key scientific goals, both within our Galaxy and from the far reaches of extragalactic space. 
They bring messages about naturally occurring particle acceleration throughout the Universe in environments so extreme they cannot be reproduced on Earth for a closer look and provide a window into beyond-the-Standard-Model Physics. Gamma-ray astrophysics is so complementary with collider work that particle physicists and astroparticle physicists are often one in the same, thus their facilities are vital tools in elucidating the mysteries of beyond-the-Standard-Model physics and astroparticle physics and for the discovery of new physics~\cite{Engel:2022bgx}.

While photons at different energies provide different pieces to each scientific puzzle, with no image being able to be completed with only a single input, the GeV-to-TeV-and-beyond energy regime hosts a highly successful set of current experiments, such as HAWC~\cite{HAWC:2019xhp}, VERITAS~\cite{2015ICRC...34..771P}, MAGIC~\cite{2016APh....72...76A}, H.E.S.S.~\cite{2006A&A...457..899A}, and LHAASO. Through their strict limits on PBHs, axion-like particles, CPT violation, and LIV, as well as observations of TeVatrons and PeVatrons, the discovery of gamma-ray halos, and countless other exciting scientific feats, these facilities have proven that gamma-ray facilities, especially those observing at the highest energies, are a force to be reckoned with. Complementary with these experiments and carrying on their heavy scientific loads in the next decade are two proposed future experiments: the Southern Wide-field Gamma-ray Observatory (SWGO)~\cite{Albert:2019afb, Hinton:2021rvp, Schoorlemmer:2019gee} and the Cherenkov Telescope Array (CTA)~\cite{CTAConsortium:2017dvg}. Building upon lessons learned from the current observatories and their predecessors, these facilities will have unprecedented sensitivity to the highest energies and are critical to carrying on the legacy of science at the forefront of particle and astroparticle physics.

Notably, current MeV and GeV gamma-ray facilities are expected to end before 2030 with no long-term plan to fill that gap in coverage that will impact, intrinsically, MeV and GeV science as well as make it impossible to collaborate with other wavelengths and messengers, effectively ending multimessenger science as we currently conceive of it. The timeline of current and proposed photon, gravitational-wave, neutrino, and cosmic-ray facilities is shown in Fig.~\ref{fig:gantt_chart}. 

\begin{figure}[ht!]
    \includegraphics[width=0.5\textwidth, trim = 14cm 0cm 14cm 0cm, clip]{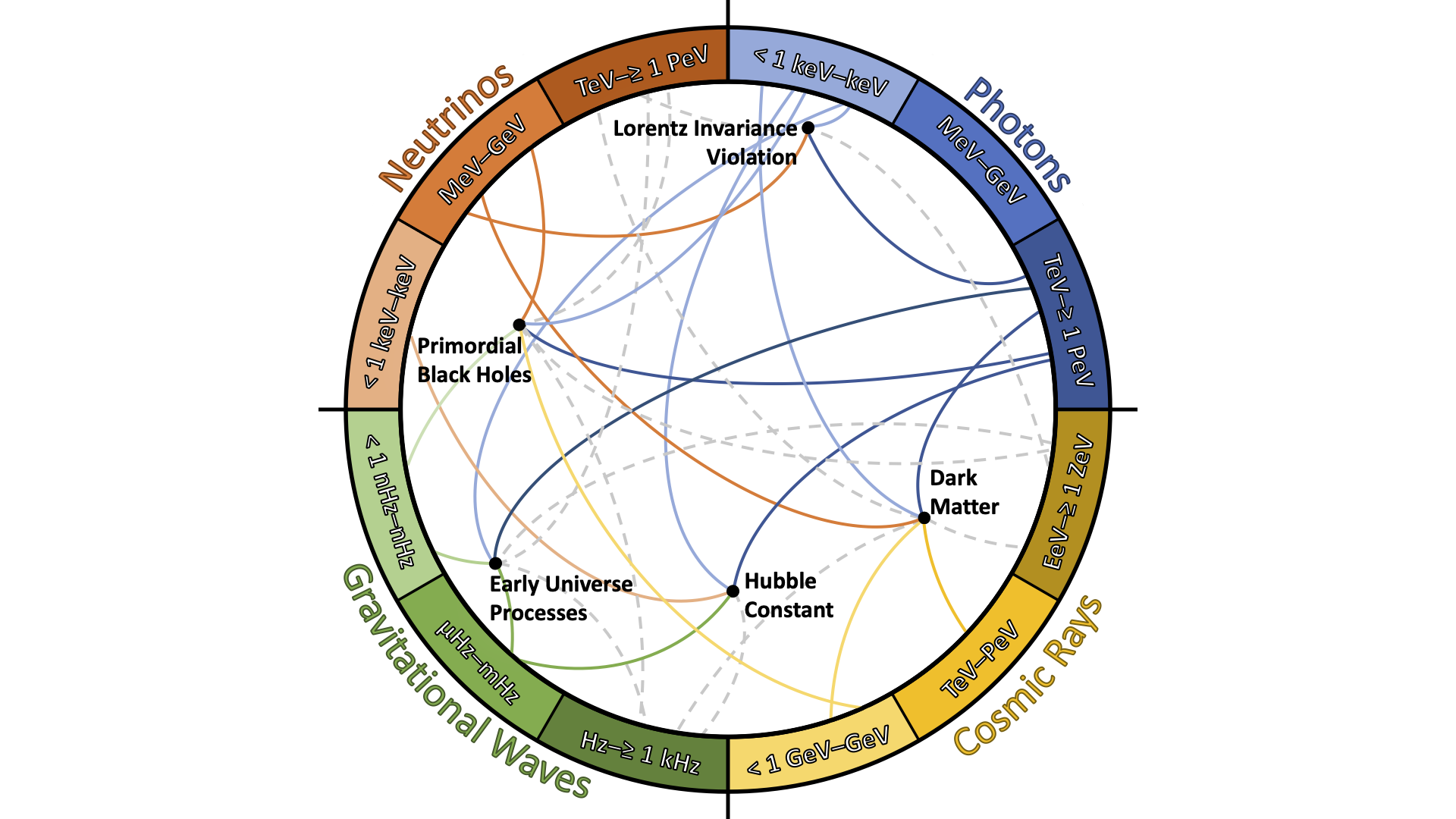}
    \includegraphics[width=0.5\textwidth, trim = 14cm 0cm 14cm 0cm, clip]{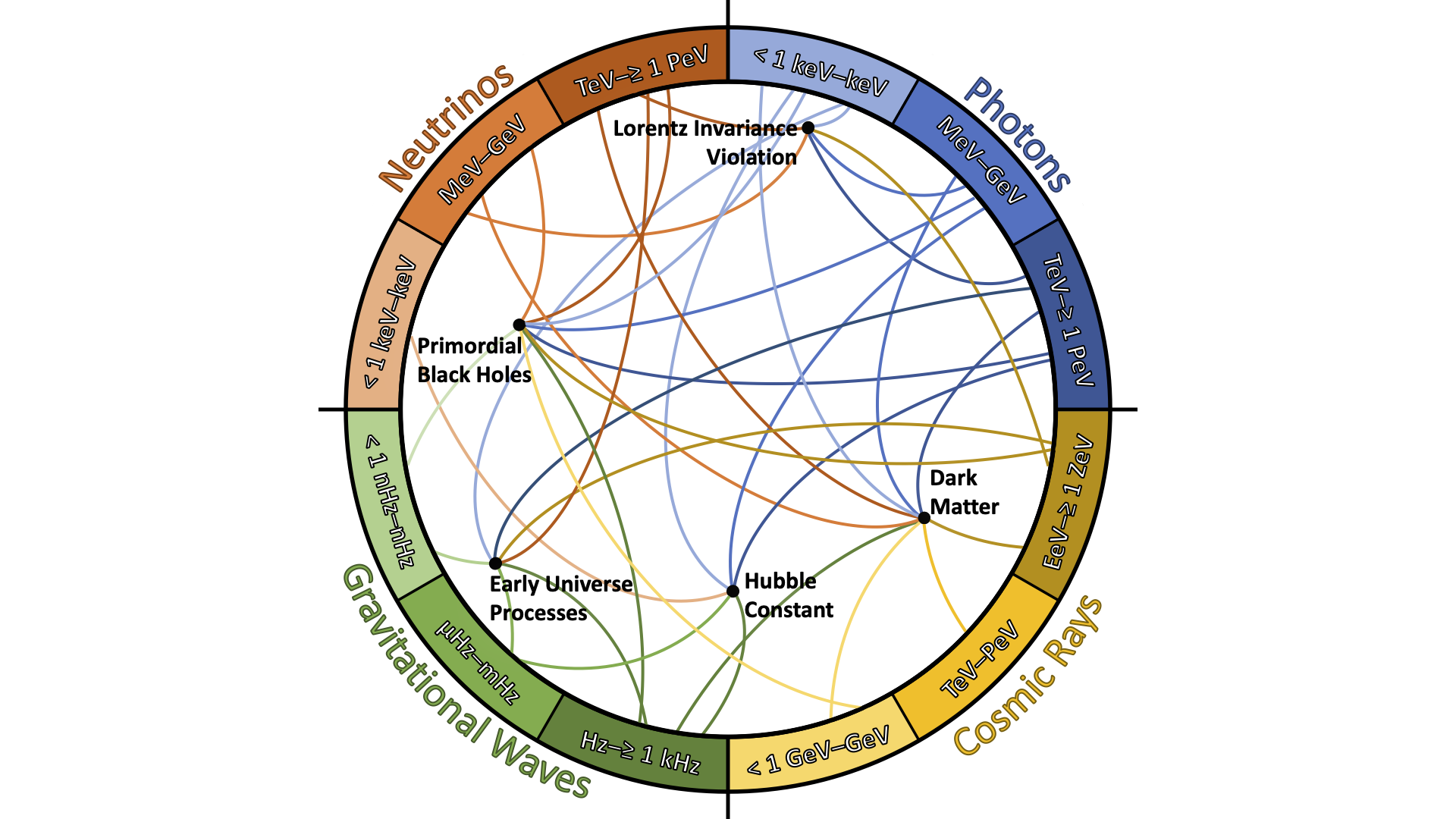}
    \includegraphics[width=0.5\textwidth, trim = 14cm 0cm 14cm 0cm, clip]{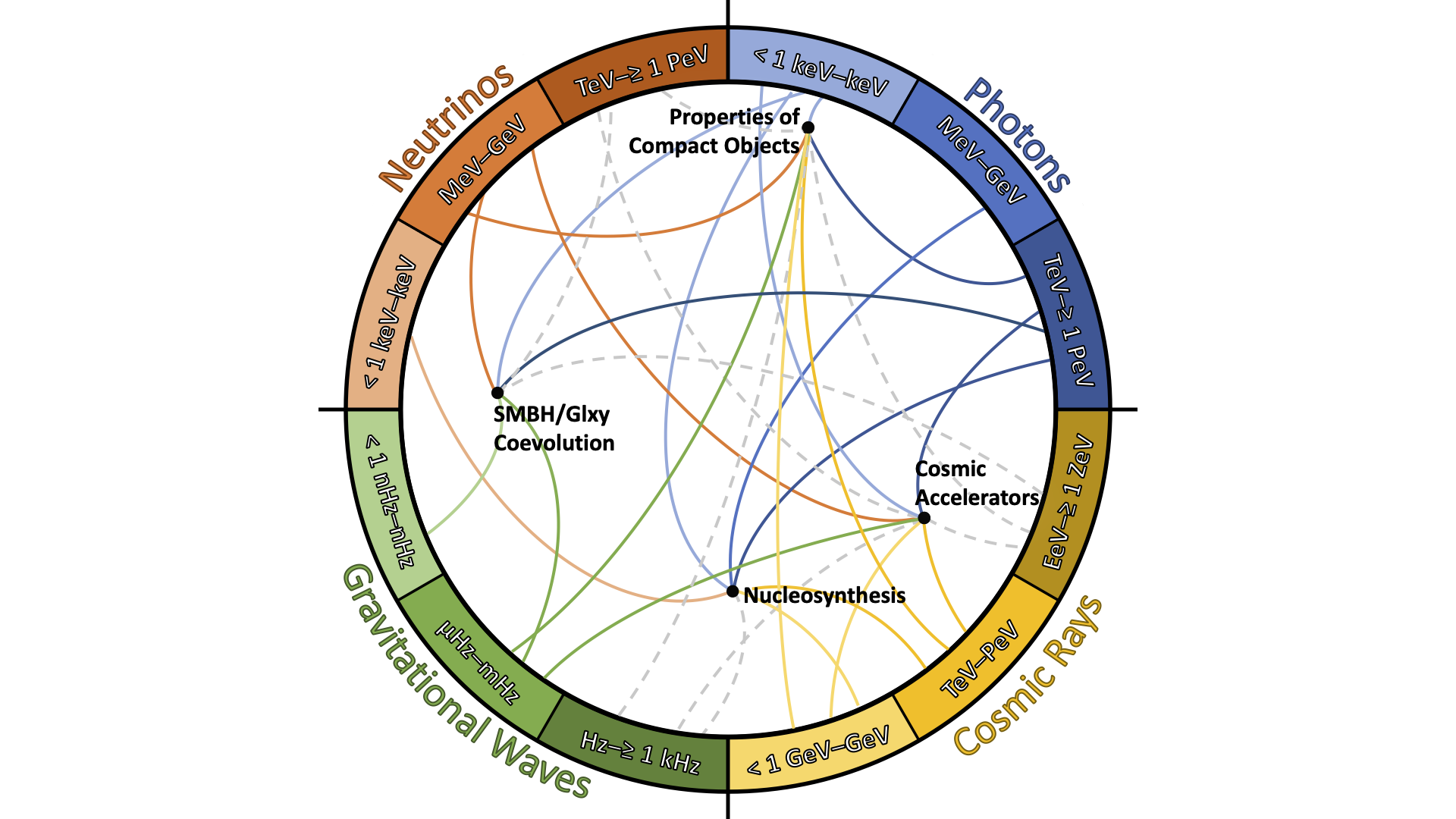}
    \includegraphics[width=0.5\textwidth, trim = 14cm 0cm 14cm 0cm, clip]{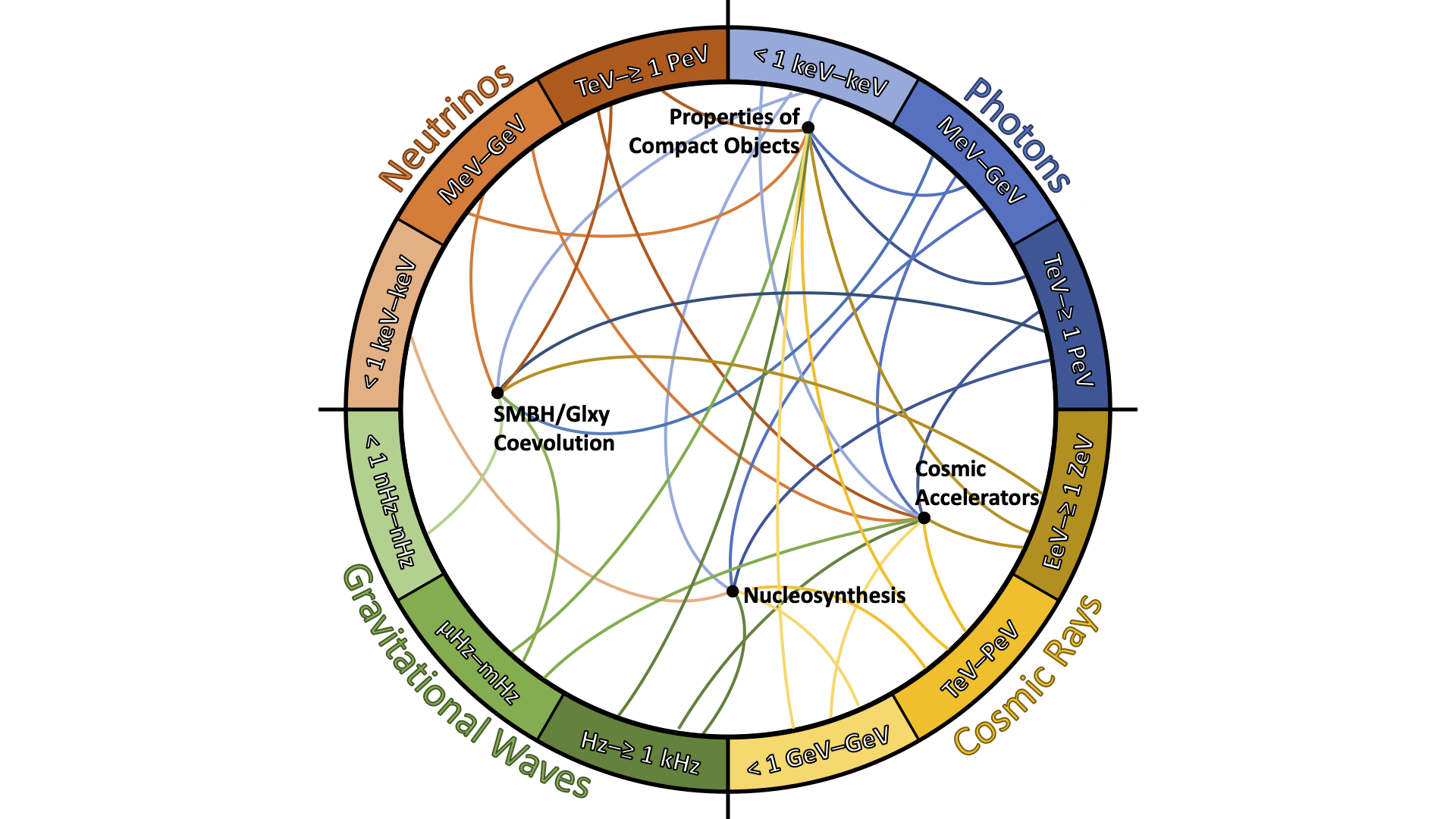}
    \caption{{\it Top panels:} Connections between messengers and fundamental physics topics. {\it Bottom panels:} Connections between messengers and particle astrophysics topics. {\it Left panels:} Future multimessenger landscape with current facilities that are planned to continue operating and future facilities that are already funded. {\it Right panels:} Future multimessenger landscape with enhanced capabilities provided by proposed facilities. From Ref.~\cite{Engel:2022yig}.}
    \label{fig:chord_plots}
\end{figure}

The loss of instrumental coverage in the MeV--GeV gap has broad implications for the goals of fundamental physics through the study of astronomical objects. Gamma rays are pivotal in the study of every major physics question in the coming decade. The lack of planned funding for this photon band should be truly alarming to those who have borne witness to the magnitude of recent multimessenger discoveries. The possible connections between fundamental physics questions, the astronomical objects through which they are studied, and observations that probe them by messenger and energy are shown in Fig.~\ref{fig:chord_plots}, where we note the potential loss of scientific excellence if key instrument classes are not prioritized over the next decade. For details, see Refs.~\cite{Engel:2022yig, Engel:2022bgx}.

There are many facility concepts in progress to improve cost and sensitivity for the gamma-ray band in this decade. A key area of investment for the future of multimessenger astrophysics is gamma-ray detector technology. Many aspects of instrumentation and software pipelines for cosmic gamma-ray detectors are nearly identical to those used in colliders, making this technology development extremely relevant to the broader particle physics community. For details, see Ref.~\cite{Engel:2022bgx}.

\begin{figure}[htb!]
  \centering
  \includegraphics[width=\textwidth]{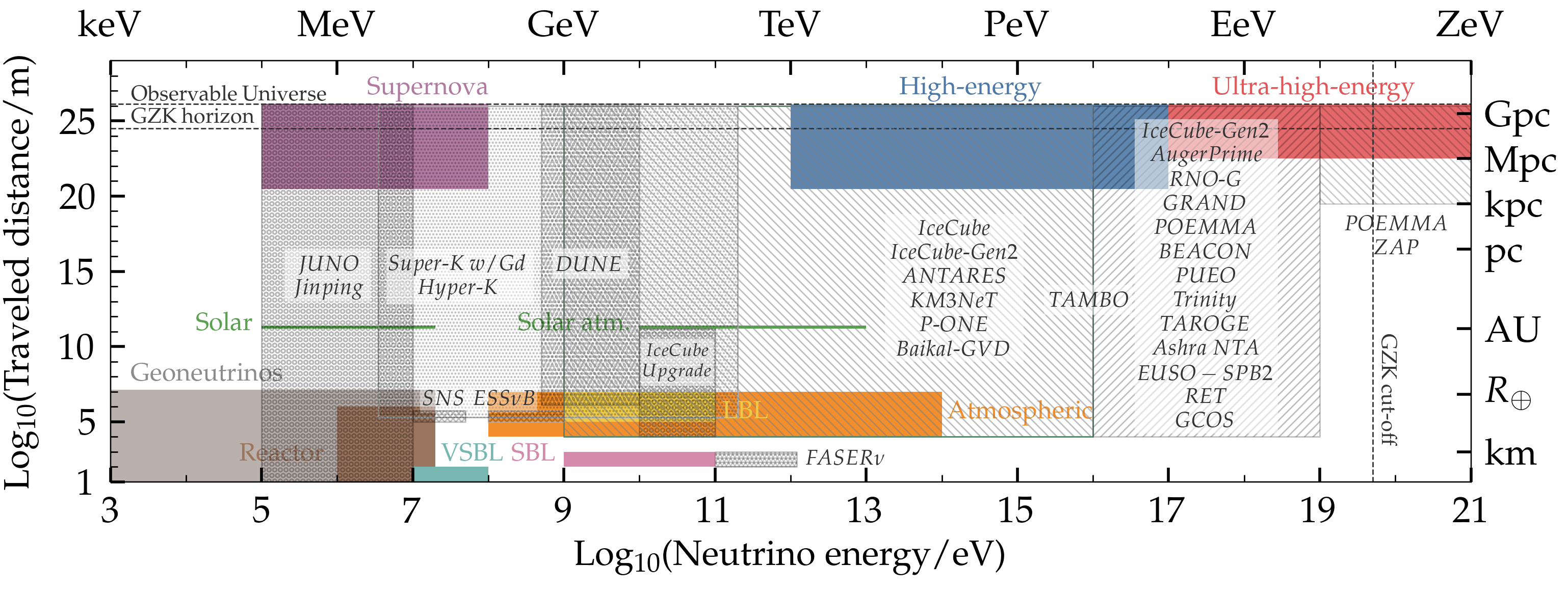}
  \caption{Distribution of neutrino sources in energy and distance traveled to the detector, and experiments aimed at detecting them  
  that are presently in different stages of planning, design, and construction.
  From Ref.~\cite{Ackermann:2022rqc}.}
   \label{fig:scales}
\end{figure}

\paragraph{Neutrino facilities} The rich experimental program of neutrino-detection facilities is encapsulated in Fig.~\ref{fig:scales} and summarized in Ref.~\cite{Ackermann:2022rqc}.  The next decade will result in the construction of multiple high-energy neutrino detectors spanning complementary regions of the sky, with differing sensitivity to different energy ranges between TeV and EeV, and complementary flavor-identification capabilities. 

The neutrino oscillation program is entering a precision era, where the known parameters are being measured with an ever increasing accuracy. The IceCube Upgrade will provide the first precision measurement of the number of tau neutrinos appearing as a result of these oscillations~\cite{Ishihara:2019aao}. A measurement inconsistent with the poorly constrained current theory would be a smoking gun pointing to undiscovered types of neutrinos or to new physics.

\noindent
\begin{minipage}{0.48\textwidth}
The wide range of neutrino energies and traveled distances allow us to explore neutrino properties, their interactions, and fundamental symmetries across a wide breadth of parameter space. 
\vskip20pt

    Since neutrinos are neutral and weakly interacting, they carry information about the physical conditions at their points of origin; at the highest energies, even from powerful cosmic accelerators at the edge of the observable Universe. Due to the fact that they travel unscathed for the longest distances---up to a few Gpc, the size of the observable Universe---even tiny effects can accumulate and become observable. The potential for searches of beyond-SM physics in a wide energy range is illustrated in Fig.~\ref{fig:models} and summarized in Refs.~\cite{Ackermann:2022rqc, Arguelles:2022xxa, Abraham:2022jse}.
\end{minipage}
\noindent 
\hfill
\begin{minipage}{0.48\textwidth}
  \centering
    \captionsetup{type=figure}
  \includegraphics[width=0.9\textwidth]{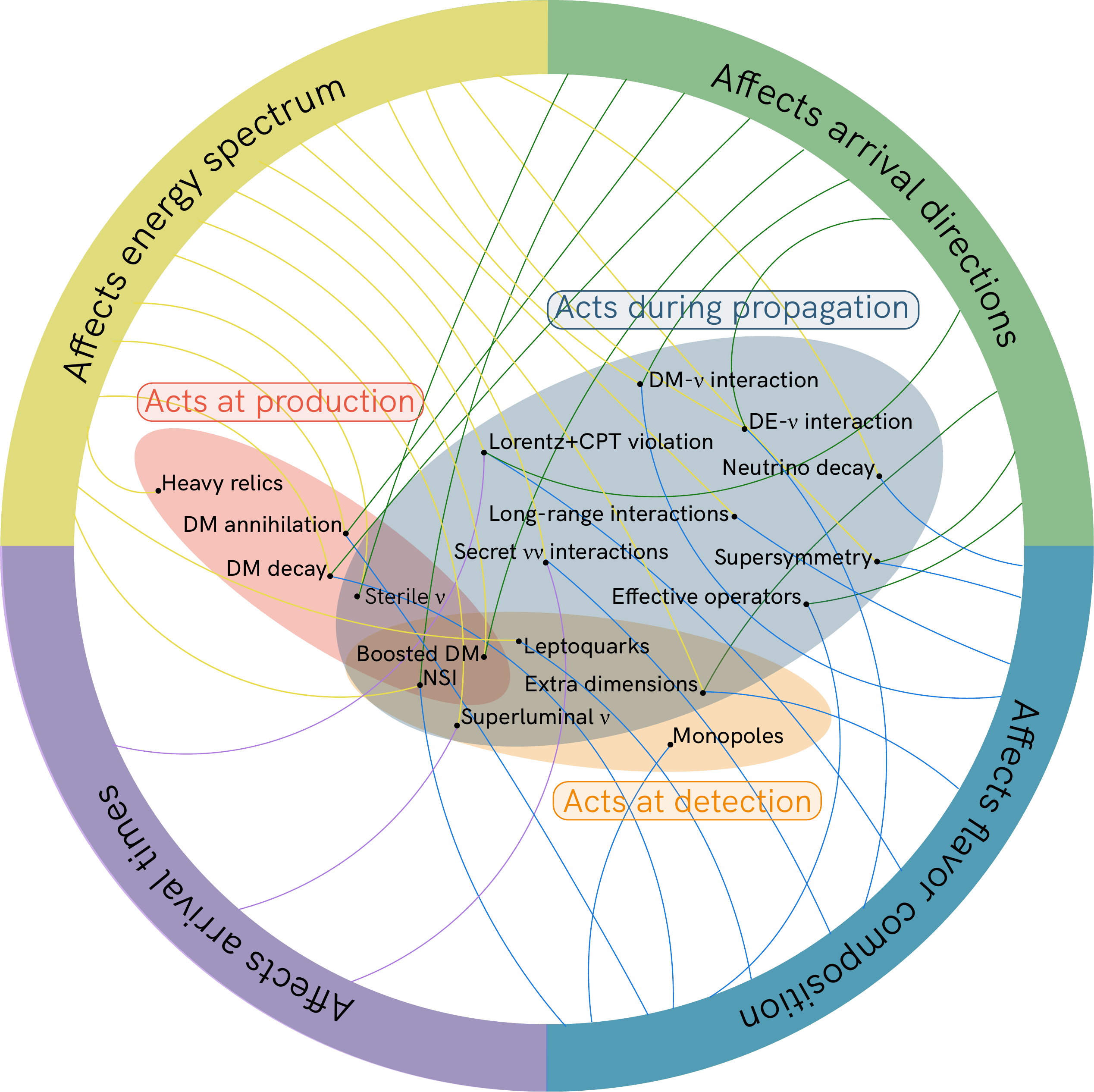}
    \captionof{figure}{\label{fig:models}Models of new neutrino physics and other new physics classified according to the stage at which they act---at production, during propagation, and at detection---and what feature they affect---energy spectrum, arrival directions, flavor composition, and arrival times. From~Ref.~\cite{Arguelles:2019rbn}.}
\end{minipage}

\begin{figure}[!htb]
     \centering
     \includegraphics[width=\textwidth]{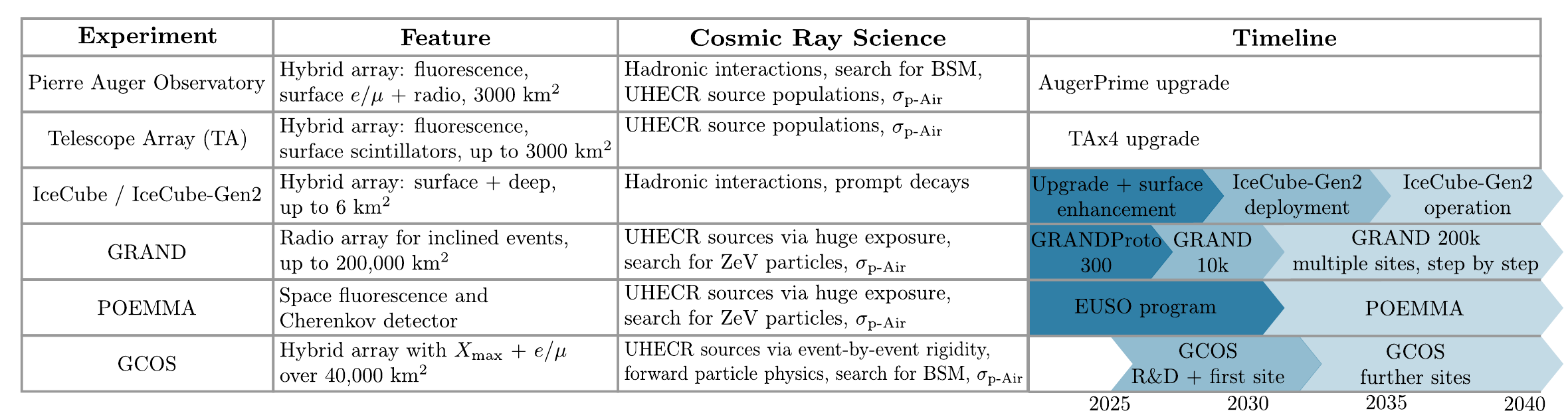}
     \vspace{-2mm}
     \caption{Upgraded and next-generation UHECR experiments with their defining features, scientific goals relevant to the APS DPF, and timeline. From Ref.~\cite{Coleman:2022abf}.}
     \label{fig:UHECRroadmap}
    \vspace{-2mm}
 \end{figure}

\paragraph{Ultra-high-energy cosmic-ray experiments (UHECR)} In the coming decade, UHECR experiments will employ the three major detection techniques: extensive surface detector arrays, high-resolution air-fluorescence detectors, and radio detectors. The UHECR particle physics roadmap is specified in Fig.~\ref{fig:UHECRroadmap} and summarized in Ref.~\cite{Coleman:2022abf}. 


To address the paradigm shift arising from the results of the current generation of experiments, three upgrades are either planned or already underway. TA$\times$4, a 4-fold expansion of the Telescope Array, will allow for Auger-like exposure in the Northern Hemisphere with the aim of identifying (classes of) UHECR sources and further investigating potential differences between the Northern and Southern skies~\cite{TelescopeArray:2021dri}. AugerPrime, the upgrade of Auger, focuses on achieving sensitivity to the cosmic-ray baryonic composition for each shower, measured by its upgraded surface detector through multi-hybrid observations~\cite{PierreAuger:2016qzd}. IceCube-Gen2, IceCube’s planned upgrade, will include an expansion of the surface array to measure cosmic rays with energies up to a few EeV, providing a unique laboratory to study cosmic-ray physics such as the insufficiently understood prompt particle decays in extensive air showers~\cite{IceCube-Gen2:2020qha}. It will also be used to study the transition from galactic to extragalactic sources by combining the mass-sensitive observables of the surface and deep in-ice detectors. The upgrades benefit from recent technological advances, including the resurgence of the radio technique as a competitive method, and the development of machine learning as a powerful new analysis technique.

Looking into the future ahead, the POEMMA mission~\cite{POEMMA:2020ykm} and the multi-site Giant Radio Array for Neutrino Detection (GRAND)~\cite{GRAND:2018iaj} are two instruments that will measure both ultra-high-energy neutrinos and cosmic rays. Thanks to their large exposure, both POEMMA and GRAND will be able to search for UHECR sources and ZeV particles beyond the flux suppression. The Global Cosmic Ray Observatory (GCOS), a $40, 000~{\rm km}^2$  ground array likely split into at least two locations, one or more of them possibly co-located with a GRAND site, will be a purposely built precision multi-instrument ground array~\cite{Horandel:2021prj}. Its design will need to meet the goal of $< 10\%$ muon-number resolution to leverage our improved understanding of hadronic interactions. With these capabilities, GCOS will be able to study particle and BSM physics at the Energy Frontier while determining the cosmic-ray baryonic composition on an event-by-event basis to enable rigidity-based studies of UHECR sources at the Cosmic Frontier.

As we discussed in Sec.~\ref{s:portals}, all of these UHECR experiments will have sensitivity to signals of ssuper-heavy dark matter (SHDM) and macroscopic dark quark nuggets. Indeed, UHECR observatories will offer a unique probe of the dark matter mass spectrum near the GUT scale. The origin of SHDM particles can be connected to inflationary cosmologies and their decay to instanton-induced processes, which would produce a cosmic flux of ultra-high-energy neutrinos and photons. While their non-observation sets restrictive constraints on the gauge couplings of the dark matter models, the unambiguous detection of a single ultra-high-energy photon or neutrino would be a game changer in the quest to identify the dark matter properties. In particular, as we discussed in Sec.~\ref{s:SHDM}, AugerPrime will achieve a world-leading sensitivity to indirect detection of SHDM particles by searching for SHDM decay products coming from the direction of the Galactic center~\cite{PierreAuger:2022wzk}. 
 
 In addition, AugerPrime will provide a unique probe of hadronic interaction models at center-of-mass energies and kinematic regimes
not accessible at terrestrial colliders, as well as high-resolution measurements of the proton-air inelastic cross section $\sigma_{p-{\rm Air}}$. Hadronic interaction models, continuously informed by new accelerator data, play a key role in our understanding of the physics driving the production of extensive air showers induced by UHECRs in the atmosphere. Thanks to ever-more-precise measurements from UHECR experiments, there are now strong indications that our understanding is incomplete. In particular, all of the hadronic models underestimate the number of muons produced in air showers, hinting at new particle physics processes at the highest energies. Reducing the systematic uncertainties between models and incorporating the missing ingredients are major goals at the interface of the field of UHECRs and particle physics. The on-going AugerPrime upgrade will give each surface detector muon separation capabilities, allowing for high precision air-shower measurements connected to the muon puzzle~\cite{Albrecht:2021cxw} and probes of BSM physics; see Sec.~\ref{s:muonpuzzle}. The general strategy to solve the muon puzzle relies on the accurate determination of the energy scale combined with a precise set of measurements over a large parameter space, that can together disentangle the electromagnetic and muon components of extensive air showers. A muon-number resolution of $< 15\%$ is within reach with upgraded detectors in the next decade using hybrid measurements. Achieving the prime goal of $< 10\%$ will likely require a purposely-built next-generation observatory. 

Additionally, as we discussed in Sec.~\ref{s:LIV}, UHECR experiments will provide the most restrictive bounds on violations of CPT and Lorentz invariance. Finally, the identification of the UHECR population could provide a direct probe of the species scale that could rule the cutoff energy of cosmic-ray accelerators~\cite{Montero:2022prj}; see Secs.~\ref{UHECRsources} and \ref{s:swampland}. Altogether, UHECR observatories offer an unparalleled opportunity to address basic problems of fundamental physics. 

\begin{figure}[ht!]
\centering
\includegraphics[width=0.99\textwidth]{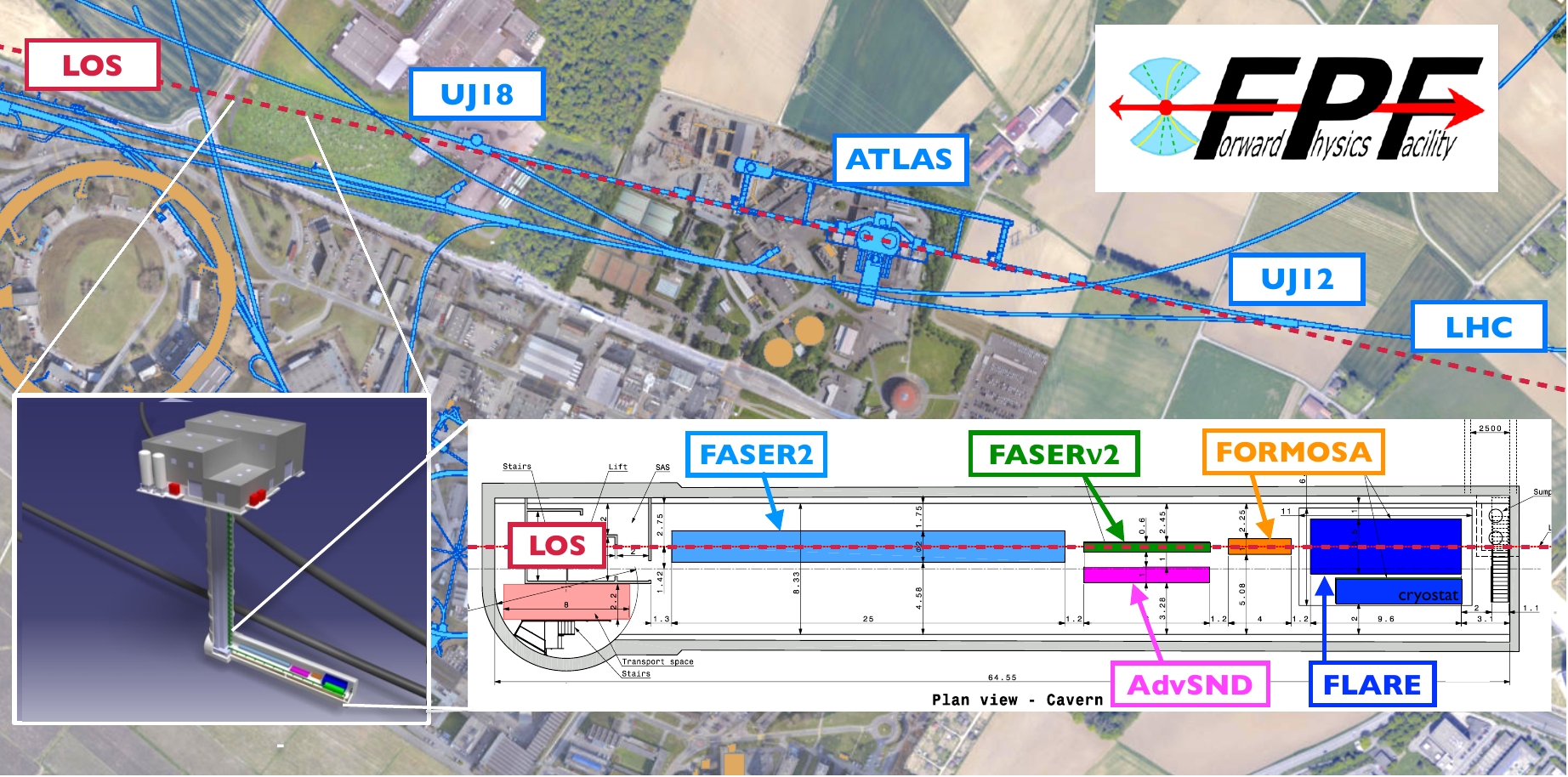}
\caption{The Forward Physics Facility, a proposed new underground cavern located near the LHC tunnel at CERN.  The FPF will house a diverse set of experiments in the far-forward region and will detect TeV-energy neutrinos, constrain forward hadron production, and probe proton and nuclear structure, with synergies with many astroparticle experiments. Adapted from Ref.~\cite{Feng:2022inv}.
}
\label{fig:FPFMap}
\end{figure}

\paragraph{Forward Physics Facility}  The Forward Physics Facility (FPF) is a proposed underground cavern at the Large Hadron Collider (LHC) at CERN that will house a suite of far-forward experiments during the High-Luminosity LHC era from $\sim$ 2030-2042~\cite{Anchordoqui:2021ghd,Feng:2022inv}. The preferred site for the FPF is along the beam collision axis, 617-682 m west of the ATLAS experiment; see Fig.~\ref{fig:FPFMap}.  FPF experiments, such as FASER$\nu$, Advanced SND, and FLArE, will detect $\sim 10^6$ neutrino interactions at TeV energies, filling the gap between current fixed-target accelerator experiments and astroparticle measurements; see Fig.~\ref{fig:scales}.  In addition, the FPF will expand our understanding of proton and nuclear structure and the strong interactions to new regimes, and carry out world-leading searches for a wide range of new phenomena.

The FPF provides opportunities for interdisciplinary studies at the intersection of high-energy particle physics and modern astroparticle physics. Cosmic rays enter the atmosphere with energies up to $10^{11}$ GeV and beyond, where they produce large cascades of high-energy particles. The development of these extensive air showers is driven by hadron-ion collisions under low momentum transfer in the non-perturbative regime of QCD. Measurements at the FPF will improve the modeling of high-energy hadronic interactions in the atmosphere, reduce the associated uncertainties of air shower measurements, and thereby help to understand the properties of cosmic rays, such as their energy and mass, which is crucial to discovering their origin. Moreover, atmospheric muons and neutrinos produced in these extensive air showers in the far-forward region are the main background for searches of high-energy astrophysical neutrinos with large-scale neutrino telescopes, including IceCube and KM3NET. The FPF will help to understand the atmospheric neutrino flux and reduce the uncertainties for astrophysical neutrino searches in the context of multi-messenger astrophysics.

\begin{figure}[thb!]
\begin{center}
    \includegraphics[width=0.50\textwidth]{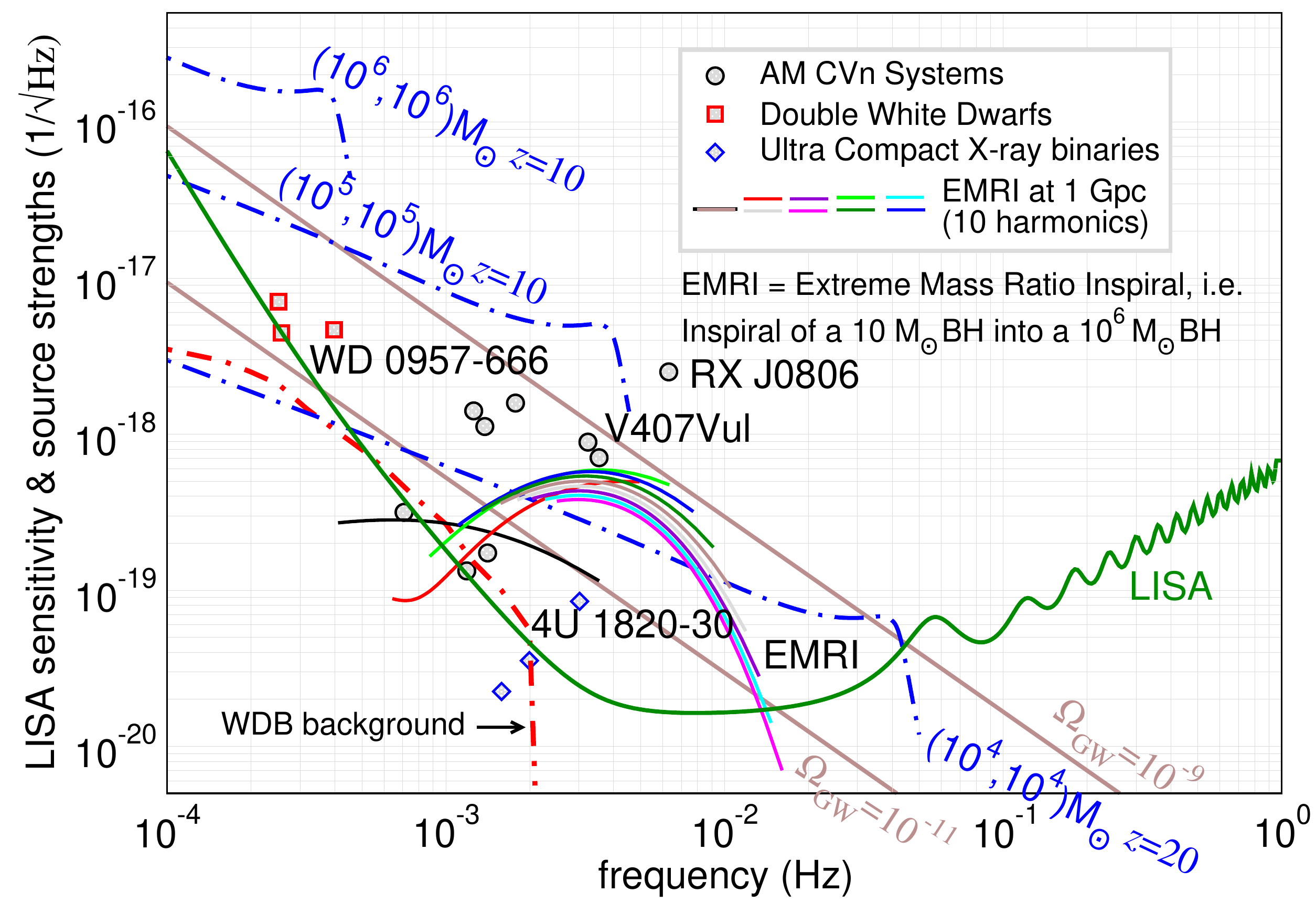}
    \includegraphics[width=0.49\textwidth]{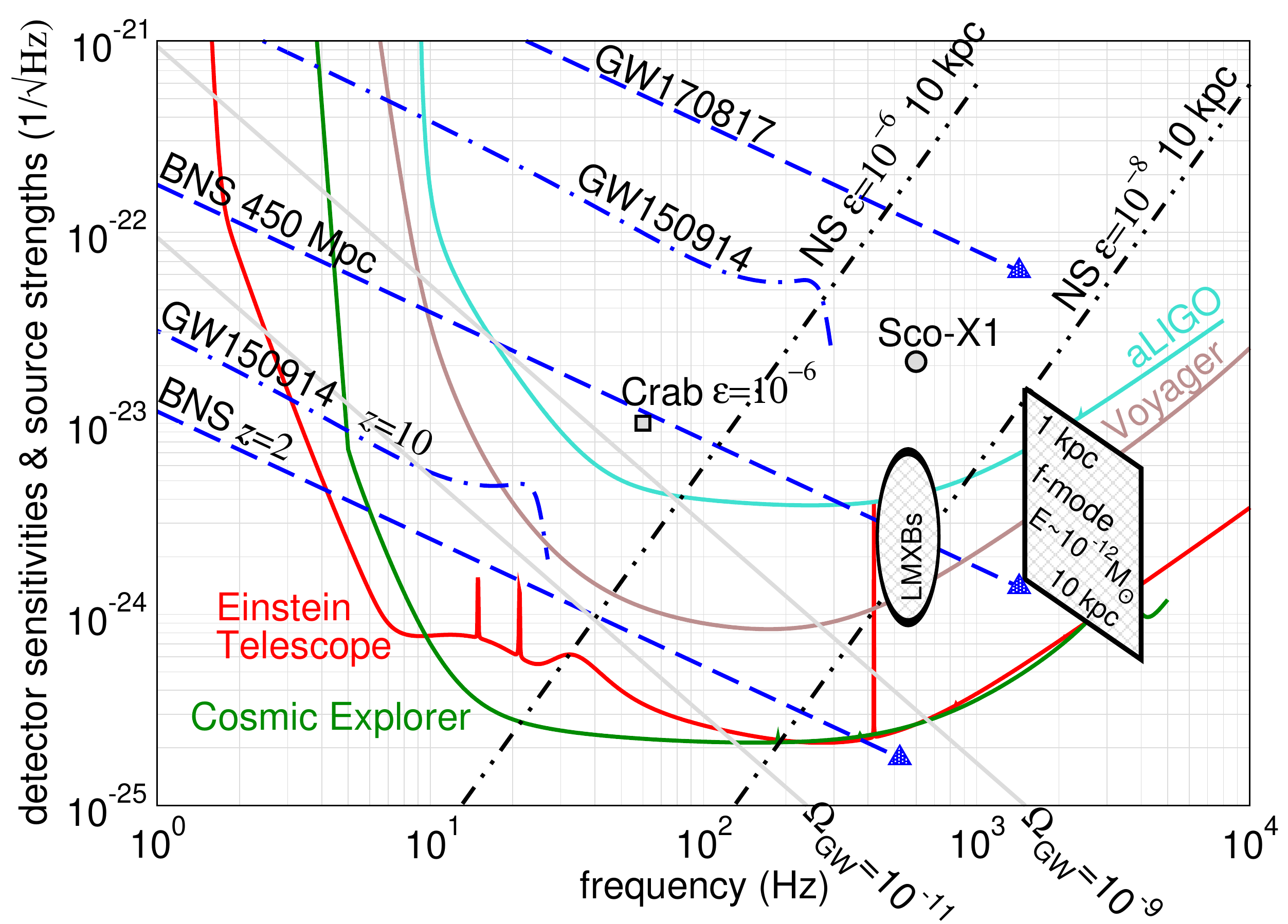}
\end{center}
    
    \caption{Strain sensitivity of various detectors and the expected signal strengths for different classes of sources plotted for the space-based LISA (left panel) and terrestrial detectors (right panel). See text for explanation of various sources plotted on the two diagrams.}
    \label{fig:sense}
\end{figure}
\paragraph{Gravitational-wave facilities} 
Gravitational wave detectors are sensitive to the signal amplitude and not energy or intensity.  Consequently, an increase in the sensitivity of a gravitational-wave detector by a factor of 10 magnifies the accessible volume of the Universe by a factor of 1000 for low redshifts, where the geometry of the Universe is approximately Euclidean. All in all, Cosmic Explorer, conceived to be ten times bigger than Advanced LIGO, can observe essentially the entire universe for mergers of black holes and neutron stars.

The next generation of gravitational-wave observatories can explore a wide range of fundamental physics phenomena throughout the history of the Universe. These phenomena include access to the Universe's binary black hole population throughout cosmic time, to the Universe's expansion history independent of the cosmic distance ladders, to stochastic gravitational waves from early Universe phase transitions, to warped spacetime in the strong-field and high-velocity limit, to the equation of state of nuclear matter at neutron-star and post-merger densities, and to dark matter candidates through their interaction in extreme astrophysical environments or their interaction with the detector itself. A comparison of the strain sensitivity of these proposed detectors is shown in Fig.~\ref{fig:sense} and summarized in Ref.~\cite{Ballmer:2022uxx}. The right plot in Fig.~\ref{fig:sense} shows the sensitivity curves of advanced LIGO (aLIGO) and the next-generation laser interferometers: Cosmic Explorer, LIGO Voyager, and Einstein Telescope. Also shown on the same plot are the spectral densities of typical sources: GW150914 and GW170817 detected by the LIGO-Virgo Scientific Collaboration, binary neutron star (BNS) mergers at 450~Mpc and redshift of 2, GW150914 if it were at $z=10$, the Crab pulsar assuming an ellipticity of $\epsilon = 10^{-6},$ strengths of rotating neutron stars at 10~kpc for ellipticities $10^{-6}$ and $10^{-8},$ the neutron star in the low-mass X-ray binary Sco-X1 and other similar systems in the Galaxy (LMXBs), stochastic backgrounds of flat power spectrum $\Omega_{\rm GW}=10^{-9}$ and $\Omega_{\rm GW}=10^{-11}$, and radiation from quakes in neutron stars that deposit an energy $E \sim 10^{-12}\,M_\odot$ in gravitational waves. The left plot shows the sensitivity curve for the Laser Interferometer Space Antenna (LISA) together with coalescences of supermassive black hole binaries of various masses, inspiral of a 10\,$M_\odot$ black hole into a $10^6\, M_\odot$ black hole at $z=1$ (EMRI), the Galactic white dwarf binary (WDB) background as well as resolvable white dwarf binaries, AM Cn systems, and ultra-compact X-ray binaries. It is assumed that continuous waves from isolated neutron stars, white dwarf binaries, and stochastic backgrounds are integrated for a year, except for Sco-X1, for which an integration time of one week is assumed. Also see Ref.~\cite{Ballmer:2022uxx} for the sensitivity curves of the proposed Neutron Star Extreme Matter Observatory (NEMO) and MAGIS atom interferometers.

\begin{figure}[hbt!]
     \centering
     \includegraphics[width=\columnwidth]{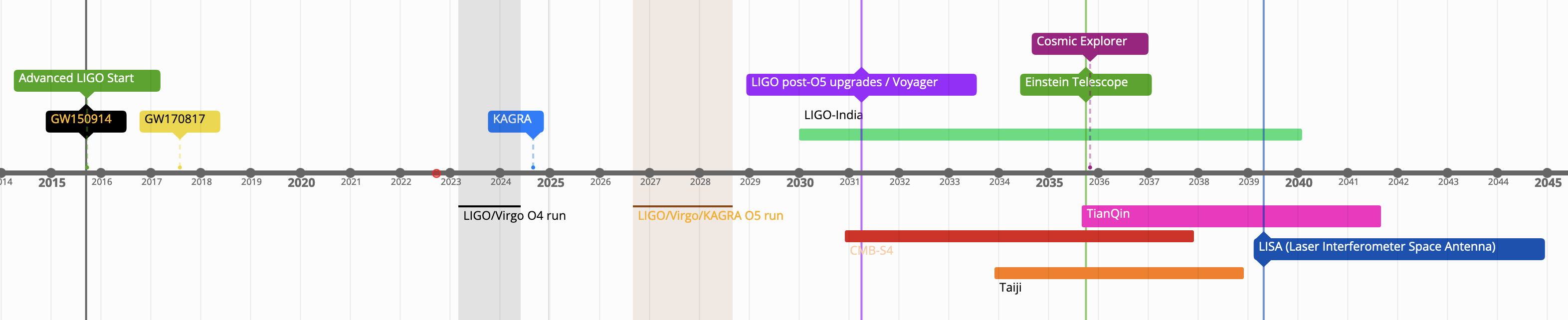}
     \caption[GW Observatories' Timeline]{Timeline of current and proposed GW observatories. The LIGO, Virgo, and KAGRA timelines are taken from the frequently updated joint run planning page~\url{https://observing.docs.ligo.org/plan}.
       The LISA launch date and mission lifetime are taken from the ESA-LISA Factsheet~\url{https://www.esa.int/Science\_Exploration/Space\_Science/LISA\_factsheet}.
     }
     \label{fig:GW-obs-timeline}
 \end{figure}

In the U.S., the proposed Cosmic Explorer observatory is designed to have ten times the sensitivity of Advanced LIGO and will push the reach of GW astronomy towards the edge of the observable Universe (redshift $z \sim 100$)~\cite{Reitze:2019iox, Evans:2021gyd}. Binary neutron star mergers at cosmological distances will  be observable with Cosmic Explorer and LIGO Voyager. A network consisting of Cosmic Explorer in the U.S. and Einstein Telescope in Europe would detect $\sim 10^5$ binary neutron star mergers per year, with a median redshift of $\sim 1.5$ (close to the peak of star formation) and a horizon of $z \sim 9$~\cite{Borhanian:2022czq}. Approximately 200 of these binary neutron stars would be localized every year to better than one square degree, enabling followup with telescopes with small fields of view. The improved low-frequency sensitivity of next-generation detectors allows them to detect and localize sources prior to merger. A rough timeline of the various gravitational-wave detectors is given in Fig.\,\ref{fig:GW-obs-timeline}. The current plan in the US is to maximize the observation in the LIGO Facilities until Cosmic Explorer is observing. While there will be some breaks to further improve the sensitivity, actual observing time will be prioritized in coordination with the other terrestrial detectors of that epoch: LIGO-India, Virgo, and KAGRA.

In order to realize the full potential of current and future observatories improved waveform models would be needed to meet the greater sensitivity of next generation observatories. A new generation of numerical-relativity codes capable of achieving greater accuracy, smaller systematic bias and larger computational speeds, should be developed \cite{Foucart:2022iwu}.  At the same time, it is important to harness analytical tools from high-energy physics, e.g. scattering amplitudes and effective field theory, and develop a framework for computing gravitational-wave signals from binary black holes and neutron stars \cite{Buonanno:2022pgc}. The synergy between the gravitational-wave and high-energy physics communities will help build waveform models that will be more accurate and mitigate systematic bias.

\paragraph{Cosmological probes} The present tensions and discrepancies among different cosmological measurements, in particular the $H_0$ tension as the most significant one, offer crucial insights in our understanding of the Universe. In the near future, we expect precise measurements of the expansion and growth history over a large range of experiments. In Table~\ref{tab:cosmo1}, we provide a list of all these multi-frequency/multimessenger experiments together with the most influential probes and space missions from the last two decades. In Table~\ref{tab:cosmo2}, the experiments are grouped by their ``driving science'': detection of the redshifted 21 cm line in neutral hydrogen, BAO, redshift space distortion (RSD), cosmic chronometers (CC), CMB, distance ladder, fast radio bursts (FRB), GW, quasars, redshift drift, spectral distortions (SDs), supernovae (SNe), time-delay cosmography, time-lag cosmography, varying fundmental constants, and weak lensing (WL). A detailed description of these experiments is provided in Ref.~\cite{Abdalla:2022yfr} and in the CF6 report.

\begin{table}[ht]
\caption{{Cosmological probes. From Ref.~\cite{Abdalla:2022yfr}. }}
\label{tab:cosmo1}
\begin{center}
\scalebox{0.6}{
\begin{tabular}{|c|l|l|l|}
\hline
Acronym & Experiment & Website & Status \\
\hline
4MOST &  4-metre Multi-Object Spectroscopic Telescope & \href{https://www.eso.org/sci/facilities/develop/instruments/4MOST.html}{https://4MOST} & expected 2023\\
ACT & Atacama Cosmology Telescope &
\href{https://act.princeton.edu}{https://act.princeton.edu} & ongoing \\
ANDES & ArmazoNes high Dispersion Echelle Spectrograph & \href{https://elt.eso.org/instrument/ANDES/}{ https://ANDES} & planned\\
ATLAS Probe & Astrophysics Telescope for Large Area Spectroscopy Probe & \href{https://atlas-probe.ipac.caltech.edu/}{https://atlas-probe} & proposed \\
BAHAMAS & BAryons and HAloes of MAssive Systems & \href{https://www.astro.ljmu.ac.uk/~igm/BAHAMAS}{https://BAHAMAS} & 2017-2018\\
BICEP & Background Imaging of Cosmic Extragalactic Polarization & \href{http://bicepkeck.org/}{http://bicepkeck.org} & ongoing
\\
BINGO & Baryon Acoustic Oscillations & \href{https://bingotelescope.org/}{https://bingotelescope.org} & planned
\\[-4pt]
 & from Integrated Neutral Gas Observations & &
\\
BOSS & Baryon Oscillations Spectroscopy Survey & \href{https://cosmology.lbl.gov/BOSS/}{https://BOSS} & ongoing \\
CANDELS &  Cosmic Assembly Near-infrared Deep  &\href{https://www.ipac.caltech.edu/project/candels}{https://candels}&\\[-4pt]
 &   Extragalactic Legacy Survey  & & \\
CCHP & Carnegie-Chicago Hubble Project & \href{https://carnegiescience.edu/projects/carnegie-hubble-program}{https://carnegiescience.edu}&\\
CE & Cosmic Explorer & \href{https://cosmicexplorer.org}{https://cosmicexplorer.org}& planned\\
CFHT & Canada-France-Hawaii Telescope & \href{https://www.cfht.hawaii.edu}{https://cfht.hawaii.edu}& ongoing\\
CHIME & Canadian Hydrogen Intensity Mapping Experiment & \href{https://chime-experiment.ca/en}{https://chime-experiment.ca} & ongoing \\
CLASS & Cosmology Large Angular Scale Surveyor & \href{https://sites.krieger.jhu.edu/class/}{https://class} & ongoing \\
CMB-HD & Cosmic Microwave Background-High Definition & \href{https://cmb-hd.org}{https://cmb-hd.org} & proposed\\
CMB-S4 & Cosmic Microwave Background-Stage IV & \href{https://cmb-s4.org}{https://cmb-s4.org} & planned 2029-2036\\
COMAP & CO Mapping Array Pathfinder & \href{https://comap.caltech.edu}{https://comap.caltech.edu} & ongoing\\
DECIGO & DECi-hertz Interferometer Gravitational wave Observatory & \href{https://decigo.jp/index_E.html}{https://decigo.jp} & planned \\
DES & Dark Energy Survey & \href{https://www.darkenergysurvey.org}{https://darkenergysurvey.org} & ongoing \\
DESI & Dark Energy Spectroscopic Instrument & \href{https://www.desi.lbl.gov}{https://desi.lbl.gov}& ongoing\\ 
dFGS & 6-degree Field Galaxy Survey & \href{http://www.6dfgs.net}{http://6dfgs.net} & 2001-2007\\
eBOSS & Extended Baryon Oscillations Spectroscopy Survey & \href{https://www.sdss.org/surveys/eboss/}{https://eboss} & 2014-2019\\
ELT & Extremely Large Telescope & \href{https://elt.eso.org}{https://elt.eso.org} & planned 2027 \\
ESPRESSO & Echelle SPectrograph for Rocky Exoplanets & \href{https://www.eso.org/sci/facilities/paranal/instruments/espresso.html}{https://espresso.html} &ongoing \\[-4pt]
& and Stable Spectroscopic Observations & &\\
ET & Einstein Telescope & \href{http://www.et-gw.eu}{http://www.et-gw.eu} & planned \\
{\it Euclid} & {\it Euclid} Consortium & \href{https://www.euclid-ec.org}{https://www.euclid-ec.org} & planned 2023 \\
Gaia& Gaia & 
\href{https://sci.esa.int/web/gaia/}{https://gaia} & ongoing\\
GBT& Green Bank Telescope & 
\href{https://greenbankobservatory.org/science/telescopes/gbt/}{https://greenbankobservatory.org} & ongoing\\
GRAVITY& General Relativity Analysis via VLT InTerferometrY& 
\href{https://www.mpe.mpg.de/ir/gravity}{https://gravity} & ongoing\\
GRAVITY+& upgrade version of GRAVITY& \href{https://www.mpe.mpg.de/ir/gravityplus }{https://gravityplus} & planned\\
HARPS & High Accuracy Radial-velocity Planet Searcher & \href{https://www.eso.org/sci/facilities/lasilla/instruments/harps.html}{https://harps.html} & ongoing\\
HIRAX & Hydrogen Intensity and Real-time Analysis eXperiment & \href{https://hirax.ukzn.ac.za}{https://hirax.ukzn.ac.za} & planned \\
HIRES & HIgh Resolution Echelle Spectrometer &\href{https://www2.keck.hawaii.edu/inst/hires/}{https://hires}&ongoing\\
H0LiCOW & $H_0$ Lenses in Cosmograil's Wellspring & \href{https://shsuyu.github.io/H0LiCOW/site/}{https://H0LiCOW} &\\
HSC & Hyper Suprime-Cam & \href{https://hsc.mtk.nao.ac.jp/ssp/survey}{https://hsc.mtk.nao.ac.jp} & finished\\ 
HST & Hubble Space Telescope & \href{https://www.nasa.gov/mission_pages/hubble}{https://hubble} & ongoing\\ 
KAGRA&Kamioka Gravitational wave detector&\href{https://gwcenter.icrr.u-tokyo.ac.jp/en/organization}{https://kagra}&expected 2023\\
KiDS & Kilo-Degree Survey & \href{http://kids.strw.leidenuniv.nl}{http://kids} & ongoing\\
JWST & James Webb Space Telescope & \href{https://jwst.nasa.gov/content/webbLaunch/index.html}{https://jwst.nasa.gov} & ongoing\\
LIGO & Laser Interferometer Gravitational Wave Observatory & \href{https://www.ligo.caltech.edu}{https://ligo.caltech.edu} & ongoing \\
LIGO-India &Laser Interferometer Gravitational Wave Observatory India&\href{https://www.ligo-india.in}{https://ligo-india.in}& planned\\
LiteBIRD & Lite (Light) satellite for the studies of B-mode polarization  & \href{https://www.isas.jaxa.jp/en/missions/spacecraft/future/litebird.html}{https://litebird.html} & planned \\[-4pt]
& and Inflation from cosmic background Radiation Detection & &  \\
LISA & Laser Interferometer Space Antenna & \href{https://lisa.nasa.gov}{https://lisa.nasa.gov} & planned\\
LGWA & Lunar Gravitational-Wave Antenna &\href{http://socrate.cs.unicam.it/index.php}{http://LGWA} & proposed\\
MCT & CLASH Multi-Cycle Treasury & \href{https://www.stsci.edu/~postman/CLASH/}{https://CLASH} & \\
MeerKAT & Karoo Array Telescope & \href{https://www.sarao.ac.za/science/meerkat/}{https://meerkat} & ongoing\\
NANOGrav & North American Nanohertz Observatory for Gravitational Waves & \href{http://nanograv.org/}{http://nanograv.org/} & ongoing\\
OWFA & Ooty Wide Field Array & \href{http://rac.ncra.tifr.res.in/ort.html}{http://ort.html} & planned\\
OWLS & OverWhelmingly Large Simulations & \href{https://virgo.dur.ac.uk/2010/02/12/OWLS}{https://OWLS} &\\
Pan-STARRS & Panoramic Survey Telescope and Rapid Response System & \href{https://panstarrs.stsci.edu}{https://panstarrs.stsci.edu} & ongoing \\
PFS & Subaru Prime Focus Spectrograph & \href{https://pfs.ipmu.jp}{https://pfs.ipmu.jp} & expected 2023
\\
{\it Planck} & {\it Planck} collaboration & \href{https://www.esa.int/Science_Exploration/Space_Science/Planck}{https://www.esa.int/Planck} & 2009-2013
\\
POLARBEAR & POLARBEAR & \href{http://bolo.berkeley.edu/polarbear/}{http://polarbear} & finished \\
PUMA &Packed Ultra-wideband Mapping Array & \href{http://puma.bnl.gov}{http://puma.bnl.gov} & planned
\\
{\it Roman}/WFIRST & Nancy Grace {\it Roman} Space Telescope & \href{http://roman.gsfc.nasa.gov}{http://roman.gsfc.nasa.gov} & planned\\
{\it Rubin}/LSST & {\it Rubin} Observatory Legacy Survey of Space and Time & \href{https://www.lsst.org}{https://lsst.org} & expected 2024-2034\\
SDSS & Sloan Digital Sky Survey & \href{https://www.sdss.org}{https://sdss.org} & ongoing\\
SH0ES & Supernovae $H_0$ for the Equation of State & \href{https://archive.stsci.edu/proposal_search.php?id=10802\&mission=hst}{https://SH0ES-Supernovae} & \\
SKAO & Square Kilometer Array Observatory & \href{https://www.skatelescope.org}{https://skatelescope.org} & planned\\
Simons Array & Simons Array & \href{http://bolo.berkeley.edu/polarbear/}{http://simonarray} & in preparation\\
SLACS & Sloan Lens ACS & \href{https://web.physics.utah.edu/~bolton/slacs/What\_is\_SLACS.html}{https://SLACS.html} & \\
SO & Simons Observatory & \href{https://simonsobservatory.org}{https://simonsobservatory.org} & expected 2024-2029\\
\hline
\end{tabular}}
\end{center}
\end{table}
  
\begin{table}[ht]
\begin{center}
\scalebox{0.6}{
\begin{tabular}{|c|l|l|l|}
\hline
Acronym & Experiment & Website & Status \\
\hline
SPHEREx  & Spectro-Photometer for the History of the Universe, Epoch of Reionization, &\href{https://www.jpl.nasa.gov/missions/spherex}{https://spherex} & expected 2025\\
  &   and Ices Explorer & &\\
SPIDER & SPIDER & \href{https://spider.princeton.edu/}{https://spider} & planned \\
SPT & South Pole Telescope & \href{https://pole.uchicago.edu}{https://pole.uchicago.edu} & ongoing\\
STRIDES & STRong-lensing Insights into Dark Energy Survey & \href{https://strides.astro.ucla.edu}{https://strides.astro.ucla.edu} & ongoing \\
TDCOSMO & Time Delay Cosmography & \href{http://www.tdcosmo.org}{http://tdcosmo.org} & ongoing\\
uGMRT & Upgraded Giant Metre-wave Radio Telescope & \href{https://www.gmrt.ncra.tifr.res.in/}{https://gmrt.ncra.tifr.res.in} & ongoing \\
UNIONS & The Ultraviolet Near- Infrared Optical Northern Survey & \href{https://www.skysurvey.cc}{https://skysurvey.cc} &  \\
UVES & Ultra Violet Echelle Spectrograph & \href{https://www.eso.org/public/teles-instr/paranal-observatory/vlt/vlt-instr/uves/}{https://uves} & ongoing\\
VIKING & VISTA Kilo-degree Infrared Galaxy Survey & \href{http://horus.roe.ac.uk/vsa/}{http://horus.roe.ac.uk/vsa/} & ongoing\\
Virgo & Virgo& \href{https://www.virgo-gw.eu}{https://virgo-gw.eu}& ongoing\\ 
VLA & Very Large Array & \href{https://public.nrao.edu/telescopes/vla/}{https://vla} & ongoing \\
VLBA & Very Long Baseline Array & \href{https://public.nrao.edu/telescopes/vlba/}{https://vlba} & ongoing \\
VLT  &Very Large Telescope & \href{https://www.eso.org/public/teles-instr/paranal-observatory/vlt/}{https://vlt} & ongoing \\
WFC3 & Wide Field Camera 3 & \href{https://www.stsci.edu/hst/instrumentation/wfc3}{https://wfc3} & ongoing \\
WMAP & Wikilson Microwave Anisotropy Probe & \href{https://map.gsfc.nasa.gov}{https://map.gsfc.nasa.gov} & 2001-2010\\
YSE & Young Supernova Experiment & \href{https://yse.ucsc.edu}{https://yse.ucsc.edu} & ongoing \\
ZTF & Zwicky Transient Facility & \href{https://www.ztf.caltech.edu}{https://ztf.caltech.edu} & ongoing \\
\hline
\end{tabular}}
\end{center}
\end{table}

\begin{table*}[ht]
\caption{Cosmological probes grouped by their driven science. From Ref.~\cite{Abdalla:2022yfr}.}
\label{tab:cosmo2}
\begin{center}
\scalebox{0.65}{
\begin{tabular}{|l|l|l|}
\hline
Science & Facilities \\
\hline
21 cm & BINGO, CHIME, GBT, HIRAX, MeerKAT, OWFA, PUMA , SKAO, uGMRT\\
BAO and RSD & 4MOST, BINGO, CHIME, COMAP, DESI, Euclid, HIRAX, PFS, {\it Roman}, {\it Rubin}, SKAO, SPHEREx \\
CC & ATLAS, Euclid, SPHEREx\\
CMB & ACT, BICEP/Keck, CMB-HD, CMB-S4, LiteBIRD, SO, SPT \\
Distance ladder & ELTs, Gaia , GBT, JWST, LIGO, {\it Roman}, {\it Rubin}, VLA, VLBA \\
FRB & CHIME \\
GW & Cosmic Explorer, DECIGO , ET, LGWA, LIGO/Virgo/KAGRA/LIGO-India, LISA, Taiji, TianQin\\
Quasars & GRAVITY+ \\
Redshift drift & ANDES, ELTs, SKAO\\
SDs & SuperPIXIE \\
SNe & {\it Rubin}, {\it Roman}, YSE, ZTF  \\
Time Delay cosmography & Euclid, Pan-STARRS, {\it Roman}, {\it Rubin}, SKAO, ZTF\\
Time Lag cosmography & {\it Rubin} \\
Varying fundamental constant & ANDES, ELTs, ESPRESSO\\
WL & 4MOST, CFHT,DES, Euclid, HSC, KiDS,  Pan-STARRS , {\it Roman}, {\it Rubin}, SKAO, UNIONS\\
\hline
\end{tabular}}
\end{center}
\end{table*}

\section{Connections to other Snowmass Frontiers}
\label{s:opportunities}

Seeking the fundamental nature of matter and associated mysteries bridges the Theory, Accelerator, Energy, Instrumentation, Neutrino, Computational, and Cosmic Frontiers, thus connecting astroparticle physics and accelerator-based particle physics. Ergo, the study of astroparticle physics can have significant implications in the search for physics beyond the SM at the LHC and future colliders. Correspondingly, LHC experiments provide the laboratory for measurements relevant to understand the subtleties of astroparticle physics. We have provided specific examples of this synergy in Secs.~\ref{s:astropart_1} and \ref{s:muonpuzzle}, where we discussed the relation between the Forward Physics Facility (FPF) with neutrino telescopes and cosmic-ray observatories. This has been discussed in more detail in Refs.~\cite{Feng:2022inv, Coleman:2022abf}. All of these specific examples are, of course, also related to the Computational Frontier as explained in Ref.~\cite{Coleman:2022abf}. 

There is also a strong synergy between cosmological and laboratory searches in new physics~\cite{Abazajian:2022ofy}. The relation between the Cosmic and Neutrino Frontiers has been emphasized in Sec.~\ref{s:nuosc}, with typical examples ranging from measurements of neutrino oscillation parameters to understanding the properties of neutrino masses~\cite{Abraham:2022jse, Ackermann:2022rqc, Arguelles:2022xxa} and bounds on the neutrino mass sum inferred from cosmological observations~\cite{Abdalla:2022yfr}.

The phenomenological implications of the Swampland program provide a strong connection between the Theory and the Cosmic Frontiers. The swampland conjectures seem to pose an interesting challenge for inflation, particle phenomenology, and the cosmological hierarchy problem. In Sec.~\ref{s:swampland}, we briefly related some of these topics, which are discussed at length in Refs.~\cite{Abdalla:2022yfr, deRham:2022hpx, Achucarro:2022qrl}. At the same time, the interface between early Universe cosmology and fundamental theories of particle physics ties up the Cosmic and Energy Frontiers~\cite{Flauger:2022hie, Agrawal:2022rqd}. Finally, searches for signals of particle dark matter and light relics provide the connector between the Accelerator, Energy, and Cosmic Frontiers~\cite{Feng:2022inv, Dienes:2022zbh, Dvorkin:2022jyg, Ando:2022kzd}.

Beyond the fundamental scientific complementarity between studying terrestrial and astroparticle physics, there is a deep connection between the instrumentation built for each subfield. The future of multimessenger astrophysics hangs on the development of new gamma-ray detector technology in the MeV and GeV range because all of the modern multimessenger co-detections involve gamma-rays in this range, and no new, long-term facilities are currently planned to replace those we may lose over the next decade. Gamma-ray detectors, discussed in Sec.~\ref{s:experiments} and further in  Refs.~\cite{Engel:2022yig,Engel:2022bgx}, are developed and built using techniques, materials, and understanding from particle physics, and vice versa. The instrumentalists who build each type of detector move fluidly from one field to the other, giving particle physicists the opportunity to work on smaller experiments in astroparticle physics, and passing new technology development back into larger particle-physics experiments. Due to the interconnected nature of gamma-ray and collider detector technology development, this is a key area for collaborative investment across agencies in the coming decade, and could define multimessenger astrophysics for many decades to come. 

\section{Diversity, Equity, Inclusion, and Accessibility}
\label{s:DEIA}

Peak scientific excellence for any country begins and develops in key government investment targeted at actions available today to produce tangible advancement tomorrow. While we often think of these investments in terms of technology development, flagship facilities, and the returns they enable, as a nation, we cannot afford to overlook investment in our scientific workforce. While this conversation must include broader generational aspects of fair compensation, scientific literacy in public education, and reasonable access to higher education without a lifetime of debt, there are also specific barriers to some people whose abilities would benefit national excellence in science, who are too often excluded for untenable reasons, collectively referred to as Diversity, Equity, Inclusion, and Accessibility (DEIA).

Diversity, equity, inclusion, and accessibility are  fundamental elements of a modern and innovative workplace, school, or team. Cosmic Probes of Fundamental Physics and the Cosmic Frontier, more generally, are well poised to bring new ideas from the extensive literature on DEIA topics to the fore of the broader physics community. It is necessary to be mindful of cultural bias and the impact of personal experiences on student recruiting, training and opportunities, as well as on the retention of more senior trainees and experts. Prioritizing the recruitment and retention of bright minds over the shape or color of the body they come in or their socioeconomic background is of vital and imminent importance to innovation and excellence. This is an argument for providing the educational, mentoring, and community support to individuals in achieving their highest potential because that raises, rather than lowers the bar for academic, scientific, and competitive achievement for the nation as a whole. 

For funding agencies, one of the most impactful changes that should be made over the next decade is to keep track of demographic information for collaborations they fund and for PIs (and other key leadership roles) specifically. That demographic information should include gender and ethnicity at a minimum, but may also include career stage, sexuality, institution type, and other items they might find relevant. The idea here being that there is demonstrated gender bias in awards and leadership roles as well as suspected racial bias, especially in small and mid-sized awards, which is where an earlier career person might start to build up their grant portfolio. Keeping those statistics and making them public in aggregate will allow the community to push for other changes and measure if they are effective, e.g., dual anonymous reviews or inclusion of DEIA service in science grants. At present, DEIA considerations are not a component of funding decisions for the grant host, but they sneak into the results through bias prone review processes. Advisors who tell their students for their (the students') own benefit not to spend time on activities that will not pan out as part of their career are not categorically in the wrong. The system they are advising for needs to support the work of tracking, studying, and supporting DEIA goals, so we ask for that support. The first step is the most important: track the demographics to map where the money goes. In the future, we hope to see hiring and performance reviews for scientists and researchers to include a component evaluating their service to DEIA in the same way that we often consider other services to the community, like mentoring and serving as a reviewer. 

Recommendations for building a culture of equitable access and success for marginalized members in today’s Particle Physics Community have been presented in various Snowmass whitepapers and seminars~\cite{Engel:2022yig, Assamagan:2022ztm, Georgi:2022jfv}.

\clearpage

\bibliographystyle{JHEP}
\bibliography{cf7}

\end{document}